\documentclass{aa}  
\usepackage{graphicx}
\usepackage{txfonts}
\usepackage{natbib}

\bibpunct{(}{)}{;}{a}{}{,}
\newcommand{\fma}[1]{\mbox{$#1$}}
\newcommand{\ltsim}{\raisebox{-0.5ex}{$\;\stackrel{<}{\scriptstyle \sim}\;$}}
\newcommand{\gtsim}{\raisebox{-0.5ex}{$\;\stackrel{>}{\scriptstyle \sim}\;$}}
\newcommand{\unit}[1]{\ifmmode \:\mbox{\rm #1}\else \mbox{#1}\fi}
\newcommand{\mone}{\fma{^{-1}}}

\newcommand{\ha}{H$\alpha$}
\newcommand{\hb}{H$\beta$}
\newcommand{\hy}{H$\gamma$}
\newcommand{\hd}{H$\delta$}
\newcommand{\he}{H$\epsilon$}
\newcommand{\hax}{H$\alpha$~}
\newcommand{\hbx}{H$\beta$~}
\newcommand{\hyx}{H$\gamma$~}
\newcommand{\hdx}{H$\delta$~}
\newcommand{\hdabs}{H$\delta_{abs}$~}
\newcommand{\hahb}{\ha/\hb}
\newcommand{\hahbx}{\ha/\hb~}
\newcommand{\ewha}{\ensuremath{\mathrm{EW}_{\mathrm{H}\alpha}}}
\newcommand{\ewhaem}{\ensuremath{\mathrm{EW}_{\mathrm{H}\alpha,em}}}
\newcommand{\ewhd}{\ensuremath{\mathrm{EW}_{\mathrm{H}\delta}}}

\newcommand{\ewhax}{\ensuremath{\mathrm{EW}_{\mathrm{H}\alpha}}~}
\newcommand{\ewhaemx}{\ensuremath{\mathrm{EW}_{\mathrm{H}\alpha,em}}~}
\newcommand{\ewhdx}{\ensuremath{\mathrm{EW}_{\mathrm{H}\delta}}~}
\newcommand{\ewhdabsx}{\ensuremath{\mathrm{EW}_{\mathrm{H}\delta,abs}}~}
\newcommand{\ewhyx}{\ensuremath{\mathrm{EW}_{\mathrm{H}\gamma}}~}
\newcommand{\hi}{H~{\sc i}}

\newcommand{\kms}{\unit{km~s\mone}}

\newcommand{\m}{$\cal M$}
\newcommand{\msun}{$\cal M$$_\odot$}
\newcommand{\msunyr}{$\cal M$$_\odot$~yr\mone}

\newcommand{\ml}{$\cal M$/$L$}
\newcommand{\mlx}{$\cal M$/$L~$}

\newcommand{\mlvx}{$\cal M$/$L_V$~}

\begin{document}
   \title{Local starburst galaxies and their descendants}

   \subtitle{Statistics from the Sloan Digital Sky Survey}

   \author{%
          Nils Bergvall\inst{1}  \and
          Thomas Marquart\inst{1} \and
          Michael J.~Way\inst{1,2} \and
	  Anna Blomqvist\inst{1} \and
          Emma Holst\inst{1} \and
          G\"oran \"{O}stlin\inst{3} \and
          Erik Zackrisson\inst{1}
          }

   \offprints{N.~Bergvall}

   \institute{
     Department of Physics and Astronomy, Uppsala University,
     Box 515, SE-751~20 Uppsala, Sweden\\
     \email{nils.bergvall@astro.uu.se}
     \and
     NASA Goddard Institute for Space Studies, 2880 Broadway, New York, New York,
     10029, USA
     \and
     Department of Astronomy, Stockholm University, SE-106~91 Stockholm, Sweden.
}

   \date{Received ; accepted }

  \abstract
  % context heading (optional) {} leave it empty if necessary  
   {}
 % aims
   {Despite strong interest in the starburst phenomenon in extragalactic
   astronomy, the concept remains ill-defined. Here we use a strict definition
   of ``starburst" to examine the statistical properties of starburst galaxies
   in the local universe. We also seek to establish links between starburst galaxies, post-starburst (hereafter postburst) galaxies and active galaxies.}
 % methods   
 {Data were selected from the Sloan Digital Sky Survey
 DR7. We
 apply a novel method to treat dust attenuation and derive star
 formation rates, ages and stellar masses assuming a two-component stellar population model. Dynamical masses are calculated from the width of the \hax line. These masses agree excellently with the photometric masses. The mass (gas+stars) range is $\sim$ 10$^9$--10$^{11.5}$ \msun. As a selection criterion for starburst galaxies we use the birthrate parameter, $b$=SFR/$<$SFR$>$, requiring that $b$$\geq$3. For postburst galaxies we use the equivalent width of \hdx in absorption, with the criterion \ewhd$_{,abs}$$\geq$6\AA.}
 %results 
 {We find that only 1\% of star-forming galaxies are starburst galaxies. They contribute 3-6\% to the stellar production and are therefore unimportant for the local star formation activity. The median starburst age is 70 Myr, roughly independent of mass, indicating that star formation is
 mainly regulated by local feedback processes. The $b$-parameter strongly depends on burst age. Values close to $b$=60 are found at ages $\sim$10 Myr, while almost no starbursts are found at ages $>$ 1 Gyr. The median baryonic burst mass fraction of sub-$L^*$ galaxies
 is 5\% and decreases slowly towards high masses. The median mass fraction of the
 recent burst in the postburst sample is 5-10\%. A smaller fraction of the
 ``postburst" galaxies, however, originates in non-bursting galaxies.  The age-mass distribution of the postburst progenitors (with mass fractions $>$3\%) is bimodal with a break at log$\cal M$(\msun)$\sim$10.6 above which the ages are doubled. The starburst and postburst luminosity functions (LFs) follow each other closely until $M_r$$\sim$-21, when AGNs begin to dominate. The postburst LF continues to follow the AGN LF while starbursts become less significant. This suggests that the number of luminous starbursts is
 underestimated by about one dex at high luminosities, because of having large amounts of dust and/or being
 outshone by an AGN. It also indicates that the starburst phase preceded the AGN phase. Finally, we look at the conditions for global gas outflow caused by stellar feedback and find that massive starburst galaxies are susceptible to such outflows.}
 %conclusions 
   {}

   \keywords{galaxies: evolution -- galaxies: luminosity function, mass function -- galaxies: starburst -- galaxies: star formation -- galaxies: statistics -- galaxies: stellar content}

   \maketitle
%
%________________________________________________________________

\section{Introduction}
\label{sec:intro}

The starburst concept was established about three decades ago
\citep{1980ApJ...238...24R,1981ApJ...248..105W}. Originally it concerned
nuclear starbursts, but later, in connection to the results from
objective--prism surveys, a ``starburst galaxy" came to refer to global
starbursts in sub--L$^*$ galaxies. Today, one paper per day or 5\%
of all extragalactic papers in refereed journals contain the word ``starburst"
in the abstract. However, despite this seemingly very familiar concept, we still
lack a commonly accepted definition of what a starburst is. This is a bit
worrisome since starbursts are associated with several important processes in
the evolution of a galaxy. Among these are the ignition of activity in AGNs
\citep{1988ApJ...325...74S,2003MNRAS.346.1055K,2004ASPC..320..205K}, an intense
production of super star clusters
\citep{1994ApJ...433...65O,1995AJ....110.2665M,1995ApJ...446L...1O,1996ApJ...466L..83H,2003MNRAS.343.1285D,2010MNRAS.407..870A}
and morphological transformations due to rapid gas consumption and
stellar and/or AGN driven global superwinds
\citep{1985Natur.317...44C,1990ApJS...74..833H,2005ApJ...620L..79S,2005ApJ...635L..13S,2015A&A...576L..13B}.
It has been suggested that starburst galaxies are also responsible for opening channels
for Lyman continuum radiation to leak out and contribute to the cosmic
reionisation at redshifts below z$\sim$11
\citep[e.g.][]{1990ApJ...350....1M,1990ApJ...348..371S,1991ApJ...376L..33M,2011ApJ...730....5H}. 

There is a problem, however, in the sense that inconsistencies in the definition
of a starburst have led to considerable confusion in the field. For example, at
low redshifts a high \hax emission line equivalent width, and at high redshift a high star formation
rate (hereafter SFR) are often taken as evidence for starbursts, although this
is not necessarily the case  \citep[see below
and][]{2007ApJ...660L..43N,2007A&A...468...33E,2007ApJ...670..156D,2010ApJ...714L.118D}.
This problem can only be remedied by applying more accurate constraints on the
definition of starbursts. The present investigation adds to several previous
efforts by other groups
\citep[e.g.][]{2003MNRAS.341...33K,2004MNRAS.351.1151B,2006ApJ...642..702K,2006MNRAS.370.1677O,2006ASPC..352..225B,2008MNRAS.385.1903L,2009ApJ...698.1437K,2009ApJ...692.1305L,2012MNRAS.426..549S}
to investigate the galaxy content in the local universe and derive a more
consistent view on starbursts from various aspects. Here we formulate our
favoured definition of a starburst that we then apply to galaxies in the
Sloan Digital Sky Survey (SDSS). Our definition is similar to that used in other studies but differs mainly in the way we apply it to the data. Throughout
this paper we assume $\Omega_\Lambda=0.7,\Omega_m=0.3$ and $h=0.7$.

\medskip

The main objectives of the present investigation are to:

\begin{enumerate}
\item identify starburst and post-starburst (hereafter called postburst) galaxies and derive their luminosity functions (hereafter LFs),
\item derive lifetimes of the bursts and masses of the burst and `host' components,
\item examine the evolutionary link between starburst and postburst galaxies,
\item have a closer look at the relation between starburst galaxies and galaxies with AGNs.
\end{enumerate}

Several statistical
investigations of SDSS star-forming galaxies have already been published
\citep[e.g.][]{2003ApJ...599..971H,2003MNRAS.341...33K,2003MNRAS.341...54K,2003ApJ...584..210G,2004MNRAS.355..874N,2004MNRAS.351.1151B,2007MNRAS.381..263A,2007ApJS..173..267S,2008ApJ...681.1183K,2010MNRAS.408.2115M,2012ApJ...754..144T,2013ApJ...776...63F,2013ApJ...778...10S,2014A&A...561A..33I},
and we discuss how some of these results relate to this work. We focus on
galaxies of low to intermediate masses, below L$^*$, but will also have a brief look at the relation between AGNs, postbursts, and starburst galaxies at high luminosities.

\section{Sample selection}
\label{sec:selection}

Our study concerns the statistical properties of starburst and postburst galaxies selected from the SDSS data release no. 7 (DR7). Most of the galaxies are at low redshifts, z$<$0.15. Below we describe the selection criteria, the spectral modelling, and its limitations, as well as potential problems with biases and aperture effects.

\subsection{Defining a starburst}
\label{subsec:starburst}

We begin this discussion with an effort to lay down a definition of a starburst that can be used to quantitatively investigate its impact on galaxy evolution. How to define a starburst has been widely discussed over the years \citep[see e.g.][and references therein]{2009ApJ...698.1437K}. Firstly, starbursts have high {\sl star formation efficiencies} and consume their fuel faster than normal galaxies \citep[see e.g.][]{1998ARA&A..36..189K}. Therefore, to qualify as a starburst, the global gas consumption time scale has to be significantly shorter than a Hubble time. This requirement is generally accepted by the astronomical community. The problem appears in the next step. Imagine an evolved galaxy with very low gas content. Even if this gas is consumed in star formation during a short period, we would hardly regard it as a starburst galaxy unless it was not spectacular in some sense. Therefore, in the definition of starburst galaxies, we must also raise the question of the importance of starbursts for the evolution of a galaxy. It may be a fairly simple procedure to derive the SFR and find out if it is extreme enough to qualify as a starburst, but what we also want to know from a galaxy evolutionary aspect, is how much gas is consumed during the starburst phase and if the starburst is powerful enough to produce a massive blowout of the gas in the central regions. This information cannot be obtained from one parameter, and we explain below how we chose to work on this problem.

A starburst leads to an increase in the SFR that is significantly higher than the mean past SFR. From the literature we find that there are strongly different opinions about what this means. To make our way of reasoning clear, we make an effort to classify different starburst phenomena from our point of view:

\begin{itemize}
\item{If star formation in a galaxy is significantly (at least a factor of three)  higher than in the past, we call the phenomenon a ``global starburst". }
\item{If the increased star formation activity affects a local region of the galaxy with a size less than a few 100 pc, such as in 30 Dor of the LMC, we call it a ``local starburst region".}
\item{A special type of starburst occurs in the centres of more massive galaxies, typically with baryon masses exceeding 10$^{10}$ \msun.  This type of starburst is called a ``nuclear starburst", or sometimes a ``circumnuclear starburst" \citep{1981ApJ...243..756B,1983ApJ...268..602B}. It is restricted to a region in the centre with a size of $<$ 1 kpc and is thought to be triggered by accretion of inflowing gas driven by bars or non-axisymmetries in the disc \citep{2014ASPC..480..211C}. A prototype of such a galaxy is NGC7714 \citep{1981ApJ...248..105W}.} 
\end{itemize}

Now we have {\sl qualitatively} defined starburst galaxies, but how can we {\sl quantify} the concept, such that we can apply useful criteria in the selection of starburst galaxies for a statistical study? Historically, the first use of the designation {\sl starburst} referred to quite a dramatical global increase in SFR. In the optical, starbursts were associated with the Haro, Markarian or Zwicky compact galaxies, among others \citep[e.g][]{1983ApJ...268..602B,1989ApJS...70..447S,1991A&AS...91..285T,2000A&ARv..10....1K,2003ApJS..147...29G}. In the infrared the focus was on luminous infrared galaxies (LIRGs) or ultraluminous infrared galaxies (ULIRGs) \citep{1985MNRAS.214...87J,1996ARA&A..34..749S,1996SSRv...77..303M}, all being massive galaxy mergers in an advanced stage. Today the starburst concept has become watered--down and is even used sometimes for late type galaxies in general. Even the well known `starburst' galaxy NGC 4038--4039, ``The Antennae", just barely qualifies as a global starburst if we use the soft starburst criterion based on the birthrate parameter discussed below \citep{1998ApJ...504..749N,2001ApJ...548..172G}. 

The character of the starburst also changes with galaxy mass. It has become evident that global starbursts in massive galaxies are rare. Starbursts in these galaxies are almost exclusively of the ``nuclear starburst" type and involve a moderate amount of the total gas mass. Exceptions are the LIRGs and ULIRGs. The high dust obscuration of the latter makes it difficult to conclude if these galaxies are powered by starbursts or AGNs without supplementing the optical observations with observations in other wavelength bands. In some cases, however, \citep[e.g.][]{1998ApJ...507..615D}, it can be concluded that strong starbursts are active in the central regions. As we go towards higher luminosities, the dominance of AGNs as the major power sources increases. However, we will argue below however that AGNs and starbursts are always tightly connected.

The problem with applying corrections for dust obscuration in star-forming galaxies based on data from the optical region is that it increases continuously with mass, and gradually becomes severe. This is another reason why we focus our discussion on galaxies with luminosities around L$^*$ or fainter.  We quantify the influence of dust as a function of mass below.

Here we refer to the outcome of what we call
``our model". This is described in more detail in Sect. \ref{model}. In our model we assume that there are two stellar populations -- a {\sl burst}
population and an older {\sl host} population. Among the properties we
derive are the ages and masses of these two populations. The stellar
population spectra used as input to the model were obtained from an in-house spectral evolutionary model (SEM) by
\citet{2001A&A...375..814Z}.  It has a stellar mass range of 0.08--120 \msun,
adopts a Salpeter mass function and includes a nebular component based on
CLOUDY \citep{1996hbic.book.....F, 1998PASP..110..761F}.  Reliability issues
related to the modelling are discussed in the $Appendix$ at the end of the
paper.

To quantify the strength of the starburst we choose the often used {\it birthrate parameter} \citep{1983ApJ...272...54K},

\begin{equation}
$b$=SFR/<SFR>
\end{equation}

\noindent i.e. the ratio between the present SFR and the mean SFR over the
lifetime of the galaxy. This is our primary criterium to characterize the
star formation activity. We call a galaxy with $b\geq$3 a {\sl starburst
galaxy}. The $b$-parameter should not be confused with the {\it burst strength,
``b"}, defined by \citet{1978ApJ...219...46L} which gives the ratio of the burst
mass over the mass of the old population. This is an alternative
characterization of a starburst and we discuss a similar parameter below,
the {\it burst mass fraction} or $burst$ $strength$, $f_{burst}$=${\cal M}_{burst}/{\cal M}_{tot}$.

The SFR used to calculate the $b$-parameter is derived from the \hax flux. Using the relation obtained from \citet{2001A&A...375..814Z}, assuming a 20\% solar metallicity and constant SFR, we can derive the SFR of a stellar population with a Salpeter initial mass function (hereafter IMF): 

\begin{equation}
SFR_{Salpeter} = \frac{L(H\alpha)}{1.51\times10^{34}} {\cal M_{\odot} }yr^{-1}
\end{equation}

\noindent where the \hax luminosity $L$(\ha), after correction for dust
attenuation, is given in Watts. The SFR derived from this relation is 16\% lower than what is obtained from \citet{1998ARA&A..36..189K}. We assume that our relation is valid for all galaxies, thereby disregarding metallicity dependence and possible fluctuations in the SFR which we consider to have marginal influence on the result.

As was shown by \citet{2001ApJ...550..212B} and
given more support by \citet{2007ApJS..173..315S} and
\citet{2008AJ....136.2648D}, spectrophotometric masses based on a pure Salpeter
mass function results in a {\m}/L systematically higher than a {\m}/L based on
maximum disc fits to galaxy rotation curves. The Kroupa IMF
\citep{2001MNRAS.322..231K} seems to give better agreement
\citep{2008AJ....136.2648D}. Therefore it was proposed to modify the results
based on the Salpeter IMF by simply multiplying the derived masses by a
factor of 0.7, corresponding to an IMF with a reduced number of stars below
0.35\msun. As we show below in Sect. \ref{masses}, this modification is
also supported by our data. During the rest of the paper we adopt this so
called ``diet" Salpeter IMF as we calculate masses, \mlx ratios and SFRs:

\begin{equation}
SFR = SFR_{Salpeter}\times0.7
\end{equation}

The birthrate parameter is our main tool to define what a starburst galaxy
is. A strong starburst in the ``classical" sense would have $b$$\gtrapprox$10.
Such cases exist in the local universe but are rare objects  \citep[see
e.g.][]{o01}. Note that the more massive low--redshift starburst galaxies seem
extremely rare. We only find one ULIRG within 100 Mpc -- Arp 220. Within this
volume we find of the order of one hundred thousand normal galaxies. 

The starburst criterion we use here, $b$$\geq$3, was also discussed e.g. in the investigation of star-forming galaxies in the SDSS by \citet{2004MNRAS.351.1151B} and also in
\citet{2003MNRAS.341...33K} in a study of descendants of starburst galaxies. A similar approach is found in the recent analysis of Herschel data (see below). We comment on their results in relation to ours below. In addition to the $b$--parameter we also discuss the mass fraction of the burst population, $f_{burst}$. This parameter is closely related to the strength of the \hdx absorption lines in postburst galaxies which is discussed below.

\subsection{Connecting starbursts with their postburst descendants}
\label{subsec:connecting}

A few 10$^7$ years after a major starburst epoch has ceased, the ageing population of the starburst will produce a characteristic postburst spectral signature with strong Balmer lines in absorption.  We use this signature to define our postburst sample. The only constraint we use here is that the equivalent width of \hdx in absorption is \ewhd$<$--6\AA, the negative sign indicating that we are dealing with an absorption line (henceforth we use this convention: negative sign means absorption line and positive sign means emission line).  We want to point out however that this criterion does not guarantee that the progenitor was a starburst. A galaxy with a constant SFR and $b\sim$1 lasting a few 100 Myrs can also produce a postburst spectrum. 

\begin{figure}[t!] 
  \centering
 \resizebox{\hsize}{!}{\includegraphics{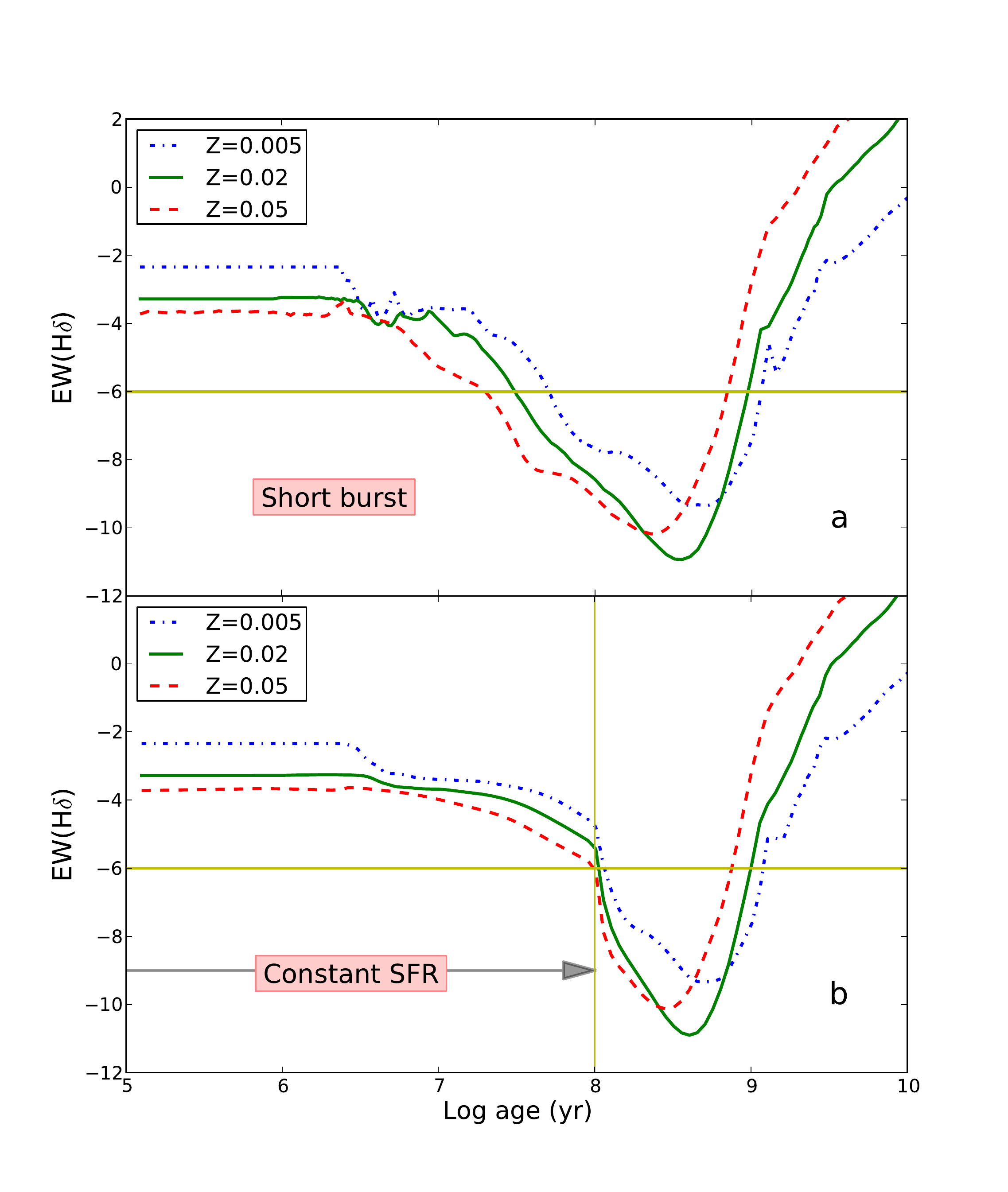}}
  \caption{\ewhdx in absorption as function of burst age in two stellar populations of different star formation histories according to \citet{2003MNRAS.344.1000B}. No nebular emission is included in the model. The upper diagram $a$ shows the evolution in a population that has experienced a short burst and in $b$ the SFR has been constant for 100 Myr before being abruptly shut down. The model results for three different metallicities, Z=0.005 (dashed-dot blue), Z=0.02 (solid green) and Z=0.05 (dashed red) are shown. Our selection criterion for a post-starburst galaxy (\ewhd$<$-6\AA) is indicated with the horizontal yellow line.}
  \label{bc_ewhd_age}
\end{figure}

Fig. \ref{bc_ewhd_age} shows model predictions of the evolution of H$\delta$ under two different star formation histories - an instantaneous burst (Fig. \ref{bc_ewhd_age}a) and  a burst with constant SFR during 100 Myr (Fig. \ref{bc_ewhd_age}b). Here we use the models of \citet{2003MNRAS.344.1000B} that have a slightly better temporal resolution than our model. As we see in the diagram, the postburst epoch sets in quickly after the burst has ended and lasts about 1 Gyr, about 10 times as long as the typical starburst lifetime (see below).

A subset of postburst galaxies are the so called E+A, k+a or a+k galaxies \citep{1983ApJ...270....7D,1999ApJS..122...51D}. These show strong Balmer lines in absorption but no [\ion{O}{ii}]$\lambda$3727 or \hax in emission. In our postburst sample we accept  the presence of [\ion{O}{ii}] and \hax that reveal ongoing star formation, AGN activity or shock heating but with a much lower intensity than during the burst period.

\begin{figure} % two column figure
  \centering
  \includegraphics[width=0.5\textwidth]{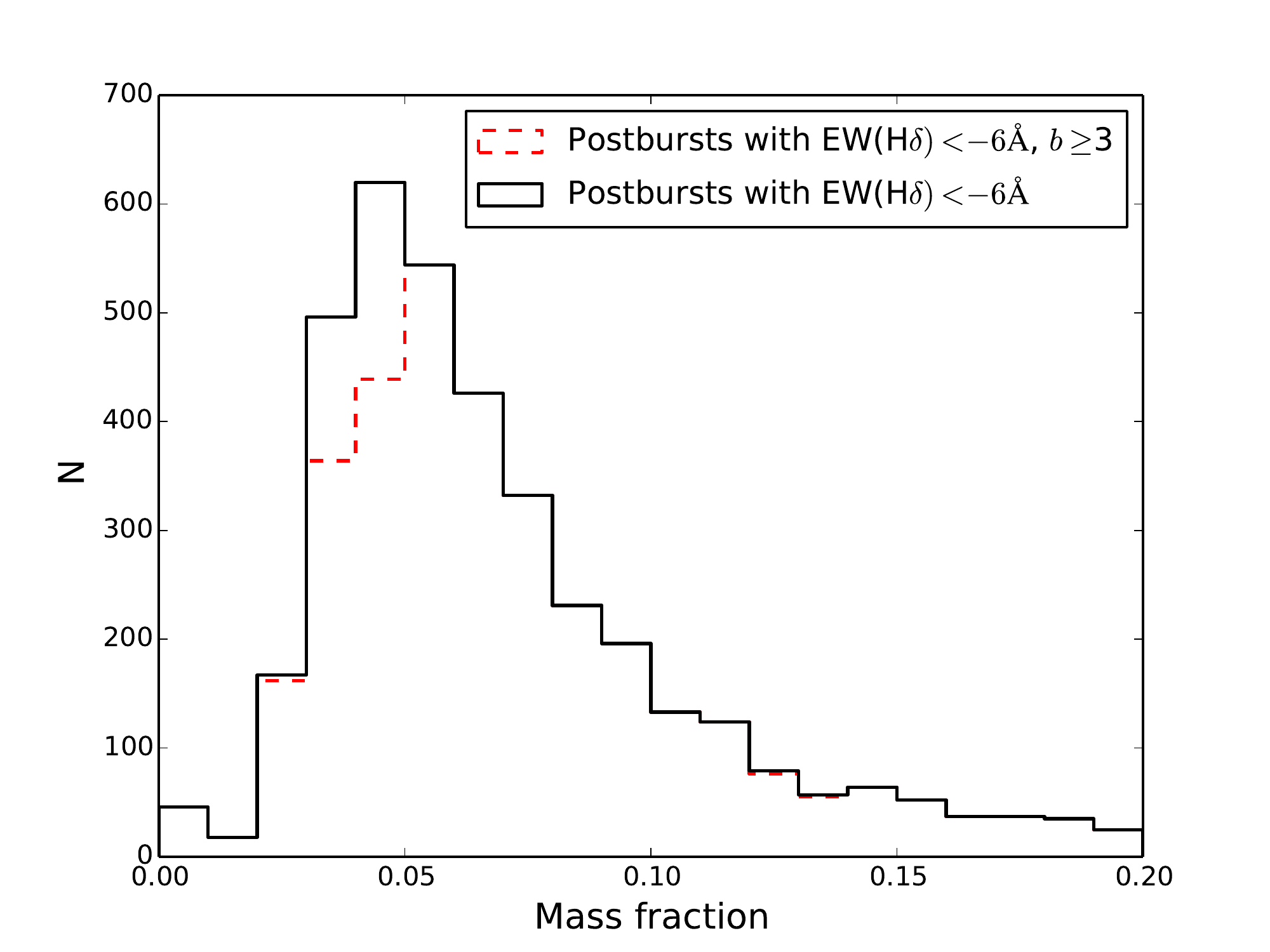}
  
  \caption{Mass fractions, $f_{burst}$, derived for the postburst sample, both with and without the $b>$3 criterion (red dashed line). The limit towards low mass fractions clearly show up as steep decline around $f_{burst}\sim$ 3\%. About 95\% of the galaxies in the larger sample have $f_{burst}\geq$ 3\%.}
  \label{hist_mf_pb}
\end{figure}

In the work by \citet{2003MNRAS.341...33K}, the past star formation history of
SDSS galaxies is investigated. As an indicator of the SF history, they use
the strength of the 4000\AA ~break, D(4000), and the H$\delta$ absorption line
index. The indices are used to estimate the burst mass fraction during the last
few Gyrs. Bayesian likelihood estimates are derived from Monte Carlo
simulations. One of the results they obtain is that 95\% of galaxies with
\ewhd$<\approx$-6\AA ~should have experienced a burst of
a strength $f_{burst}>$5\% during the last 2 Gyr.  Here we also use
the \hdx criterion to select our postburst sample. However, when we look at the
results from our analysis of the burst mass fraction in the postbursts, we end
up with slightly different results from what Kauffmann et al.~find. We
find that the burst strength in more than 95\% of the cases is $f_{burst}>$3\% 
which is lower than the $>$5\% they found in their studies. The result
is displayed in Fig. \ref{hist_mf_pb}. Two histograms are shown. One displays
postburst galaxies with \ewhd$<$-6\AA  ~(based on our remeasurements in Sect.~\ref{sec:remeas}) and the
other is based on one additional criterion, the mean $b$-parameter over the
burst period, $<$$b$$>$, demanding that  $<$$b$$>$~$\geq$3. The mean $b$-parameter is
difficult to derive for postbursts since we lack detailed information about the
duration of the burst. We assume that, in the case of exponentially
decaying bursts, the duration is equal to the timescale of the decay. With the
$b$-parameter criterion active, the total number of galaxies in the diagram is
reduced by about 10\%. The median $f_{burst}$ of the postbursts is 6.6\% and
the histograms show a steep decline towards smaller $f_{burst}$. At lower
values the \ewhdx criterion is invalidated. It is this limit that differs
between our study and Kauffmann's et al.. We discuss this problem in Sect. \ref{discussion}.

\begin{figure*} % two column figure
  \centering
  \includegraphics[width=0.9\textwidth]{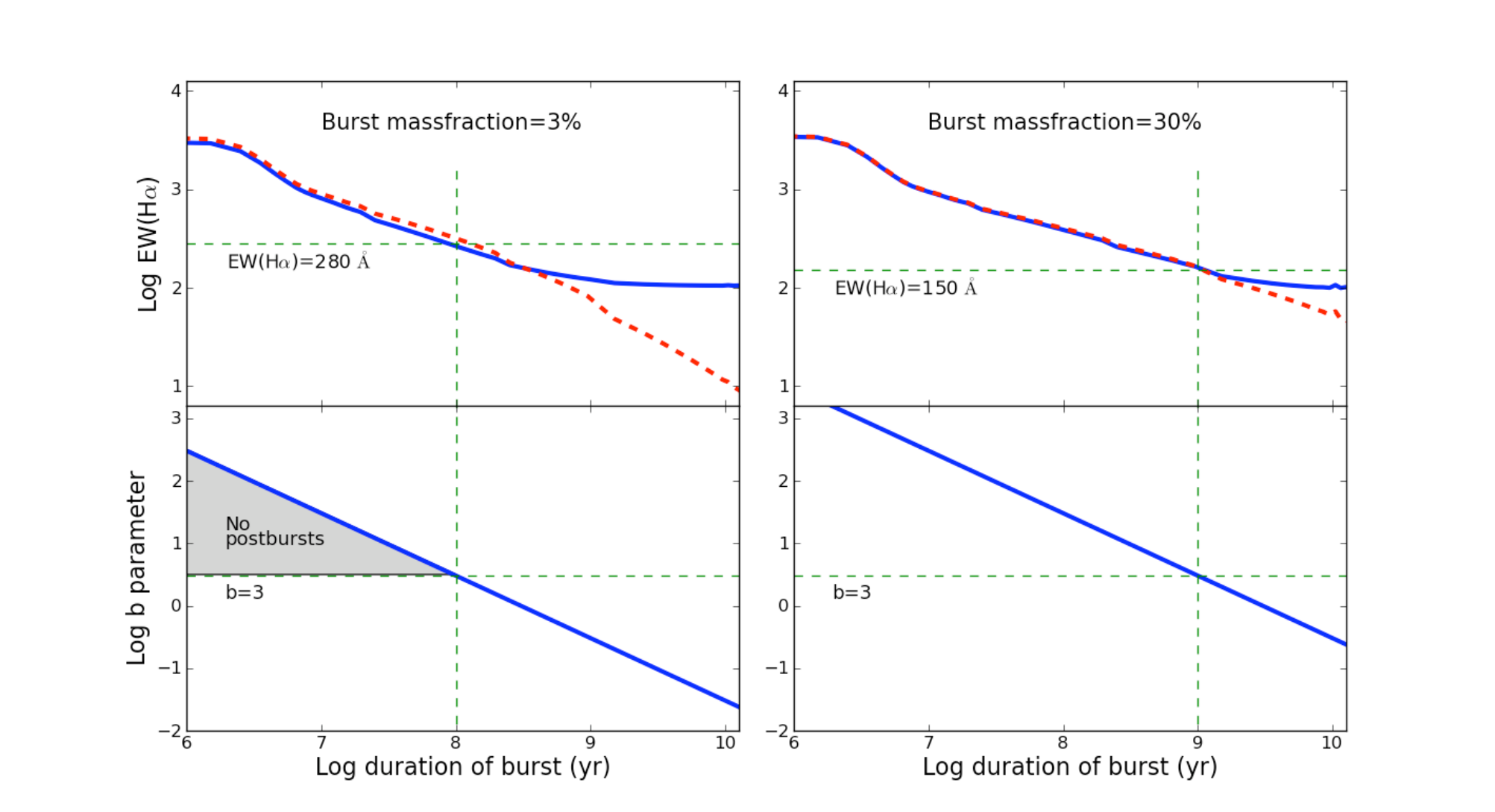}
  
  \caption{Upper diagrams: Our model predictions of \ewhaem ~as function of the duration of the burst when 3  and 30\% of the total mass is formed during a burst with constant SFR. Two different scenarios are displayed: 1) a burst in an old galaxy where the stars were formed in a single burst (hatched line) and 2) a burst in a galaxy which has had a constant SFR over a Hubble time (solid line). Lower diagrams: The evolution of the $b$--parameter. The limit of our criterion of a starburst, the `soft criterion', stating that $b \equiv \frac{SFR}{<SFR>} \ge $3, is marked with the horizontal line in the lower diagrams. The blue inclined line corresponds to mass fractions of 3 and 30\% respectively. The vertical lines indicate the maximum ages of a starburst with mass fractions 3 and 30\% respectively, to fulfil the soft starburst criterion. The horizontal lines in the upper diagrams indicate \ewhaemx at that burst duration and the corresponding \ewhaemx in numbers are given below. The shaded area indicates a region containing starbursts according to the $b$--parameter criterion but which will not be powerful enough to produce postbursts according to the \ewhdabsx$<$-6\AA ~criterion.}
  \label{ewha_dur}
\end{figure*}

We now look for criteria that will make it possible for us to identify the precursors of the postbursts.  From our model we can obtain the equivalent width of H$\alpha$
in emission (EW$_{H\alpha}$)  as function of the duration of the
burst. In Fig. \ref{ewha_dur} a we show how \ewhax varies with the duration
of the burst. It is important to be aware that the purpose of this diagram is not to show the evolution with time of \ewhax but rather how the burst impacts the galaxy spectrum if we assume the same mass fraction but different durations of the burst, as given on the abscissa. Obviously the influence of the burst weakens with increasing duration and the $b$--parameter decreases linearly with the inverse of the duration of the burst. We look at two different burst mass fractions, $f_{burst}$ = 3\% and $f_{burst}$ = 30\%.  In one case we assume that the
burst is taking place in an old galaxy formed in a single burst (age $\sim$ 14 Gyr) and in the
other case we assume that the burst is taking place in a galaxy which
has had a constant SFR over its lifetime.  We can see from Fig. \ref{ewha_dur} that if we assume that the minimum burst mass of a postburst galaxy is 3\%,  the starburst cannot be older than 100 Myr. There may be younger galaxies that qualify as starbursts from the $b$--parameter criterion (the shaded area in the diagram), but they will not show up as postbursts since $f_{burst}<$3\%. 

From a semantic aspect we find it proper to say that to qualify as a {\sl burst},���~the duration of the SF epoch should not be longer than $\sim$1 Gyr.  From Fig. \ref{ewha_dur} we see that for a duration of 1 Gyr and $b$=3, the mass fraction is 30\% and \ewhax is
=150\AA. Thus, a minimum (but not sufficient) requirement to qualify as a starburst is that \ewhax $>$150\AA. A galaxy with continuous SFR over the age of the universe
has \ewha $\sim$ 100\AA. In many studies of starbursts based on \ewha, the lower limit is often chosen to be 100\AA. Does this mean that a large portion of the galaxies in those studies should be disqualified as starbursts? It depends. As we argue below, the concept of age dependent dust attenuation will lead to a reduction of the observed strength of \hax but also the equivalent width. We estimate the maximum reduction to be a factor 2--3 (assuming a dust attenuation in the V band A$_V\lesssim$1.5$^m$ for ages $\geq$ a few Myr). Thus in order to be on the safe side, one should examine all galaxies with \ewhax above 150\AA ~divided by this factor.  Here we have chosen our selection criterion to be  \ewhax $>$ 60\AA. From this potential starburst sample we then select the final sample by applying the $b>$3 criterion.

\subsection{AGNs}
\label{subsec:AGN}

As we go towards higher luminosities the importance of AGNs in combination with centrally concentrated star formation increases. Statistically (see Fig.~\ref{lumfunction}), AGN dominated galaxies (with \ewha $>$ 60 \AA) occur as frequently as starbursts at $M_r\sim$ 
--21. If the energy production in the central region is dominated by AGN activity it will be difficult to derive information about the possible presence of a starburst. However, if we can properly link starbursts with postbursts, we have an opportunity to statistically determine how frequently starbursts occur in active galaxies of different luminosities by using postbursts as indicators of preceding bursts. 

\begin{figure}[t!]
\centering
 \resizebox{\hsize}{!}{\includegraphics{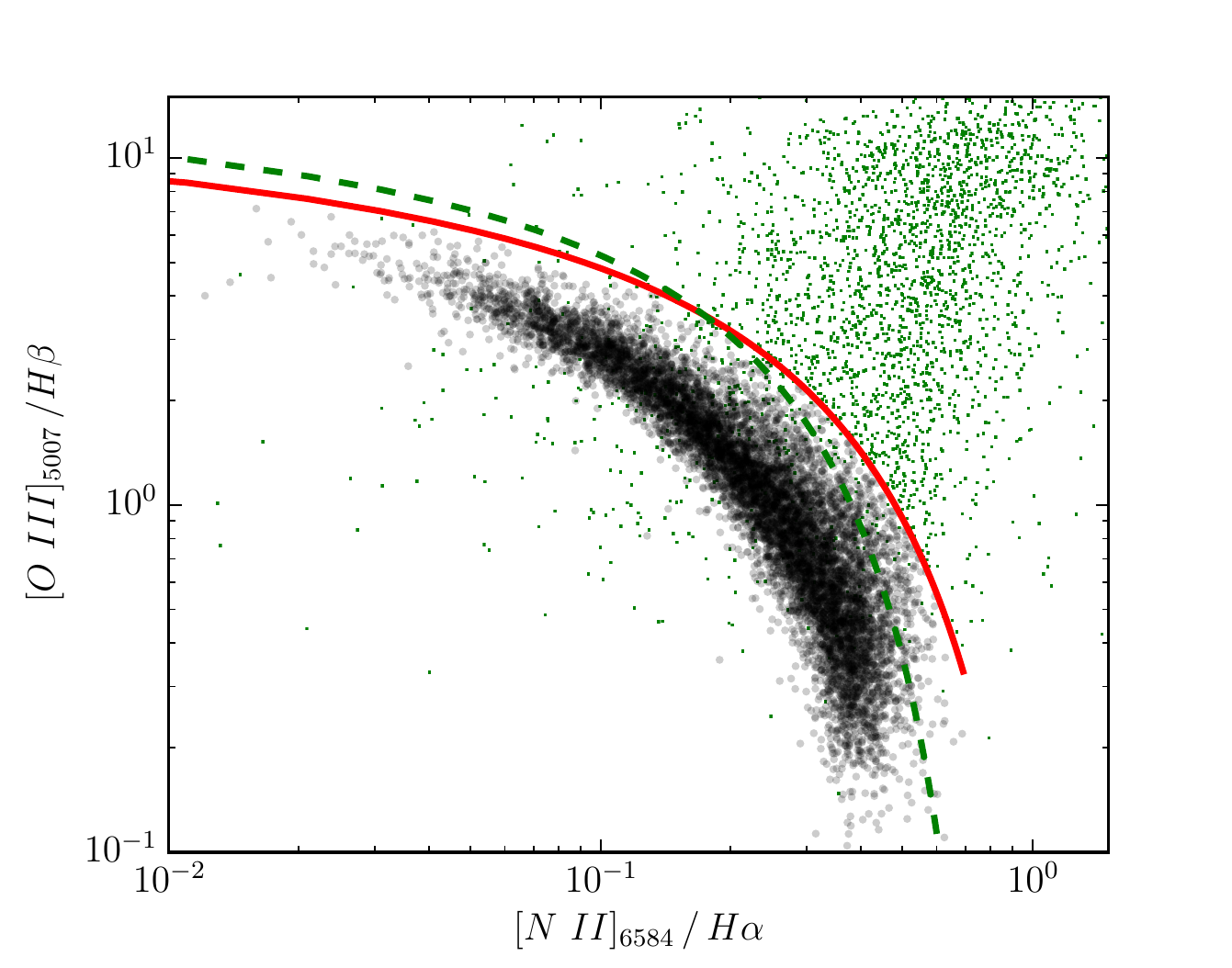}}
\caption{BPT \citep{1981PASP...93....5B} diagram of the \ewha $\geq$ 60\AA ~starburst galaxy candidate sample based on line peak intensity ratios. The solid red line is the empirical dividing line (a modified version of the equation given by \citet{2003MNRAS.346.1055K} shown in hatched green) between starburst galaxies (lower part, grey symbols) and galaxies containing an AGN contributing significantly to the flux in the strong emission lines (upper part, green dots). As a complementary AGN criterion we also used FWHM(\ha$_{em})>$540 km s$^{-1}$.}
  \label{bpt}
\end{figure}

Here we first look at the LFs of three types -- starbursts, postbursts and AGNs. To identify AGNs we used two criteria -- 1) the FWHM of the \hax emission line and  2) the position of a galaxy in the BPT diagram \citep{1981PASP...93....5B}. \citet{2003MNRAS.346.1055K} derived an equation that separates star-forming galaxies from composite (thermal+non-thermal) and active galaxies in the upper part of the BPT diagram. The line ratios given in the diagram are based on line peak intensities instead of fluxes. The consequence will be that the AGNs will be shifted slightly upwards towards the right, thus increasing the separation between starbursts and AGNs. This allows us to shift the demarcation line slightly in the same direction. Thus, as the second criterion we have used a modified version of the relation given by Kauffmann et al.. Fig. \ref{bpt} shows how our criteria separates pure starbursts from active and composite galaxies. Notice that some green dots (AGN) fall in the region below the demarcation line. These are galaxies fulfilling the second AGN criterion based on the Balmer line width.The galaxies were selected from the same sample as the starburst galaxies, i.e. all have \ewha$>$60 \AA. We summarize the criteria quantitatively in the next section.

\subsection{Summary of selection criteria}
\label{subsec:selection}

Here we present the final selection criteria for the spectral types we wish to work with. The data were collected from SDSS DR7 \citep{2009ApJS..182..543A}. Three samples were created -- a {\sl starburst sample}, a {\sl postburst sample} and an {\sl AGN sample}. Selecting an AGN sample is motivated by the need to take into account the mixture of AGN and starbursts. In addition to the spectral criteria we also chose a redshift range of 0.02$\le$z$\le$0.4. The limiting magnitude of SDSS main sample spectroscopic data is $M_r$ = 17.77 \citep{2002AJ....124.1810S} but the recommended magnitude limit for statistical studies is $M_r \sim$17.5 \footnote{http://classic.sdss.org/dr7/products/general/target\textunderscore quality.html}. We use the 17.5 limit when we derive the LF (see below), otherwise we use the 17.77 limit. At a redshift of z=0.4 an apparent magnitude of 17.5 corresponds to an absolute magnitude of $M_r$=--24. At this luminosity it will be difficult to separate starbursts from AGNs so we chose z=0.4 as an upper limit for our purpose. The lower limit was chosen to reduce the problems with deviations from the Hubble flow with the corresponding negative impact on the determination of absolute luminosities from the Hubble law and secondly, aperture biases, discussed in the next subsection.

 Our selection criteria can be summarised as follows:

\begin{itemize}\itemsep3pt
\item 0.02$<$z$<$0.4
\begin{itemize}
\item Starburst {\it candidates}. All four criteria below have to be fulfilled:
\begin{itemize}
\item \ewha$\ge$60\AA
\item FWHM$_{H\alpha,em} \le$ 540 \kms
\item log([\ion{O}{iii}]$\lambda$5007/\hb)$<$0.71/log([\ion{N}{ii}]$\lambda$6584/\ha-0.25)+1.25
\end{itemize}
\item Postburst {\it candidates}
\begin{itemize}
\item \ewhd$\le$--6\AA
\end{itemize}
\item AGNs.
\begin{itemize}
\item \ewha$\ge$60\AA ~and at least one of the following criteria:
\item FWHM$_{H\alpha,em} >$ 540 \kms
\item log([\ion{O}{iii}]$\lambda$5007/\hb)$\ge$0.71/log([\ion{N}{ii}]$\lambda$6584/\ha-0.25)+1.25
\end{itemize}
\end{itemize}
\end{itemize}

The total number of starburst {\it candidates} brighter than $M_r$=17.77 (after correction for galactic extinction) is 9167. At a limiting magnitude of $M_r$=17.5 we count 6368 galaxies. The number of starburst galaxies brighter than $M_r$=17.77 with $b>$ 3 is 1743 and the number with $f_{burst}>$3\% (no restriction on $b$) is 3215. The number of postburst {\it candidates} is 7067. After remeasurement (cf. Sec. \ref{sec:remeas}) of the \hdx line in absorption, the number drops to 4032. This is the sample we call postburst galaxies. Of these, 3804 galaxies have $f_{burst}>$3\%. The number of AGNs is 2701.

\begin{figure}[t!] 
  \centering
 \resizebox{\hsize}{!}{\includegraphics{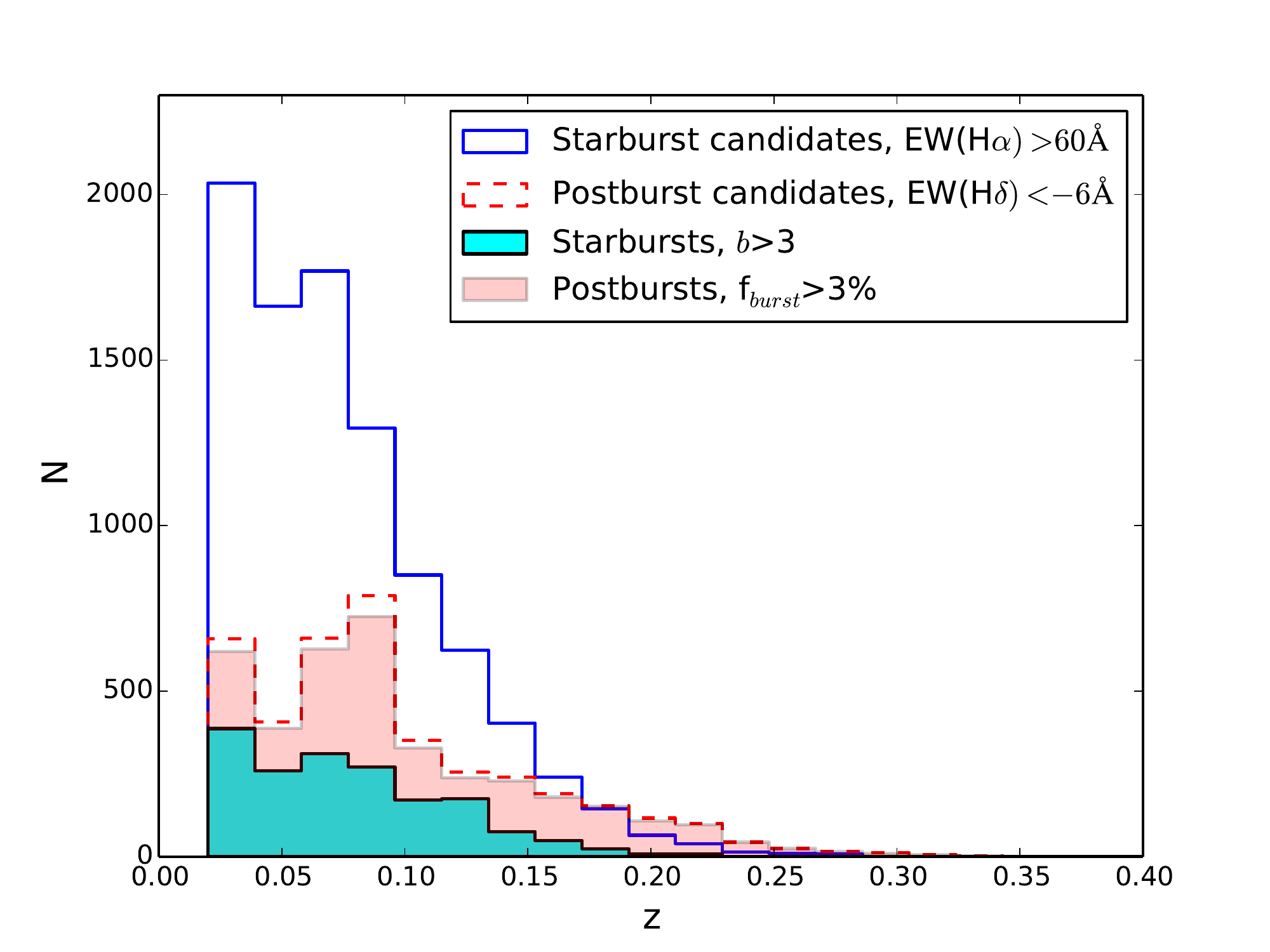}}
  \caption{Redshift distributions of the galaxies in the starburst and postburst samples.}
  \label{zhist}
\end{figure}

In the present investigation we only use spectral information from the
central 3$\arcsec$ (in most cases) of the target galaxies (the diameter of the
SDSS spectroscopic fiber aperture). Most of our objects are at low redshift.
Fig. \ref{zhist} shows a histogram of the redshifts in the starburst and
postburst samples. At z=0.02 the aperture diameter will be 1.2 kpc and at
the median redshift of the starburst population ($b$$>$3), z=0.07, the aperture is 4 kpc. At the maximum redshift we
will explore (z$\sim$0.3), the aperture corresponds to a size of about 13 kpc.
The typical scale lengths of luminous starburst galaxies is 2--3 kpc. The
effective radius is about 70\% of this, i.e $\approx$ 2 kpc. As we show in
Sect.~\ref{sec:apteff}, more
than 50\% of the light is within the aperture in about 25\% of the galaxies in the starburst $candidate$ sample,
and a slightly larger fraction in the finally selected starburst sample. At higher luminosities we are
more restricted to nuclear starbursts. In Sect. 3.4 we have a look at
systematic trends with redshift that can be caused by aperture effects and
decide whether we need to correct for such potential problems.

\subsection{The data retrieval}

To put these selection criteria into practice, we ran queries on the
CasJobs interface of SDSS DR7 \footnote{http://casjobs.sdss.org},
first selecting the Object-IDs and
corresponding SpecObjIDs that connect the photometry tables with the
spectroscopic data. In order not to skew the redshift distribution
between starburst and postburst candidates, we chose only galaxies
from the main galaxy sample \citep{2002AJ....124.1810S}, excluding for example the Luminous
Red Galaxy sample \citep{2001AJ....122.2267E} which would have
played mainly into the postburst sample. We also set the flag to avoid
blended objects and required the confidence of the
redshift--determination to be $>95\%$. Using the selected IDs, we
retrieved the different photometric and spectroscopic measurements
from the SDSS tables as needed and mentioned in the appropriate
context below. We also downloaded the corresponding SDSS spectra as
FITS files for the targets which we de-reddened and used for the model
fits.

\medskip

This is an example of a starburst candidate query used:

\medskip 

{\noindent 
SELECT g.objID, s.specObjID \newline
FROM Galaxy g, SpecObj s, SpecLine Ha \newline
WHERE g.objID = s.bestObjID AND Ha.specObjID = s.specObjID AND  Ha.LineID = 6565 AND
Ha.ew$>$60 AND s.zConf $>$ 0.95 AND s.z BETWEEN 0.02 AND 0.4 AND ((g.flags
\& 8)==0) AND (s.primtarget \& 64$ >$ 0)}

\section{Spectral analysis}
\label{sec:analysis}

\subsection{Remeasuring spectral lines}
\label{sec:remeas}

When we analysed the SDSS spectra we found a problem with the SDSS derived values of the equivalent widths of \hdx in absorption. In many cases the \hdx line was misidentified or there were problems with the continuum fit. We therefore remeasured the Balmer lines listed in Table \ref{abslines}. We did this by defining a relatively line free region on both sides of the line. Then we calculated the mean fluxes of these regions and interpolated linearly between these across the line region to define the continuum level.  In Table~\ref{abslines} we list the spectral windows used.

\begin{table}[h]
\caption{Line data. The table gives information about how the equivalent widths of the Balmer lines are derived from the spectra. A straight line is fit to the two continuum regions and the absorption/emission line is measured below/above the fit in the interval given. The values are given in \AA.}             
\label{abslines}      
\centering    
\begin{tabular}{ l l l l}   
\hline\hline    
\\   
Line & Continuum 1 & Line region & Continuum 2 \\ 
\\
\hline   
\\
Absorption lines \\
\\
\hy & 4200-4260 & 4280-4380 & 4400-4460 \\
\hd & 4005-4035 & 4066-4136 & 4158-4208 \\
\he & No useful region & 3950-3990 & 4005-4035 \\
\\
\hline    
\\
Emission lines \\
\\
\ha & 6500-6530 & 6556-6572 & 6600-6630 \\
\hb & 4760-4800 & 4853-4872 & 4925-4955 \\  
\\
\hline\hline            
\end{tabular}
\end{table}

\begin{figure}[t!]
\centering
 \resizebox{\hsize}{!}{\includegraphics{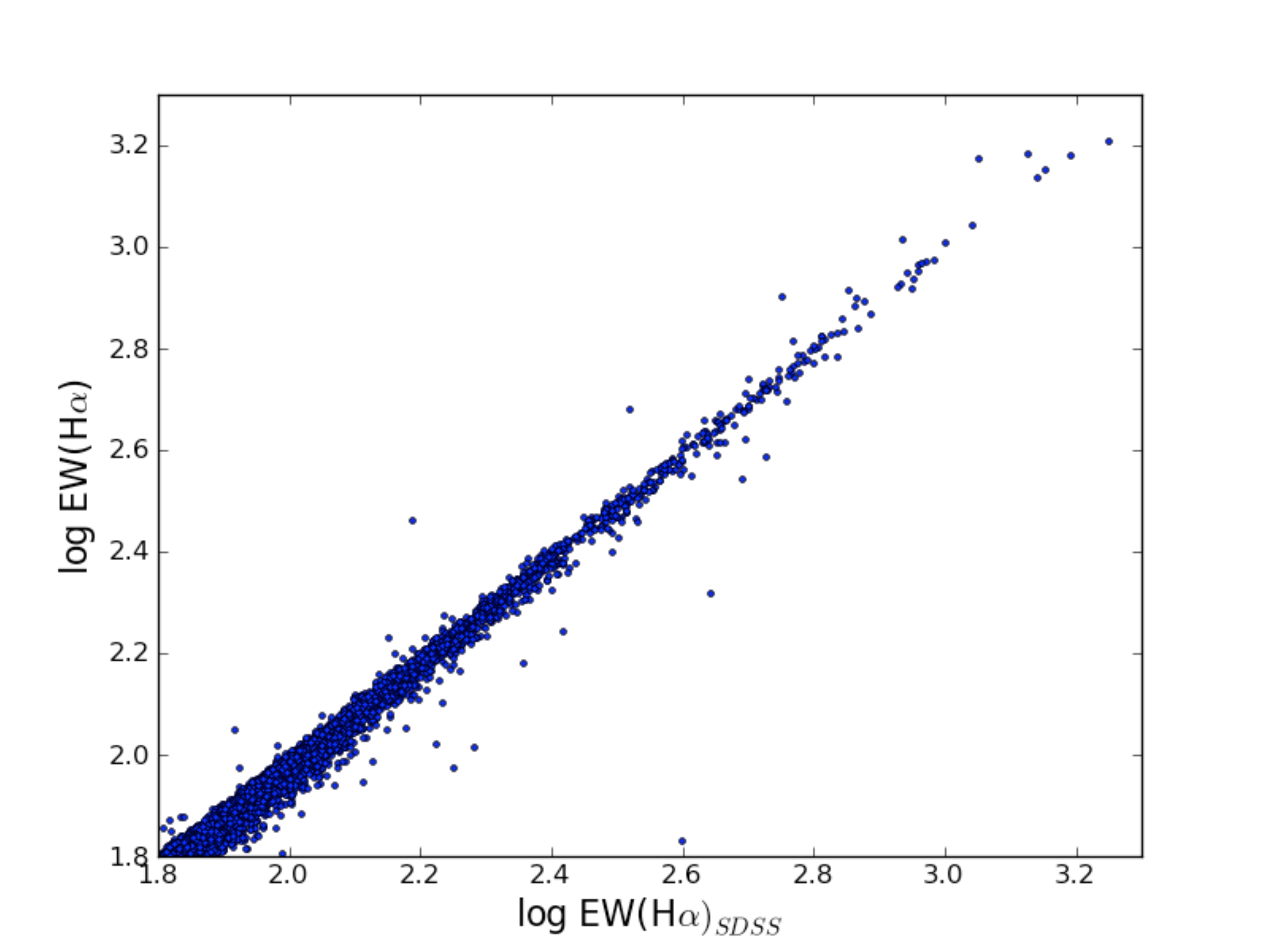}}
\caption{\hax emission line equivalent widths of our target galaxies as derived from our measurements, versus the value as obtained from the SDSS.}
  \label{ewha_ewha}
\end{figure}

\begin{figure}[t!]
\centering
 \resizebox{\hsize}{!}{\includegraphics{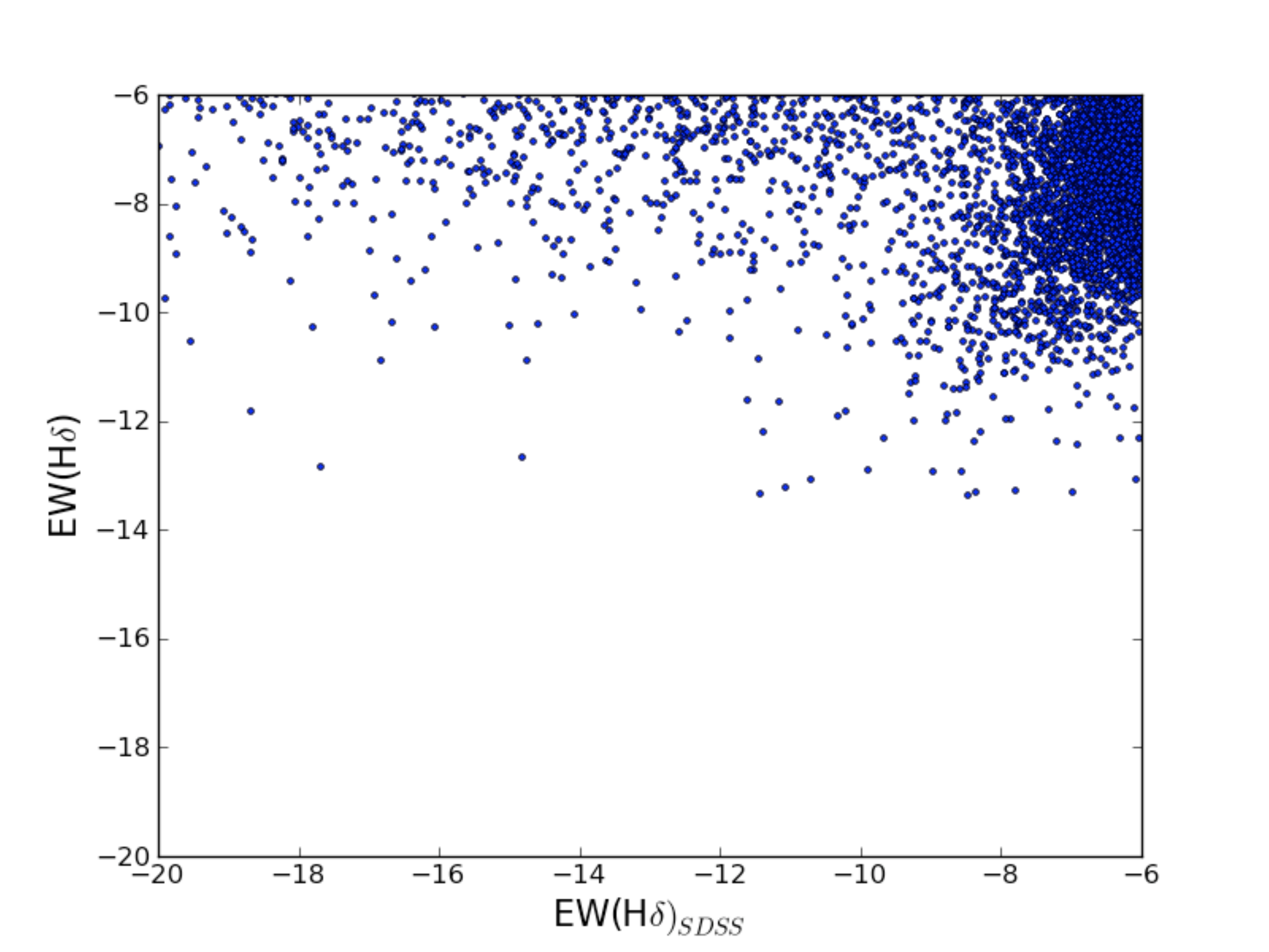}}
\caption{\hdx absorption line equivalent widths of our target galaxies as derived from our measurements, versus the value as obtained from the SDSS.}
  \label{ewhd_ewhd}
\end{figure}

\begin{figure}[t!]
\centering
 \resizebox{\hsize}{!}{\includegraphics{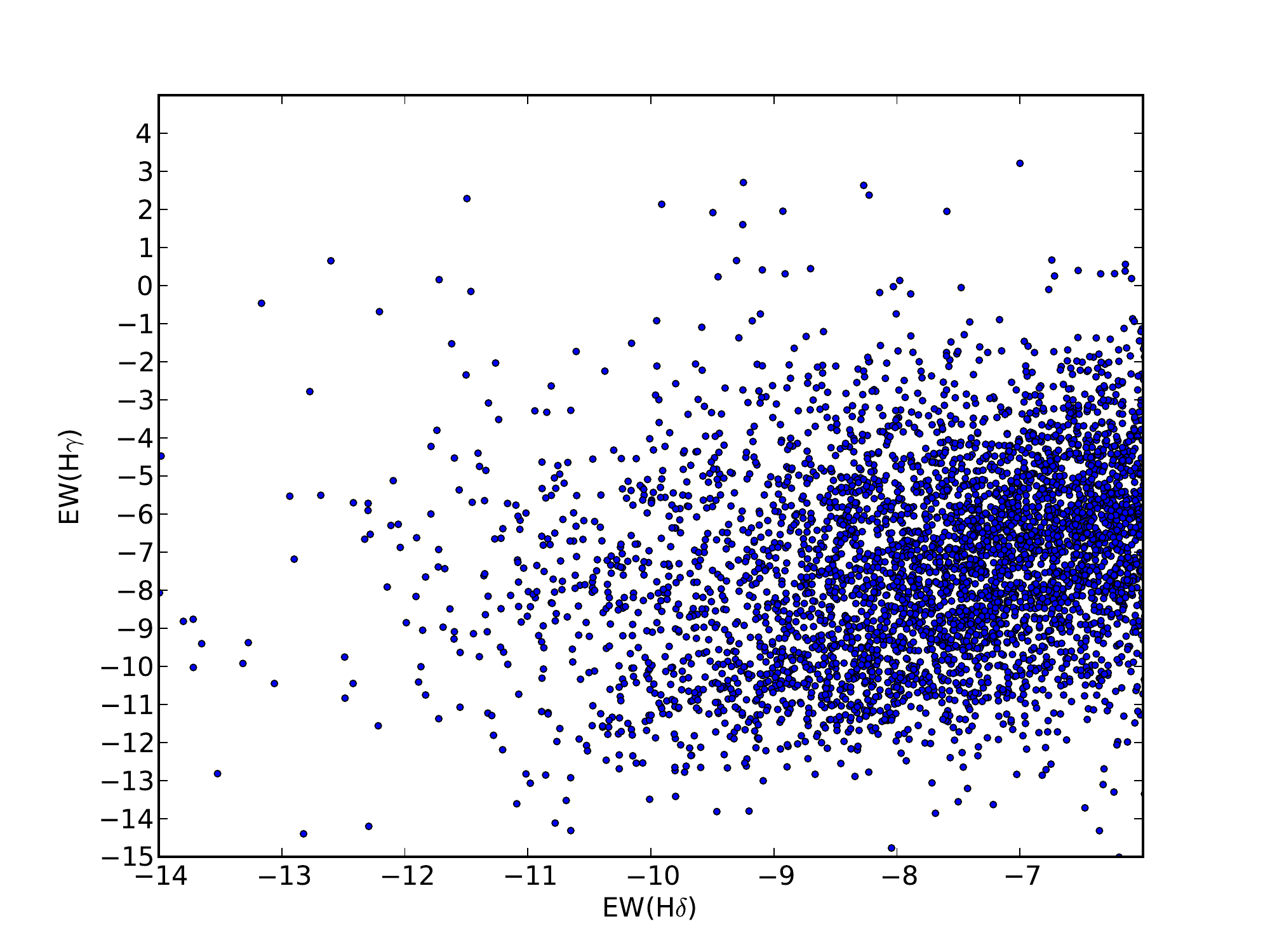}}
\caption{\hdx and \hyx absorption line equivalent widths after our remeasuring.}
  \label{ewhd_ewhg}
\end{figure}

Fig.~\ref{ewha_ewha} shows how our measurement of \ewhax in emission correlates with the data from SDSS. The agreement is good, with a trend for the SDSS values to be slightly larger ($<$20\%). However, the correlation between \ewhdx in absorption is much worse as is seen from Fig.~\ref{ewhd_ewhd}. The SDSS data contains many large negative values, inconsistent with properties of a normal stellar population. In Fig.~\ref{ewhd_ewhg} we plot \ewhdx versus \ewhyx from our remeasured data in both cases. We see a broad distribution with a clear correlation, indicating that the data are reliable, albeit with a strong noise component, mainly from the \hyx line.

\subsection{Spectral modelling}
\label{model}

In our modelling of the SDSS spectra it was assumed that the stellar population has two components -- a young and an old population. The young component may have a range of ages, from 0 to 10 Gyr while the old component has a fixed age of 10 Gyr. We chose this age to agree with the cosmic lookback time at peak cosmic  SFR \citep{2006ApJ...651..142H}. Observations of local dwarf galaxies also support the formation of most of the stars at z$>$1 \citep{2011ApJ...739....5W,2011AJ....141...68Z}. In the first tests of the model we also had a look at cases where the old population had a declining SFR over various timescales. In the mean, however, the best fits were obtained with models assuming that the old population formed over a short period of time {\ltsim} 100Myr. 

In the modelling, the programme steps through the spectral library of the young population with increasing age. At each step the young population is mixed with the old in various proportions, each time making the mix the most optimum one to agree with the observations. We have not used the nebular emission lines since the relative strengths depend strongly on the ionisation parameter, the density of the ionised gas, the filling factor and the metallicity \citep[e.g.][]{1989agna.book.....O,1991ApJ...380..140M,2000ApJ...542..224D,2001MNRAS.323..887C}. Therefore we have masked the strong emission lines hence the fit is more or less based on the pure stellar spectrum. Since the density of emission lines are strongest in the blue region of the spectrum we have, when we do the spectral fit, given the continuum double weight in the wavelength region 3800-4800 \AA. The programme chooses the solution that gives the lowest $\chi^2$ of the different fits. We have tested various combinations of models that have different star formation histories and metallicities and decided to use the restricted number of model parameters given in Table~\ref{modspectra}. The table contains the {\sl Mode} parameter. The Mode can be either a constant SFR over a time period given by $\tau$ or an exponentially declining SFR according to $SFR \propto e^{-t/\tau}$, where t is the age. Our preliminary tests favour an exponentially declining SFR instead of a constant rate. An exponentially declining SFR is also supported by models \citep{2007A&A...468...61D} (but see also \citet{2010MNRAS.402..985H} who propose a power law at the end of the burst). Each track starts at an age of 0.5 Myr. The incremental time steps along each track were interpolated to a size of 0.04 dex.

\begin{table}[h]
\caption{Basic setup of model parameters used in the old/young mixture. The models either have an exponential or constant star formation rate with $\tau$ corresponding to the timescale of the exponential decline time scale or the duration of the burst.}             
\label{modspectra}      
\centering    
\begin{tabular}{ l l l l l l l}   
\hline\hline       
Model & Age group & Mode & $\tau$ (yr) & Age (yr) & Z/Z$_{\odot}$ \\ 
\hline   
\\
Starburst & young &  exp & 3$\times$10$^7$ & 0- 10$^{10}$ & 0.2 \\
& & exp & 10$^8$ & 0- 10$^{10}$ & 0.2 \\
& & exp & 3$\times$10$^8$ & 0- 10$^{10}$ & 0.2 \\
& & exp & 10$^9$ & 0- 10$^{10}$ & 0.2 \\
& & exp & 3$\times$10$^7$ & 0- 10$^{10}$ & 0.4 \\
& & exp & 10$^8$ & 0- 10$^{10}$ & 0.4 \\
& & exp & 3$\times$10$^8$ & 0- 10$^{10}$ & 0.4 \\
& & exp & 10$^9$ & 0- 10$^{10}$ & 0.4 \\
& & exp & 3$\times$10$^7$ & 0- 10$^{10}$ & 1.0 \\
& & exp & 10$^8$ & 0- 10$^{10}$ & 1.0 \\
& & exp & 3$\times$10$^8$ & 0- 10$^{10}$ & 1.0 \\
& & exp & 10$^9$ & 0- 10$^{10}$ & 1.0 \\
\\
Postburst & young & const & 10$^7$ & 0- 10$^{10}$ & 0.2 \\
& & const & 10$^8$ & 0- 10$^{10}$ & 0.2 \\
& & const & 10$^9$ & 0- 10$^{10}$ & 0.2 \\
& & exp & 3$\times$10$^7$ & 0- 10$^{10}$ & 0.2 \\
& & exp & 10$^8$ & 0- 10$^{10}$ & 0.2 \\
& & exp & 3$\times$10$^8$ & 0- 10$^{10}$ & 0.2 \\
& & exp & 10$^9$ & 0- 10$^{10}$ & 0.2 \\

& & const & 10$^7$ & 0- 10$^{10}$ & 0.4 \\
& & const & 10$^8$ & 0- 10$^{10}$ & 0.4\\
& & exp & 3$\times$10$^7$ & 0- 10$^{10}$ & 0.4 \\
& & exp & 10$^8$ & 0- 10$^{10}$ & 0.4 \\
& & exp & 3$\times$10$^8$ & 0- 10$^{10}$ & 0.4 \\
& & exp & 10$^9$ & 0- 10$^{10}$ & 0.4 \\

& & const & 10$^7$ & 0- 10$^{10}$ & 1.0 \\
& & const & 10$^8$ & 0- 10$^{10}$ & 1.0 \\
& & exp & 3$\times$10$^7$ & 0- 10$^{10}$ & 1.0 \\
& & exp & 10$^8$ & 0- 10$^{10}$ & 1.0 \\
& & exp & 3$\times$10$^8$ & 0- 10$^{10}$ & 1.0 \\
& & exp & 10$^9$ & 0- 10$^{10}$ & 1.0 \\

\\
Both models & old & const & 10$^8$ &  10$^{10}$ & 0.2 \\
& & const & 10$^8$ &  10$^{10}$ & 0.4 \\
& & const & 10$^8$ &  10$^{10}$ & 1.0 \\
\hline                  
\end{tabular}
\end{table}

In the fit to the SDSS spectra we used two different approaches, one for the emission line galaxies and one for the postburst candidates, which we now describe.

\subsubsection{Emission line galaxies}

As an indicator of the amount of dust attenuation we derive the \hahb ~ratio from the SDSS spectrum. Then we step through the library of the model spectra from low to high ages. The library contains model spectra of  ages increasing with 1 Myr between 0--20 Myr, 10 Myr between 20--150 Myr, 20 Myr between 150--250 Myr, 100 Myr between 250 Myr--1 Gyr and thereafter 1 Gyr. For each age there exists a set of reddened spectra with a range of values of the extinction coefficient. Each of these spectra is based on a sum of spectra of younger populations, assumed to have the same or higher dust opacity than the present population, as described in Sect. \ref{sec:dust}. For each time step we thus step through the library with increasing attenuation until we obtain a value of \hahb ~that corresponds to the observed value. We interpolate linearly among the spectra until we have obtained the best agreement. Once we have this value, we automatically get a value of the extinction in the old stellar component as we describe in Sect. \ref{sec:dust}. We then mix the reddened old and young model spectra so that \ewhax agrees with the observed value.  This is the final model spectrum for this particular age. We then step to the next age. When \ewhax of the model spectrum is lower than the observed one, the process is stopped and the fit that gives the lowest $\chi^2$ is selected.

\begin{figure}[t!]
\centering
 \resizebox{\hsize}{!}{\includegraphics{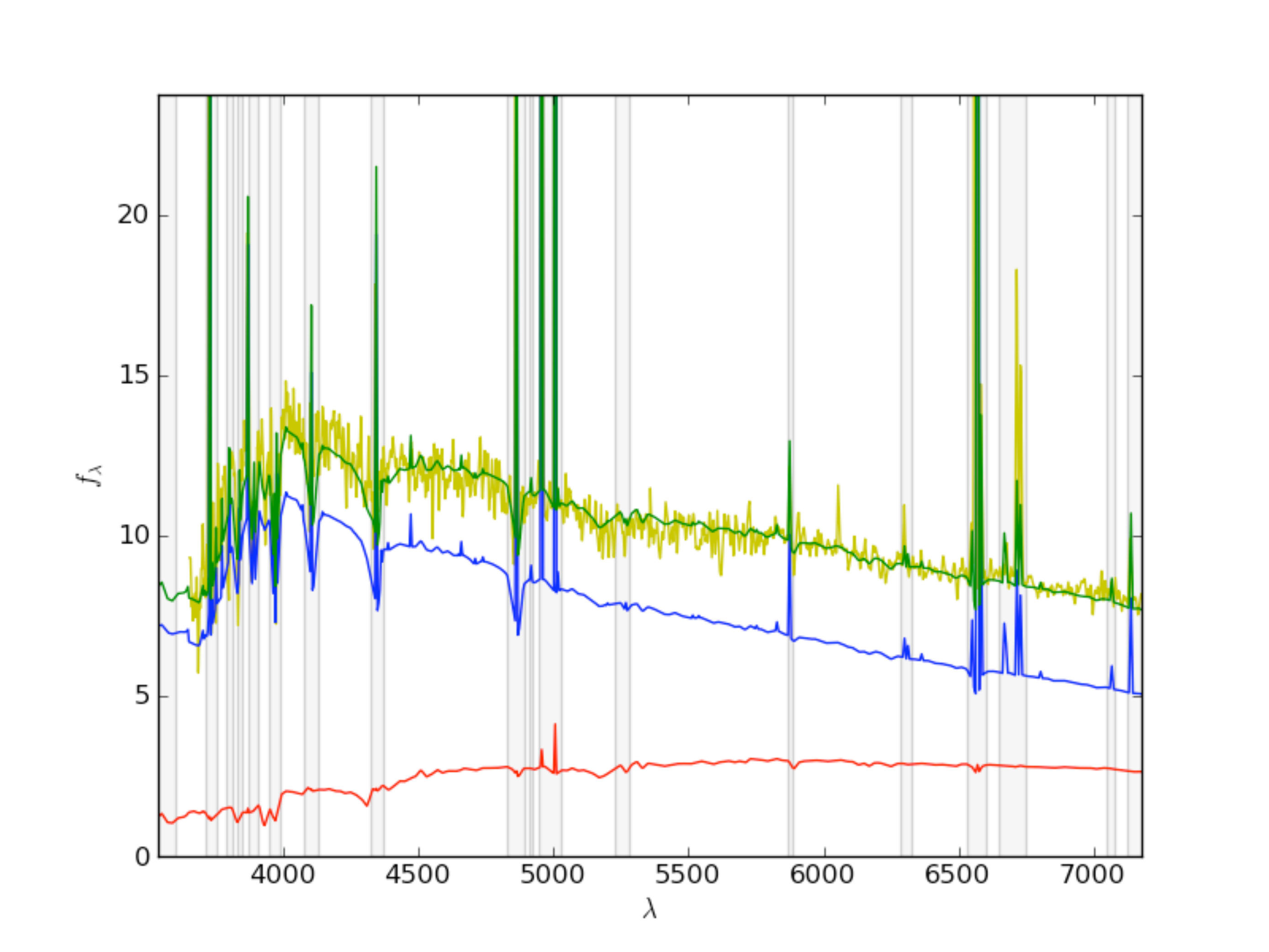}}
 \caption{An example of a model of a starburst spectrum (SDSS SpecObjId 81288133269782528). Displayed are the two model components -- a young (blue line) and an old (red line) stellar population. The sum of the model spectra (dark green line) is shown on top of the observed spectrum (light green). The observed spectrum has been smoothed with a triangular filter with a width of 7 pixels. The shaded areas indicate regions we have flagged to avoid strong emission lines in the fits. Note however that \ewhaemx is used as a separate criterion in the fitting.}
\label{starburst}
\end{figure}

As an output from the model we obtain the age, relative mass and \mlx of the burst population, the \mlx and relative mass of the old population (the age is always assumed to be 10 Gyr), the corrected \hax flux and the mean $b$--parameter. Having access to the distance to the galaxy we can also obtain masses and SFRs based on the corrected \hax luminosity. An example of the modelling is seen in Fig.~\ref{starburst}.

\subsubsection{Postbursts}

\begin{figure}[t!]
\centering
 \resizebox{\hsize}{!}{\includegraphics{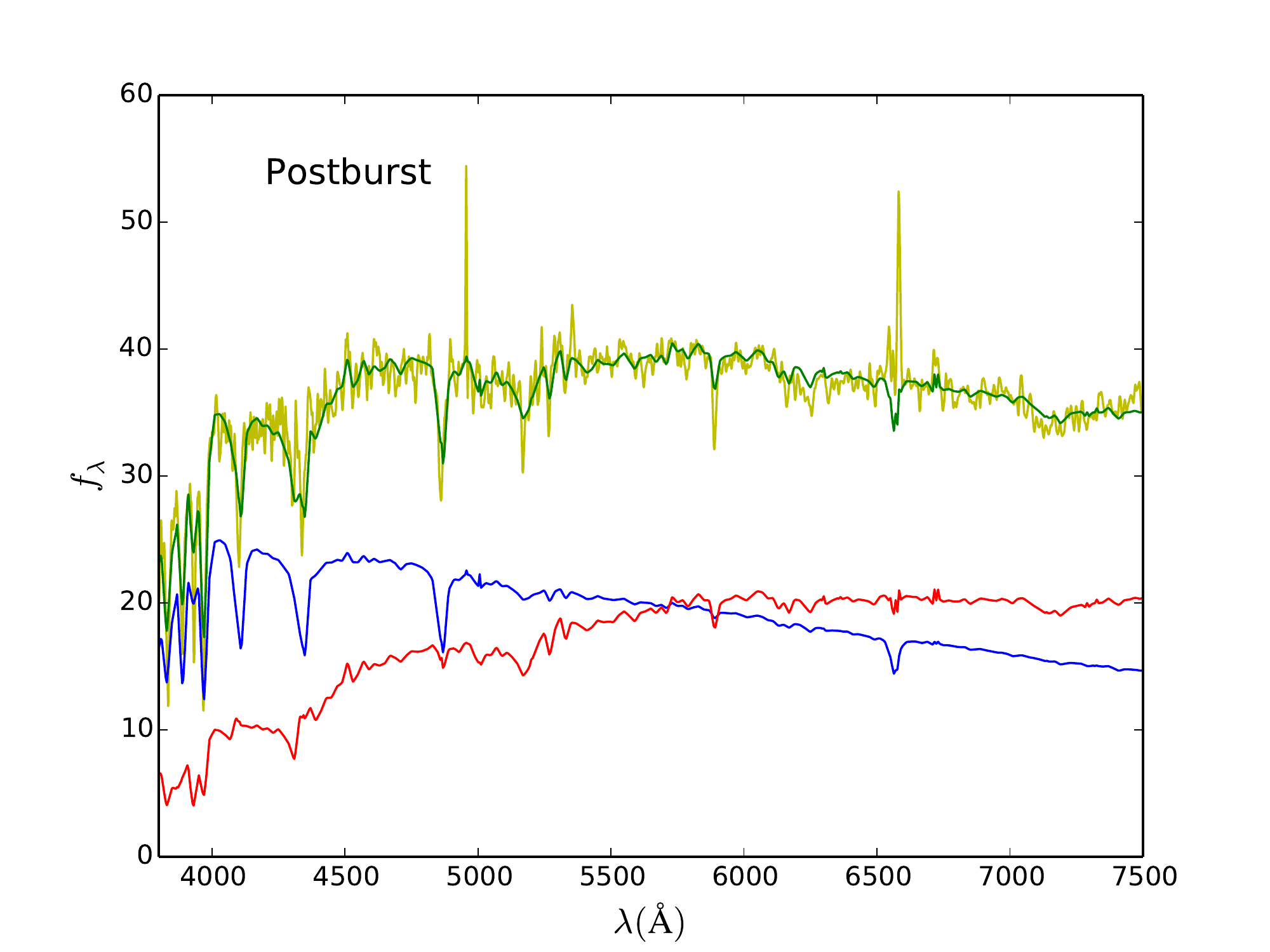}}
 \caption{An example of a model of a postburst spectrum (SDSS SpecObjId 85508603053604864). 
 Displayed are the two model components -- a young (blue line) and an old (red line) stellar population. The sum of the model spectra (dark green line) is shown on top of the observed spectrum (light green line). The observed spectrum has been smoothed with a triangular filter with a width of 7 pixels. The bright emission line just below 5000\AA ~is a remnant from the [\ion{O}{i}]$\lambda$5577 night sky line. The emission lines in the region around \hax are the [\ion{N}{ii}]$\lambda\lambda$6548,6584 lines. Since no \hax is seen (although some is probably swallowed by the stellar absorption line), most of the nitrogen lines are probably originating from a weak AGN.}
\label{postburst}
\end{figure}

The postburst spectra are modelled by mixing a young and an old population in proportions that give the same \ewhdx as measured in the SDSS spectra. A condition is of course that the young population has sufficiently strong \hdx in absorption to make this possible. As we describe in more detail in Sect. \ref{sec:dust} we then derive the dust attenuation in each time step by gradually increasing the amount of attenuation in the model until we get the best fit to the observed spectrum. This procedure is repeated time step after time step until the postburst epoch is over. The best fit is then chosen as the solution. An example of the modelling is seen in Fig.~\ref{postburst}. As we have limited possibilities to classify certain types of AGNs weakly present in postburst galaxies one should be aware that a substantial part of the postburst sample at higher luminosities may host AGNs. This starts to become a problem for galaxies brighter than $L^*$ ($M_r$=-20.83) \citep{2001AJ....121.2358B}, corresponding to log$\cal M$$_{baryonic}$(\msun)$\sim$10.6). We discuss this problem in Section  \ref{sec:lumfunction}.

\subsection{Corrections for dust attenuation}
\label{sec:dust}

If only optical spectra are available, the corrections for dust attenuation are normally based on the Balmer emission line decrement, in particular the \ha/\hbx ratio. This is very useful if the Balmer lines are strong. But even when they are, it is not straightforward to apply the corrections since what we measure is a luminosity weighted mixture of contributions from regions of different dust content. In postburst galaxies the Balmer emission lines are practically nonexistent and we have to use another method. Below we describe how we have treated the effects of dust attenuation in the two cases.

\subsubsection{Emission line galaxies}

Traditionally the spectral distribution and strengths of the emission lines have been corrected for dust attenuation using the observed \hahbx emission line ratio as compared to the dust free condition. A relevant extinction curve is then applied to make a correction for an interstellar medium (ISM) of this type of galaxy. For a long time it was assumed that the attenuation could be approximated by a foreground screen of dust whose column density was independent of  the spatial distribution, age and metallicity variations across the main body of the galaxy. However, it has been obvious for many years that this is too crude an approximation. The optical depth in young star-forming regions is about 2 times higher than in older regions \citep{1997AJ....113..162C}. In the study by \citet{2000ApJ...539..718C} the effect of dust attenuation and how it changes with time during the early star-forming period when star clusters and super star clusters are formed, is discussed in detail. This model has been applied and further developed in various analyses of spectra of star-forming galaxies \citep[e.g][]{2004MNRAS.351.1151B,2012MNRAS.421.2002P}. In this paper we adapt a similar, but largely empirical approach to improve the correction for dust in starburst galaxies. But there are two major differences between our model and that of \citet{2000ApJ...539..718C}. One is that we can correct for the influence of the temporal effect on the dust attenuation on the \ewha while the Charlot \& Fall model does not provide this information. The second important difference is that we directly connect the age of the burst population to the \ha/\hbx ratio \emph{and} the strength of the attenuation. In the work by 
\citet{2004MNRAS.351.1151B}  the correction for dust attenuation is obtained from a specific dust model while the age has to be obtained from another source.
To first approximation it is our purpose to take the changes in the extinction coefficient during the early phase of star formation into account. Thus, our model adopts a correction for dust attenuation that systematically decreases with the age of the starburst population. 

It is well known that a significant part of the star formation in starburst galaxies occurs in young stellar clusters. The fraction of stars formed in clusters increases with star formation density  \citep{2010MNRAS.405..857G} and may reach high numbers. For example, \citet{2011MNRAS.417.1904A} conclude that in the nearby well investigated luminous blue compact galaxies ESO 338-IG04 and Haro 11, the amount of stars formed in clusters is $\sim$ 50\% while the remaining stars are formed in associations and agglomerates. Here we assume that the age dependence of dust attenuation found in clusters can be applied to the young stars in the galaxy as a whole. We show below that the fits of the model spectra to the SDSS data are significantly better with this approach.

As mentioned above, it was established many years ago that the dust attenuation in star-forming galaxies is about a factor of 2 lower in the older regions than in the younger and these results have been confirmed also in studies based on SEMs \citep[e.g.][]{2007MNRAS.381..263A}. But there is strong support that the youngest  population, i.e. stars formed during the first few Myrs,  is more severely affected than the young population of a normal mean age that holds for the starburst population in our study ($\sim$ a few 10 Myr). The extinction is many times higher when a star formation region is only one million years old than a few millions years later when the ambient dusty clouds have been dispersed, while it appears to become constant over the rest of the time when the star-forming region is producing ionising photons. 

\begin{figure}[t!]
\centering
 \resizebox{\hsize}{!}{\includegraphics{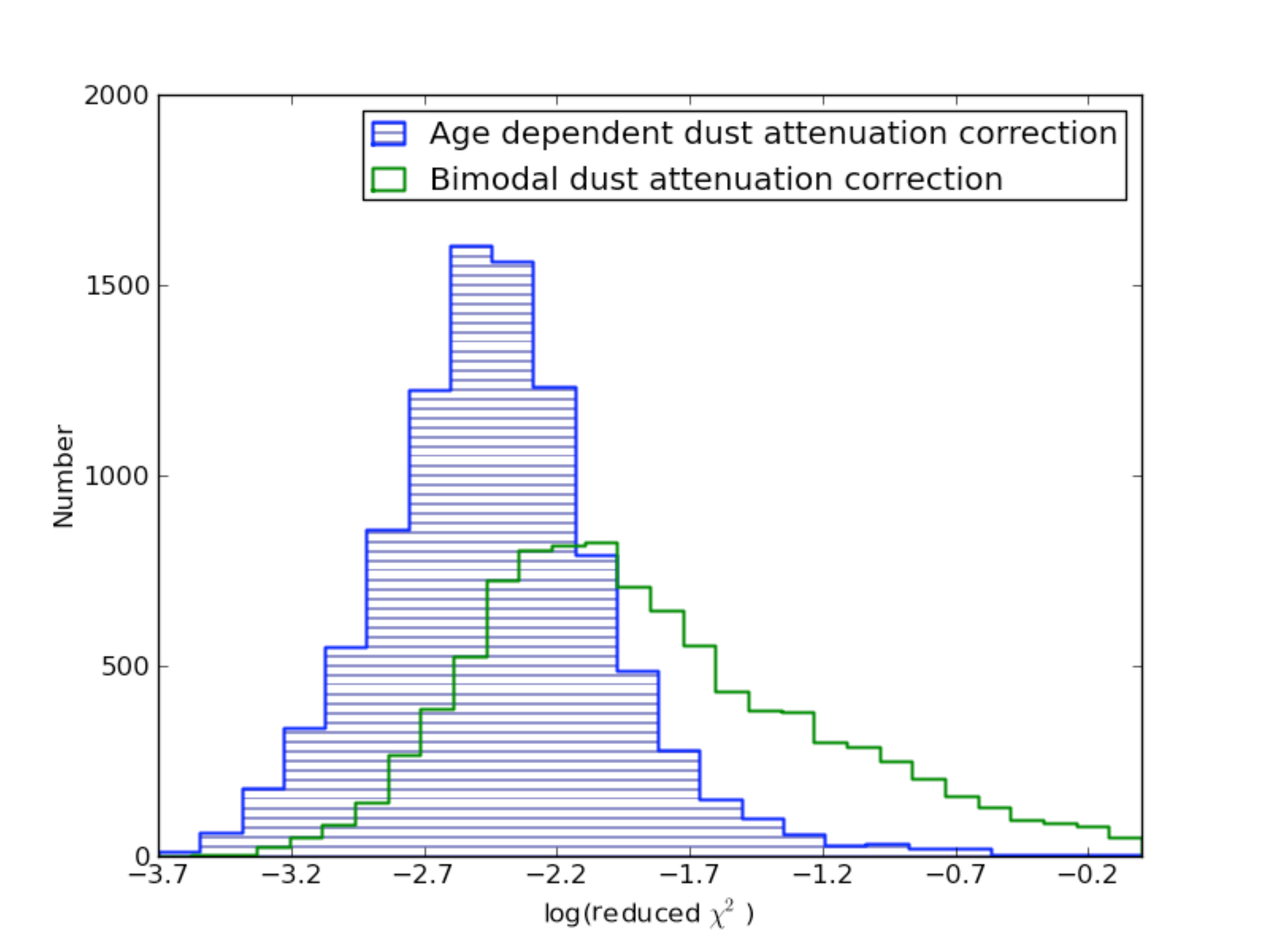}}
 \caption{ Reduced $\chi^2$ of the spectral fits adopting two different procedures for correction of dust attenuation. The hatched region shows the distribution after applying the corrections according to Tab. \ref{dust}. The other line shows the distribution after we have applied the more classic approach with an extinction coefficient that is twice as high in star-forming regions compared to passive regions. Obviously the age dependent dust attenuation correction yields the smaller reduced $\chi^2$ values and thus gives the better fit.}
\label{chi2hist}
\end{figure}

In several papers \citep{2010MNRAS.405..857G,2011MNRAS.414.1793A, 2011MNRAS.415.2388A} the dust attenuation has been derived for star clusters of different ages. The trends found in these investigations seem to give a similar relation between the amount of attenuation and age. Following these results, we have roughly adopted a relation between age and attenuation according to Table~\ref{dust} below where $A_0$ is the attenuation during the major period of star formation. We model the \hax and \hbx line strengths as a function of age for a range of attenuation values corresponding to an attenuation in the $V$ band of 0 $\le$$A_{0,V}$$\le $3.8 magnitudes. As we model the SDSS spectra we step through this library with increasing age. For each age we choose the model that gives the best agreement with the measured \hahbx ratio. This gives us $A_0$. For the case of  continuous star formation, each output with age $\geq$ 3 Myr contains stellar components with a range of values of the attenuation. In Fig.~\ref{chi2hist} we show the distribution of  the $\chi ^2$ residuals from the best fitting models of the SDSS spectra. Two cases are shown, one where we treat the dust attenuation according to Table \ref{dust} and the other where the dust attenuation in star-forming regions is twice that of non--star-forming regions. We see a clear difference between the two cases, in favour of  our approach.

\begin{table}[h]
\caption{Starburst dust attenuation for a given age range}             
\label{dust}      
\centering          
\begin{tabular}{ l l }     % 7 columns 
\hline\hline       
\\
Age & Dust attenuation (magnitudes) \\ 
\\
\hline   
\\
$<$3 Myr & (4.5-Age(Myr))$\times A_0$\\
3 Myr-100 Myr & $A_0$ \\
$>$100 Myr &$A_0$/2 \\
\\
\hline                  
\end{tabular}
\end{table}

\begin{figure}[t!]
\centering
 \resizebox{\hsize}{!}{\includegraphics{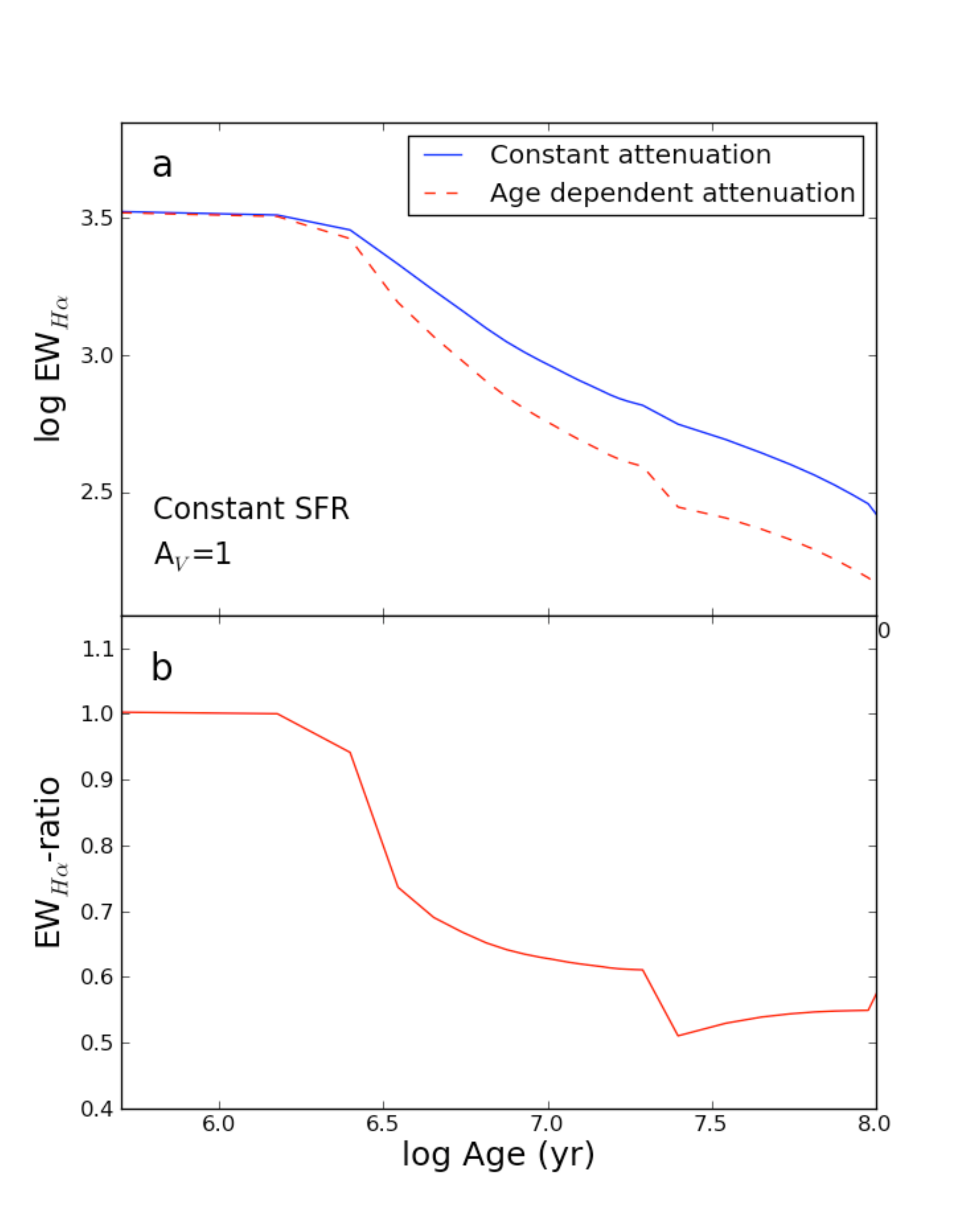}}
\caption{{\bf a)} Model of the evolution of \ewha ~in a star-forming galaxy under two different conditions for dust attenuation. The attenuation in the $V$ band during the major star formation epoch is $A_V=$1 mag. and the SFR is constant. The full drawn line shows how the equivalent width changes with age assuming constant dust attenuation. In this case \ewha ~ is identical to the dust free case. In the second case, shown in the dashed line, we assume an age dependency in the attenuation. {\bf b)}. The ratio between \ewha ~in the two cases. The figure shows that \ewha ~in the age dependent dust attenuation case drops off up to $\sim$ 50\%.}
  \label{whadiff}
\end{figure}

\begin{figure}[t!]
\centering
 \resizebox{\hsize}{!}{\includegraphics{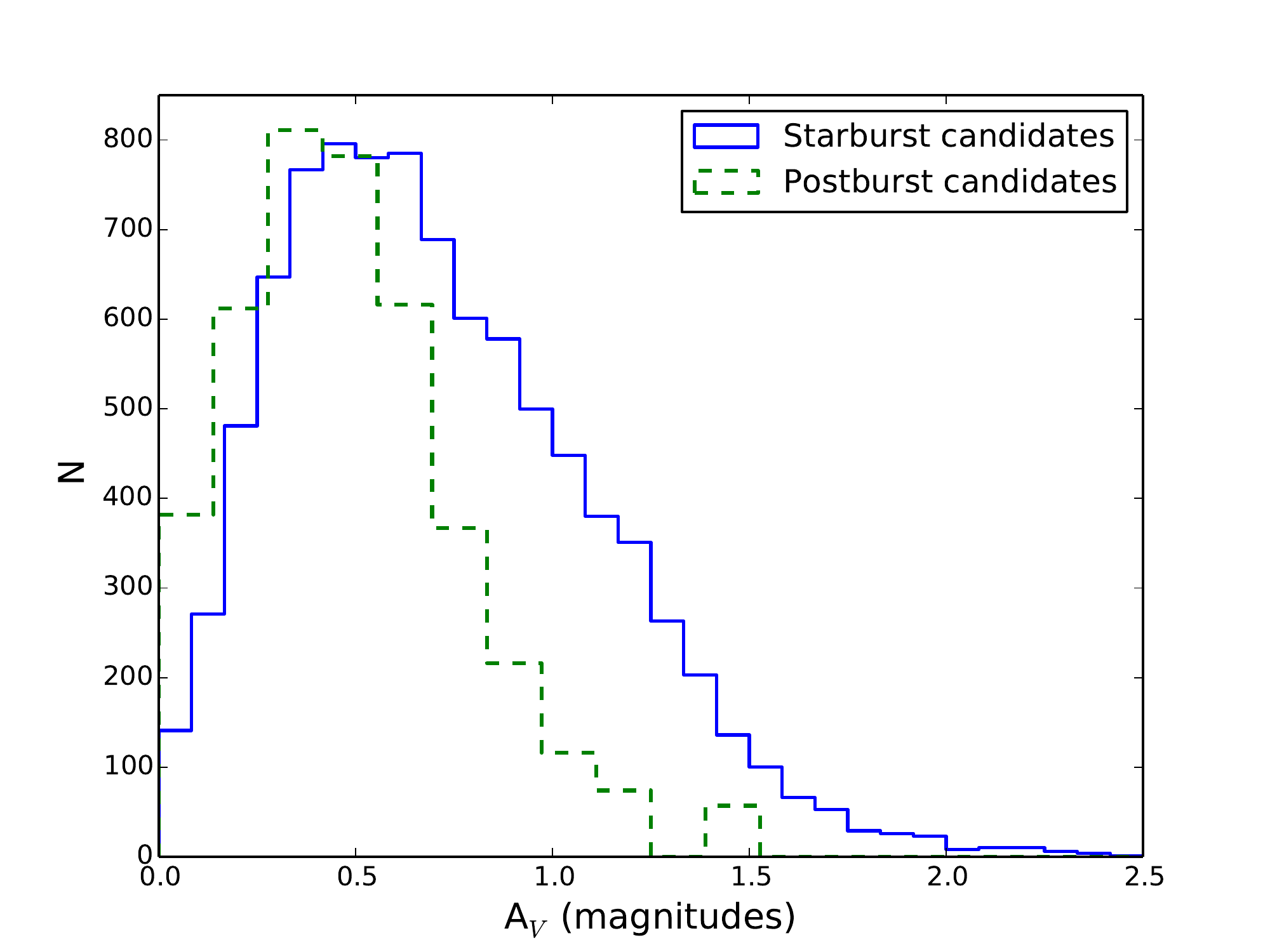}}
\caption{Luminosity weighted attenuation in the $V$ band in the samples of starburst and postburst candidates. Notice that for clarity, the bin size differs between the two samples.}
  \label{av_burstcand}
\end{figure}

The way we correct for dust attenuation has an important consequence. Since the attenuation is strongest during the epoch of highest Lyman continuum photon production, it means that \ewhax will be lower than if the extinction should be assumed to be constant during the entire SF epoch. We illustrate this is Fig.~\ref{whadiff} where we show the difference between our model and the case of constant attenuation. We see that, e.g. where $A_{0,V}$=1 mag., the case of variable attenuation will diminish \ewhax up to 50\%. When we discussed the selection criteria above we argued that a lower limit to \ewhax of 120\AA ~would be a sufficient criterion to make sure that most starbursts would be captured. If we now take the effects of age dependent attenuation into account, we need to decrease the lower limit of the \ewhax selection criterion. In Fig.~\ref{av_burstcand} we show the distribution of the attenuation in the $V$-band derived for the galaxies in the starburst sample. The median attenuation is $\sim$ 0.6 mag. An attenuation of $A_V$=1 mag. can suppress \ewhax with a {\it maximum} of 50\%. 94\% of the sub $L^*$ galaxies have $A_V$$<$1.0$^m$. At higher luminosities the dust attenuation increases to an amount that we do not properly handle here. One reason is that we will have severe problems with confusion caused by the presence of AGNs and shock heated gas caused by outflows. Another reason is that, if we wish to catch the strongly dust  enshrouded starbursts (where $A_V$$>>$1$^m$), we have to deal with weak emission lines (caused by dust attenuation) that would cause very unreliable results if we tried to use them for correction for dust attenuation. We show however that we still can reach interesting conclusions also in the high luminosity end by using the statistics of postbursts, being less affected by dust.

\subsubsection{Postburst galaxies}

In the postburst sample we cannot use the Balmer emission lines to derive the dust attenuation. Instead we have done the following. As in the case of the starburst candidates, we step through models of increasing age and select candidates that have \ewhd$\le$--6\AA. At each time step we vary the attenuation in steps of $A_V$=0.1 magnitudes in the range 0$\le $$A_V$$\le $1.5 magnitudes. For each value we make a fit. Finally we choose for each time step the case where the $\chi^2$ test gives the best result. We then apply the derived value of $A_V$ to the model spectra.  The distribution of $A_V$ is shown in Fig.~\ref{av_burstcand}. It can be compared to the results of \citet{2003MNRAS.341...33K}. We conclude that the results agree in the sense that $>$99\%  of the galaxies in both cases have $A _V$$<$1.5 magnitudes. The peak in our distribution differs by $\sim$0.2 magnitudes or 20\% if we transform the $z$ magnitudes (used by Kauffmann et al.) to V magnitudes but the medians are quite similar. We also find agreement between our derived distribution of $A_V$:s and that derived for K+A galaxies by \citet{2013MNRAS.431.2034M}. In the Appendix we also make a successful test of the reliability of our model determination of $A_V$. This makes us confident in our method of deriving the dust attenuation.

Here we assume that the dust is homogeneously distributed across the galaxy. This is a rough approximation and one may question if it is accurate enough to be applicable to the data in our statistical investigation. One could imagine that postburst cases where no or very weak emission lines are seen could perhaps host a starburst with high dust obscuration in the centre. But such cases appear to be quite uncommon. A look at the distribution of the young stellar population shows \citep{2012MNRAS.420..672S} that the blue population is mostly concentrated in the centre but over a larger volume than young stars in nuclear starbursts (normally constrained to the central 1 kpc). If there is a dusty starburst in the centre it should reveal itself by free--free emission in the radio domain. In a search for 20 cm emission from 36 E+A galaxies (\ewhd $\le$ --6\AA ~and no [\ion{O}{ii}]$\lambda$3727 or \ha)  \citet{2004A&A...427..125G} could set upper limits that excluded strong starbursts in 34 of these galaxies and moderate starbursts in the more nearby (z$\leq$0.15) cases. This was confirmed in a similar study by \citet{2012ApJ...761L..16N} of a sample of 811 K+A galaxies obtained from the FIRST survey, based on 1.4 GHz radio observations. The origin of 1.4 GHz emission is primarily thought to be synchrotron radiation from relativistic electrons, accelerated by the shocks from supernova ejecta. The signal thus can be calibrated to be used as a measure of the SFR. The results from this study are consistent with very weak or no star formation activity in general. One should also remember that a substantial part of the postburst galaxies may host AGNs which also contribute to the radio emission \citep{2007AAS...211.9722H}. It thus seems we can safely assume that the dust corrections we apply are valid for most of the stars we observe in the central region.

\subsubsection{General validity of the correction for dust attenuation}

\begin{figure}[t!]
\centering
 \resizebox{\hsize}{!}{\includegraphics{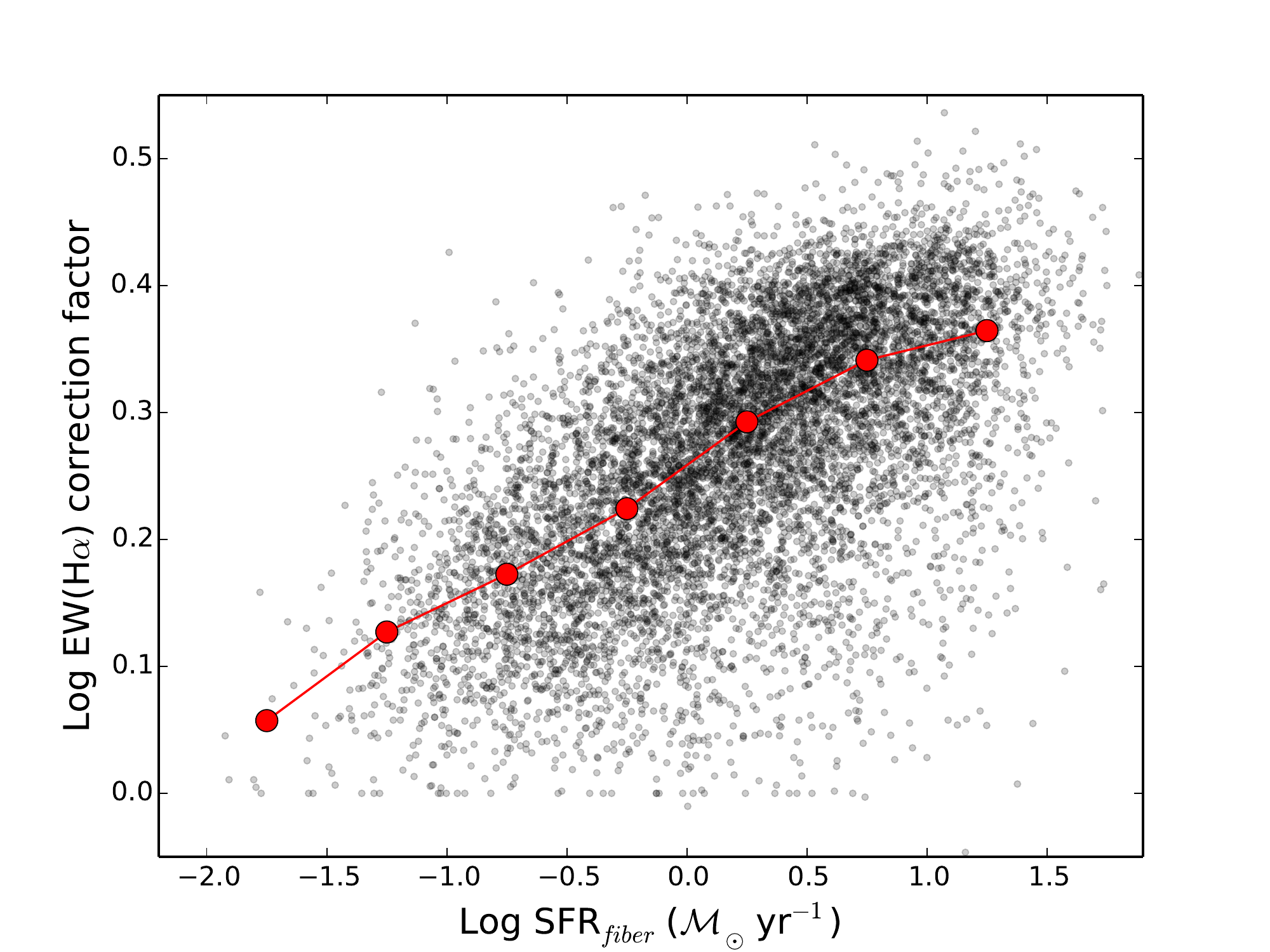}}
\caption{Dust correction factor of \ewhax vs. dust corrected SFR for the full starburst candidate The overplotted filled circles connected with solid lines are the median values in bins along the x-axis.}
\label{corrfactor}
\end{figure}

In Fig.~\ref{corrfactor} we show how the correction factor for dust attenuation of \ewhax varies with SFR.  How does this agree with other observations?  \citet{2003ApJ...597..269A} investigated the correlation between the SFR derived from the 1.4 GHz emission versus the \hax line -- the correction factor can be estimated from their diagrams. \citet{2009ApJ...703.1672K} used Spitzer 24$\mu $$m$ observations to compare with the SFR determined from \ha. Although the scatter in both their data is large, we find a good ($\sim$ 20--30\% at highest SFR) agreement between our result and theirs. At low luminosities, the SFR derived from \hax has been shown to be underestimated as compared to SFRs derived from UV fluxes \citep{2009ApJ...706..599L}. One problem may be a significant UV leakage from the \ion{H}{ii} regions \citep{2012MNRAS.423.2933R}. Yet another problem at low luminosities is the stochasticity of the IMF. However, these problems typically emerge at a SFR below 0.01 \msun yr$^{-1}$ \citep{2012ApJ...745..145D} and thus should have a minor effect on our sample (compare the range in SFR in Fig.~\ref{corrfactor}).

\subsection{Aperture effects}
\label{sec:apteff}

\begin{figure}[t!]
\centering
 \resizebox{\hsize}{!}{\includegraphics{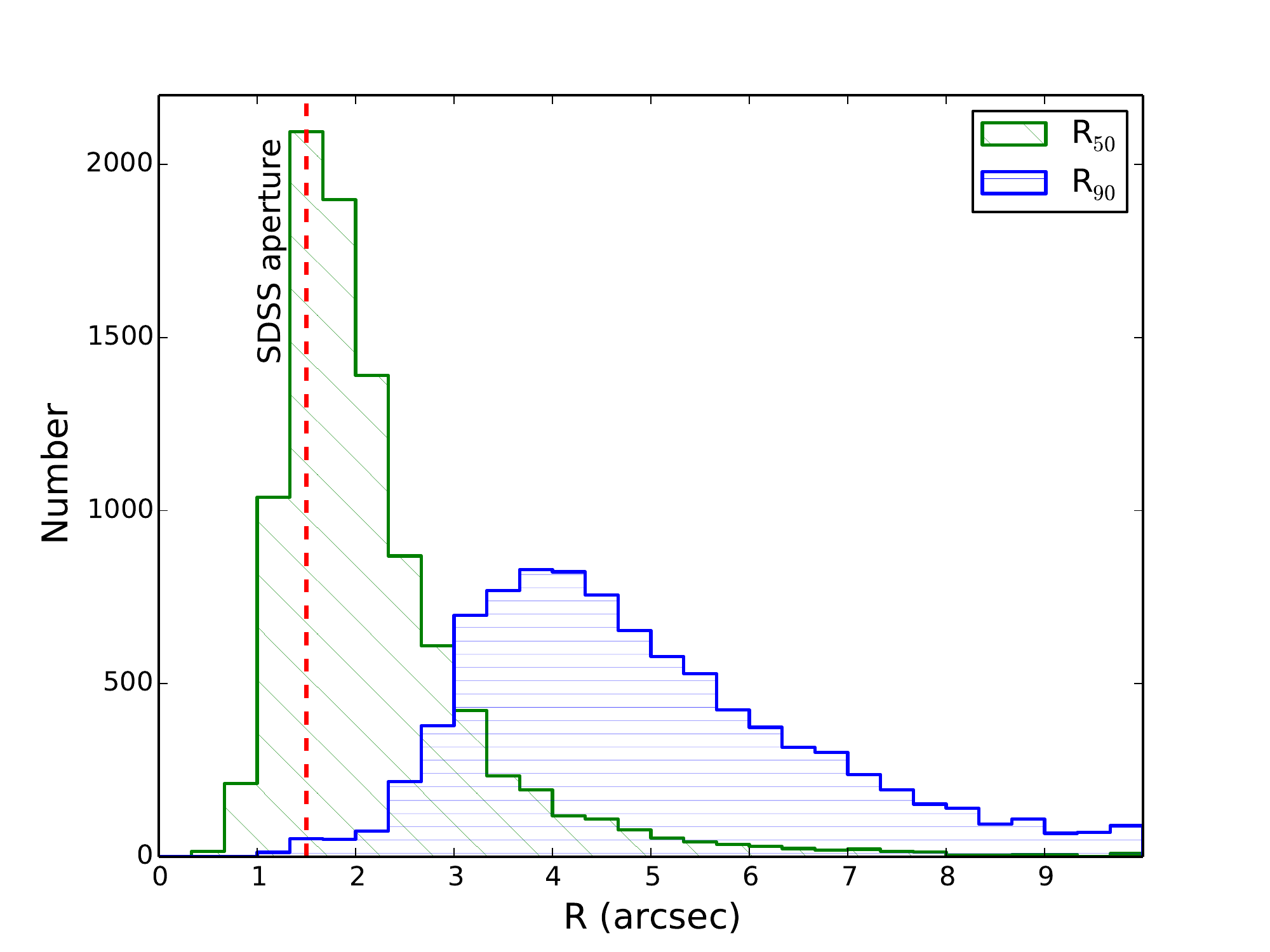}}
\caption{Radii encompassing 50\% (R$_{50}$) and 90\% (R$_{90}$) of the Petrosian flux in the $r$ band of the starburst galaxy candidates.}
  \label{hist_rpet}
\end{figure}

Due to the small 3$\arcsec$ cross section of the SDSS spectrograph fibers, one has to worry about aperture losses at small redshifts. This problem has been discussed at length by \citet{2004MNRAS.351.1151B} in their study of star-forming galaxies in the local universe (see also other references in the same paper). They applied corrections to their data, which they claim, nearly removed the bias. In the present investigation we wish to focus on low-to-intermediate luminosity starburst galaxies, hence we also want to push the limit as low as possible without introducing demonstrably large uncertainties in the derived data. However, we still chose a lower redshift limit (z$_{low}$=0.02) which is 4 times higher than the z$_{low}$=0.005 used by \citet{2004MNRAS.351.1151B}. Thus our data should be less affected by aperture losses than theirs for galaxies of the same physical sizes. Let us have a look at the aperture effects on the data based on our low redshift limit. In Fig. \ref{hist_rpet} we show the frequency distributions of the radii encompassing 50\% (R$_{50}$) and 90\% (R$_{90}$) of the Petrosian fluxes of our starburst galaxy candidates. The Petrosian flux is the flux within the Petrosian radius \citep{1976ApJ...209L...1P}. The Petrosian radius defines a region containing the flux of a galaxy with an exponential profile and about 80\% of the flux of a galaxy with a de Vaucouleurs profile. In 25\% of the cases the SDSS fiber aperture encompasses more than 50\% of the Petrosian flux (R$_{50}$). This indicates that aperture effects may be severe and we need to test the reliability of extrapolating the data outside the fiber aperture.

\begin{figure}[t!]
\centering
 \resizebox{\hsize}{!}{\includegraphics{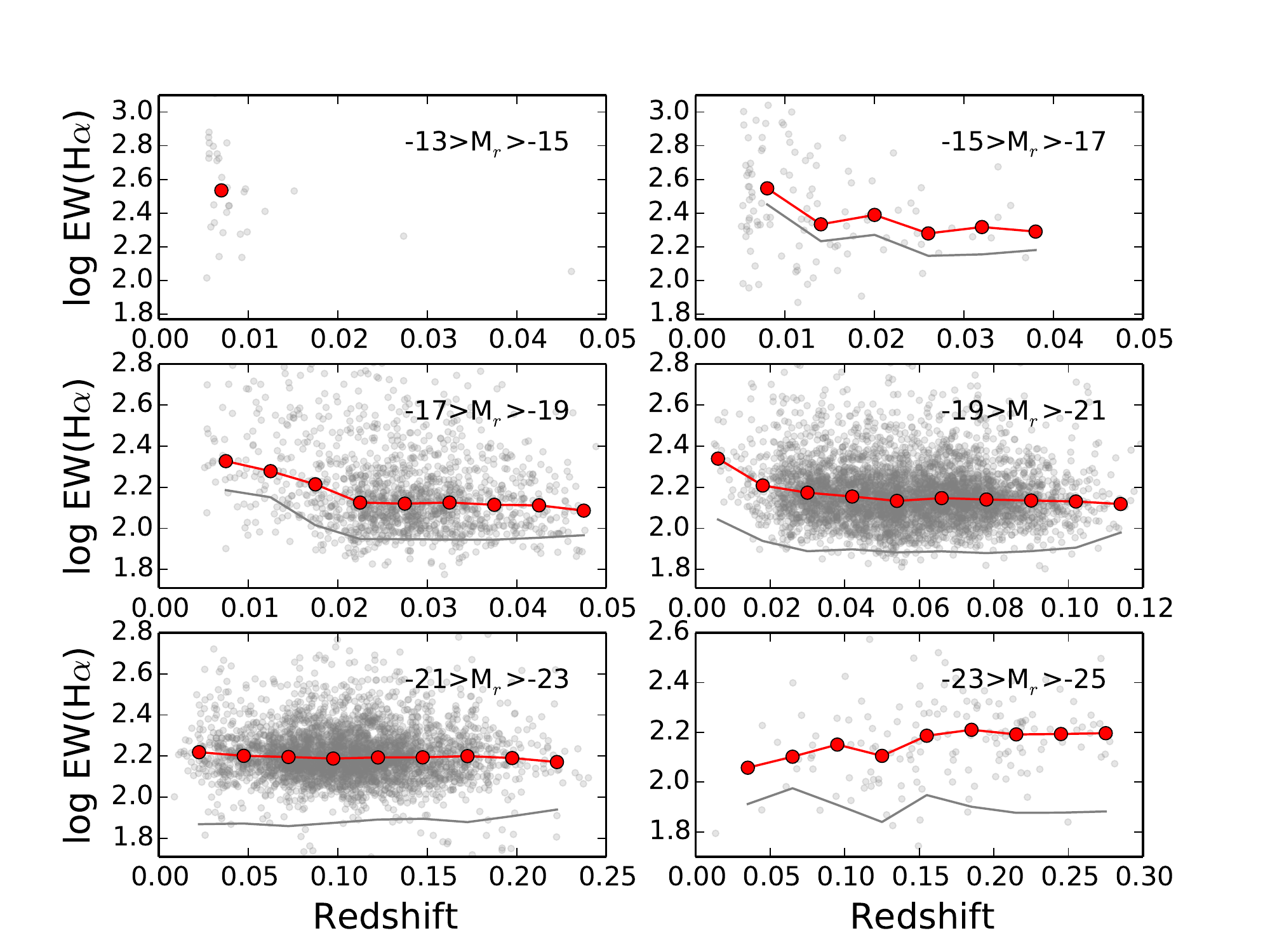}}
\caption{\ewhax versus redshift in different luminosity bins after remeasurements and correction for dust attenuation. The red solid line shows the medians of log(\ewha) in bins along the x-axis. For comparison, the faint gray line shows the medians before corrections for dust attenuation.}
  \label{ewhacorr_z}
\end{figure}

\begin{figure}[t!]
\centering
 \resizebox{\hsize}{!}{\includegraphics{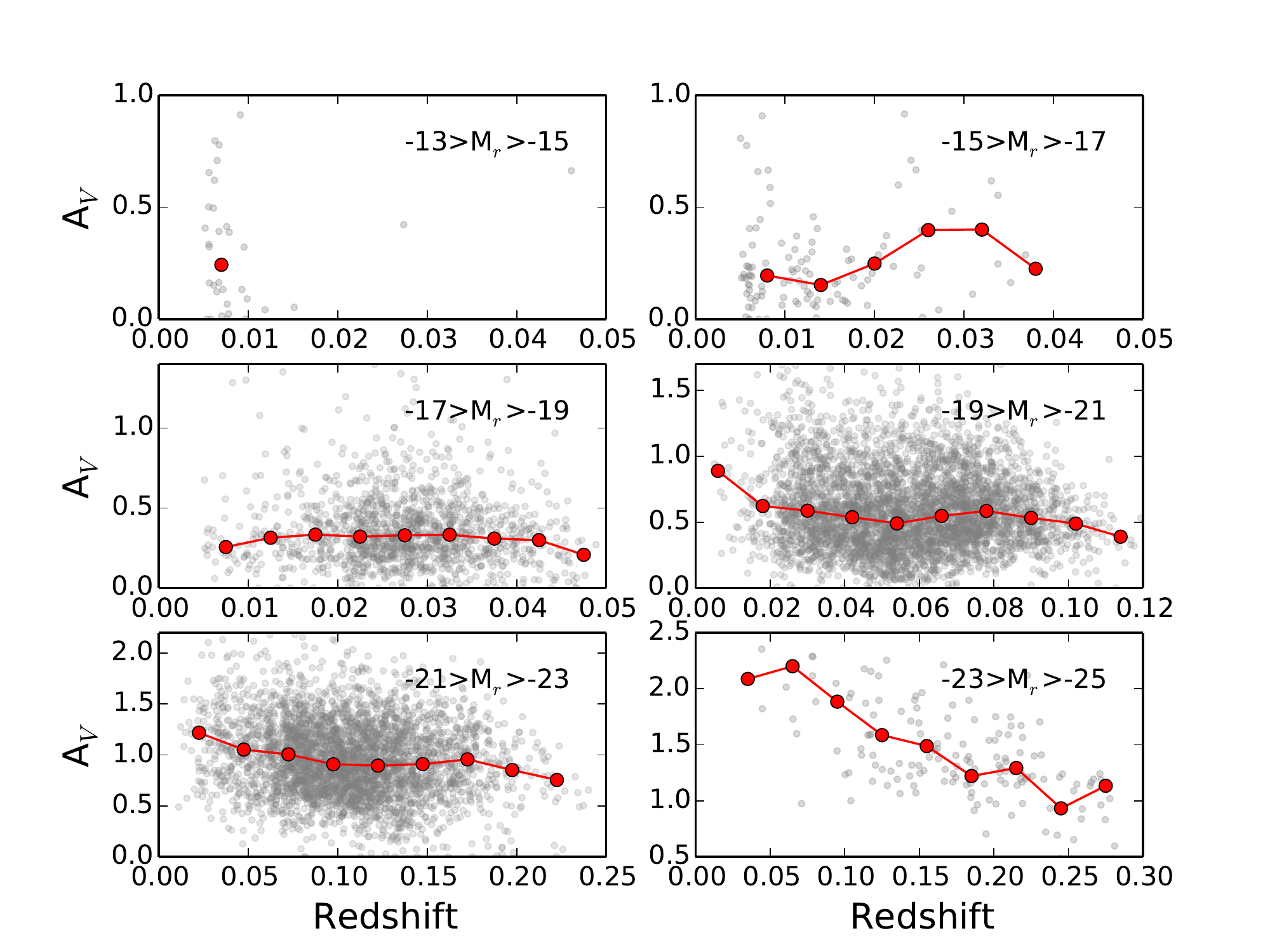}}
\caption{Dust attenuation in $V$ versus redshift in different luminosity bins. The solid red line shows the medians of $A_V$ in bins along the x-axis.}
  \label{av_z}
\end{figure}

\begin{figure}[t!]
\centering
 \resizebox{\hsize}{!}{\includegraphics{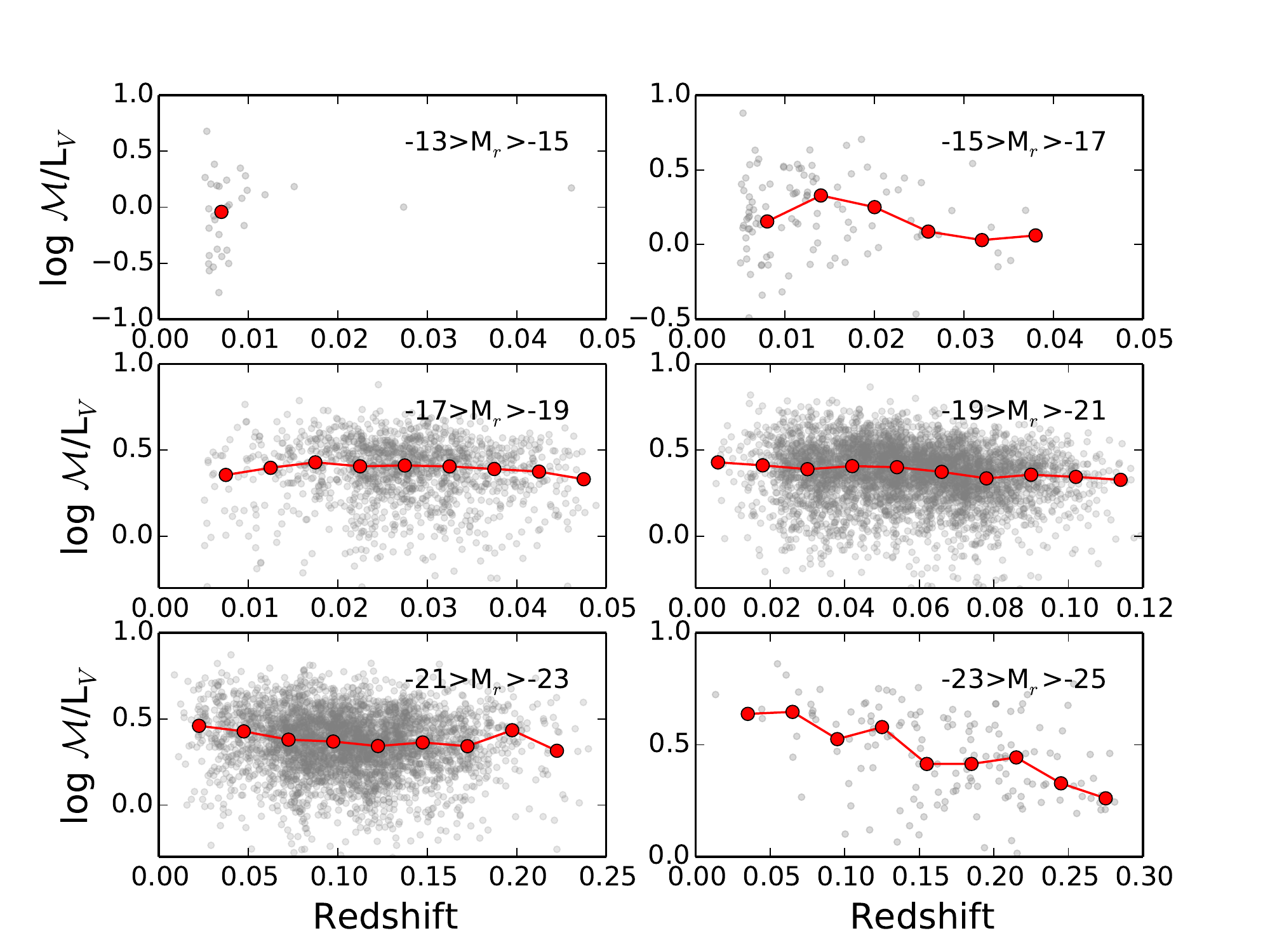}}
\caption{$\cal M$$_*$/$L_V$ in different luminosity bins as function of redshift, after correction for dust attenuation. The solid red line shows the medians of $\cal M$$_*$/$L_V$ in bins along the x-axis.}
  \label{ml_z}
\end{figure}

In Fig.~\ref{ewhacorr_z} we show in grey lines how the median values of \ewhax depend on redshift. We display the results in 6 boxes of increasing luminosity where the absolute magnitudes are derived as described below (see Sect. \ref{sec:lumfunction}). There are significant trends with redshift at z$\leq$0.02, in particular in the -17$> $M$_r$$>$-19 and -19$> $M$_r$$>$-21 box. This seems to imply that we have rather significant aperture effects at low redshifts. After remeasuring \ewhax and applying the corrections for dust attenuation we see from Fig.~\ref{ewhacorr_z} that the distributions have become somewhat flatter but that the trends at low redshifts essentially persist. Given these results it would seem prudent to use data above z=0.02, without applying other aperture corrections than a simple extrapolation from the fiber magnitudes to the total magnitudes in the $r$-band. Let us now examine two other parameters $-$ the dust attenuation in $V$, $A_V$, and the \mlvx ratio. Naively one would assume that the dust attenuation decreases with distance from the centre and thus we would expect to see $A_V$ decrease with increasing redshift for galaxies with the same luminosities. In a starburst galaxy one might also expect that the young stellar population has less and less influence at larger distances from the centre. We can see the actual situation in Fig. \ref{av_z} and Fig. \ref{ml_z}. We notice aperture effects in $A_V$ at z$<$0.02 and at the highest redshifts. In the highest luminosity bin the aperture effect is severe. At these luminosities there is also a risk that the spectra are of a composite nature, containing both a starburst component, an AGN component and a shock component. We see the same problem for the \mlx ratios. Otherwise the distributions for the majority of the data show only weak trends of the order of 0.1 dex over the redshift interval in each luminosity bin except the -23$>$M$_r$$>$-25. We therefore feel we can rely on our extrapolation from fiber data to global data for galaxies of $M_r\gtsim$-23. We do not further discuss the properties of the brightest galaxies (M$_r<$-23) here.

\section{Results}
\label{sec:results}

\subsection{The luminosity function}
\label{sec:lumfunction}

\begin{figure}[t!]
\centering
 \resizebox{\hsize}{!}{\includegraphics{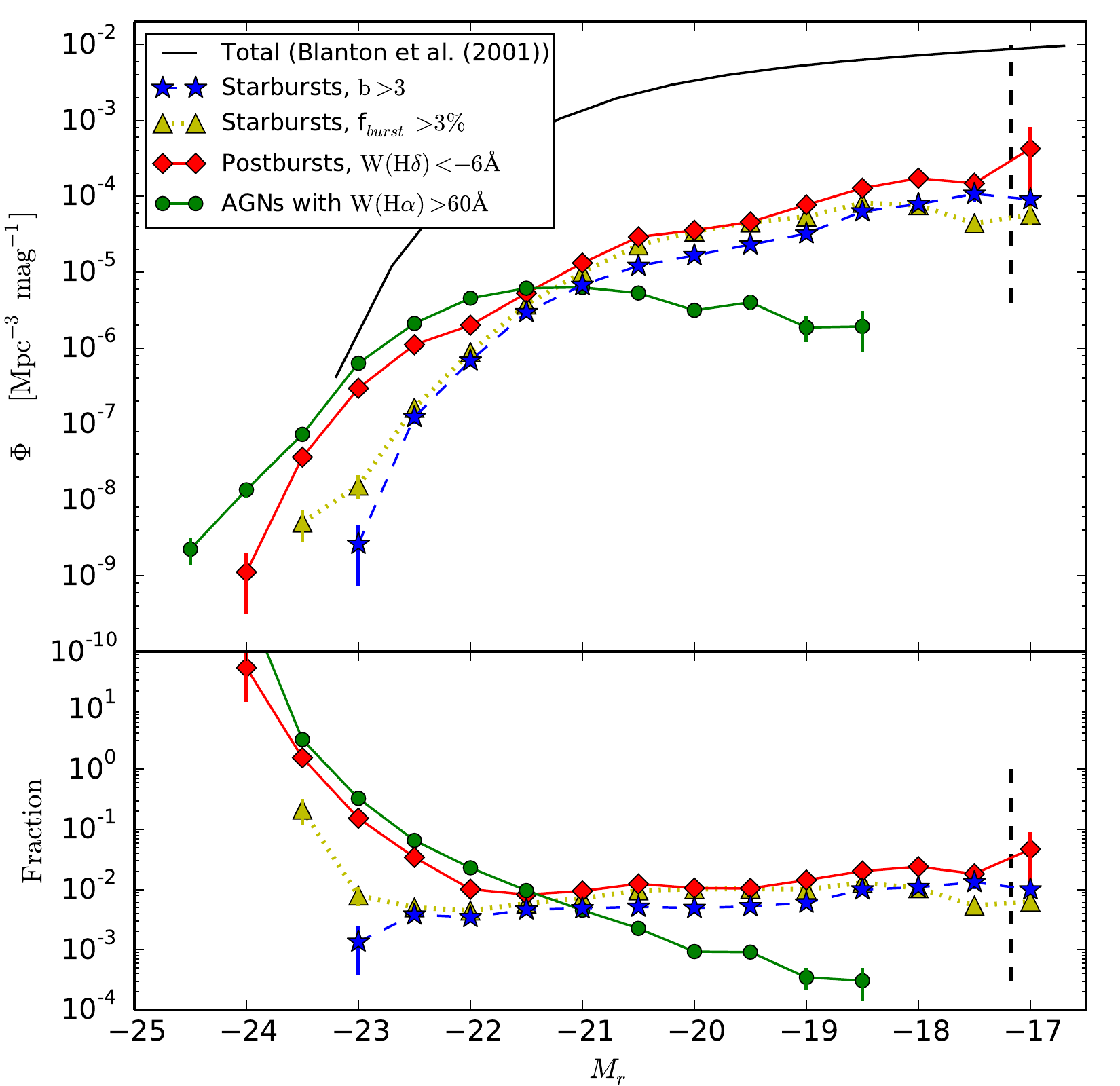}}

\caption{{\it Upper diagram:} LF of 1) starburst galaxies with a birthrate parameter $b$$>$3 2) starburst galaxies with a mass fraction of the burst $f_{burst}$$>$3\% 3) postburst galaxies 4) AGNs in the emission--line galaxy sample 5) the full galaxy LF of the local universe as derived by  \citet{2001AJ....121.2358B}. The sampling completeness limit based on the apparent spectroscopic limit $M_r$=17.5 and the redshift limit z=0.02 is indicated with a vertical hatched line. {\it Lower diagram:} The relative fraction of the different samples with regards to the total LF, as a function of absolute magnitude. The vertical bars are standard deviations. Please notice that we have assumed (for purely technical reasons) that the Blanton LF for normal galaxies can be extrapolated to M$_r$$<$-23.}
\label{lumfunction}
\end{figure}

In the derivation of the LF, $\Phi$, we use the $1/V_{max}$ method \citep{1968ApJ...151..393S}. Here we follow a procedure as described by e.g. \citet{1996MNRAS.280..235E}. The LF can then be calculated from

\begin{equation}
\Phi (M) ~\mathrm{d}M = \sum_{i=1,n} \frac{1}{V_{max,i}}
\end{equation}

where $\phi$ is the number of galaxies per volume element within a certain absolute magnitude interval $dM$. $V_{max}$ is the maximum volume within which the galaxy could be observed under all observational constraints, including magnitude and redshift limits and $n$ is the number of galaxies within the absolute magnitude interval $\mathrm{d}M$. $V_{max}$ is given by 

\begin{equation}
V_{max}=\Omega \int_{z_{min}}^{z_{max}}\frac{dV}{dz}\mathrm{d}z,
\end{equation}

where $\Omega$ is the effective area in steradians, $z_{min}$ is the lower redshift limit ($z=$0.02) and $z_{max}$ is the upper limit, corresponding to the distance at which the apparent magnitude of the galaxy would be equal to the limiting magnitude, $m_r$=17.5. $dV/dz$ is the comoving volume element per redshift interval and is given by

\begin{equation}
\frac {dV}{dz}=\frac{c}{H_0}\frac{d_L(z)^2}{(1+z)^3(1+2q_0z)^{1/2}}
\end{equation}

over one steradian. $d_L$ is the luminosity distance given by

\begin{equation}
d_L(z)=\frac{cz}{H_0}\frac{1+z+(1+2q_0z)^{1/2}}{1+q_0z+(1+2q_0z)^{1/2}}
\end{equation}

The  standard deviation per bin is 

\begin{equation}
\sigma_{\Phi}=\sqrt{\sum_{i=1,n}\frac{1}{V_i^2}}
\end{equation}

The absolute magnitude of a galaxy is calculated from

\begin{equation}
M=m-5logd_L(z)-K(z)-A_{gal}-25
\end{equation}

where $d_L(z)$ is the luminosity distance in Mpc, $K(z)$ is the $k$-correction and $A_{gal}$ is the galactic absorption in magnitudes.

The LF based on the $1/V_{max}$ method holds true for a homogeneous distribution of sources. In shallow surveys local inhomogeneities may cause problems, in particular at low luminosities. By choosing a lower redshift limit of z=0.02 we believe the most serious problems of this kind will be overcome except for the very faintest parts of the sample. Small number statistics also increases the uncertainty at low fluxes. At the high end one may have problems with evolutionary effects. We have very few objects above z=0.2. Thus the upper redshift limit ($z$=0.4) is sufficiently small to exclude evolutionary effects that may have an influence on the results, in particular in the low--mid range luminosity region.

In Fig.~\ref{lumfunction} we show the derived LF for 5 sets of data: 1) the full galaxy LF of the local universe as derived by  \citet{2001AJ....121.2358B} 2) starburst galaxies with a birthrate parameter $b$$>$3 3) star-forming galaxies with a mass fraction of the young population $>$3\% 4) postburst galaxies with \ewhd$<$-6\AA  ~5) AGNs in the emission--line galaxy sample (\ewha$_{,em}>$60\AA). We want to point out that the starburst galaxy LF starts to become obsolete at galaxies brighter than $\sim L_*$ ($M_r$-20.8; log $\cal M$(\msun)$\sim $10.6) because of the increasing influence of AGNs and severe dust obscuration.

At this point we want to  remind the reader what we mean by postburst galaxies. Strictly we would have liked to define a postburst galaxy as a galaxy that fulfills the criteria \ewhd$<$-6\AA ~and $b>$3. In Sect.~\ref{subsec:connecting} we explained the problem in determining the $b$-parameter for postburst galaxies. 92\% of the galaxies in the postburst candidate sample have $<$$b$$>$ larger than 3, if we use the derived exponential (or constant) time scale of the burst as a rough measure of the duration of the burst. This is at least an indication that the postburst galaxies, based only on the \hdx criterion are closely linked to the starburst population.  Due to the uncertainty in the determination of the $b$ parameter however, we decided to use the term postburst for galaxies only with \ewhdabsx$\leq-6$\AA, based on the remeasured values (as mentioned in Sect.~\ref{subsec:selection})

There are a few conclusions we can immediately extract from Fig.~\ref{lumfunction}. Starburst galaxies defined both from the birthrate parameter and the mass fraction criterion are less common than generally assumed. AGNs start to become significant relative to starburst galaxies at an absolute magnitude of  $M_r$ $\approx$ --21. At fainter luminosities we have more or less clean cut cases of starbursts. If we base our definition of starburst on the $b$$>$3 criterion, those galaxies contribute no more than 1.0$\pm0.1$\% to the total LF at an absolute magnitude of $M_r$=-18 and 0.6$\pm0.05$\% at  $M_r$=-19. According to \citet{2013AJ....145..101K}, about 70\% of the dwarf galaxies in the local universe are star-forming, i.e. are classified as Ir, Im or BCD. Comparing their fraction means that only 1\% of star forming dwarfs is a starburst galaxy. The relative fraction decreases only weakly with increasing luminosity and reaches a relative fraction of 0.3-0.4\% at $M_r$=-22.5. However, as is discussed in the next section, one has to keep in mind that a fraction of the starbursts in massive galaxies may be difficult to detect if an AGN is present at the same time. 

\begin{figure}[t!]
\centering
 \resizebox{\hsize}{!}{\includegraphics{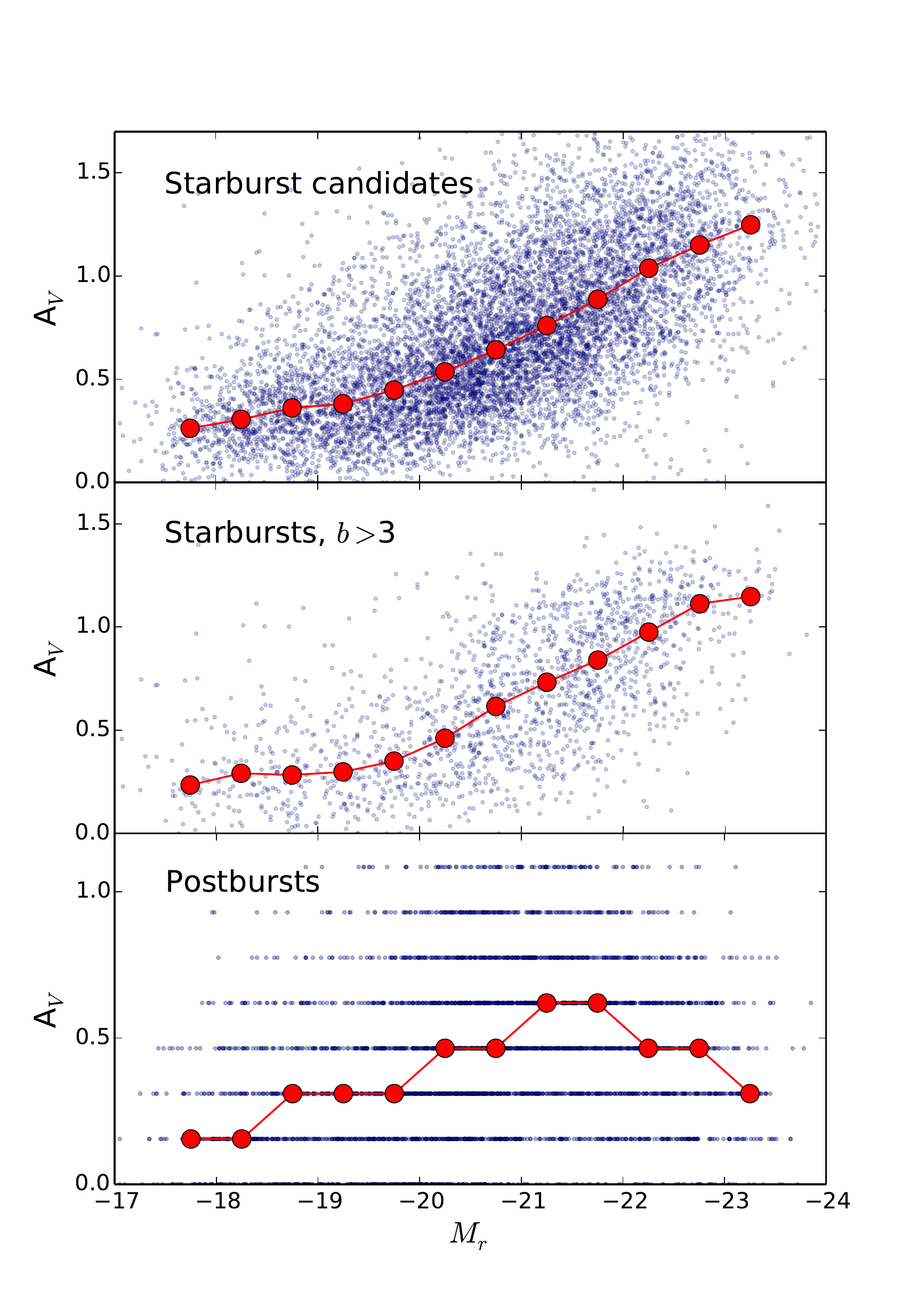}}
\caption{Dust attenuation given in $A_V$, as function of absolute magnitude. Red dots are medians. Notice the peak in $A_V$ for the postbursts which is not seen in the starburst sample.}
\label{av_absmag}
\end{figure}

We see from Fig.~\ref{lumfunction}  that the LF of postburst galaxies tightly follows that of starburst galaxies at low-intermediate luminosities. At higher luminosities the postburst LF deviates from starbursts and joins the AGN LF. We show below that the lifetimes of the starbursts are roughly independent of mass. The lifetime of postbursts are insensitive of mass if the star formation is shut down during the postburst phase. Then the postburst criterion, \ewhd, is only a function of time. We show in Sect. \ref{trends} that this is probably true at high luminosities. Here, the postbursts follow the AGNs. AGNs cannot produce postburst signatures. Therefore, considering the strong correlation between starbursts and postbursts up to $M_r$$\sim$-21.5, we argue that the starburst LF follows that of the postbursts also at higher luminosities, but that the starburst galaxies for some reason are difficult to detect. One probable reason is that the AGNs are outshining the starbursts. Another reason is that starbursts in luminous galaxies are strongly dust enshrouded while the postbursts and AGNs may be less affected. We know that the dust opacity increases with mass and luminosity.  At high luminosities the starburst phase is to a large extent hidden by dust while the postburst phase is less affected. This is clearly seen in Fig. \ref{av_absmag}. Notice also that the dust attenuation in posbursts reaches a maximum at $M_r$$\sim$-21.5. In Fig. ~\ref{lumfunction} we see that postbursts closely follow the AGNs at high luminosities. This indicates that the AGN phase has been preceded by a dusty starburst. Shortly after the starburst, the dust opacity decreases and the postburst/AGN phase begins. We can infer from the high fraction of postbursts that the number of starbursts at high luminosities have been underestimated by about one dex. We conclude that from the fact that we find a strong link between starburst and postburst LFs, it follows that there must also be a strong link between starbursts and AGNs. The AGN phase seems to appear after the starburst has ceased, when the dust has probably been destroyed or removed by radiation pressure and/or gas outflows. This scenario is also supported from observations of luminous postburst quasars \citep{2013ApJ...762...90C}.

\begin{figure}[t!]
\centering
 \resizebox{\hsize}{!}{\includegraphics{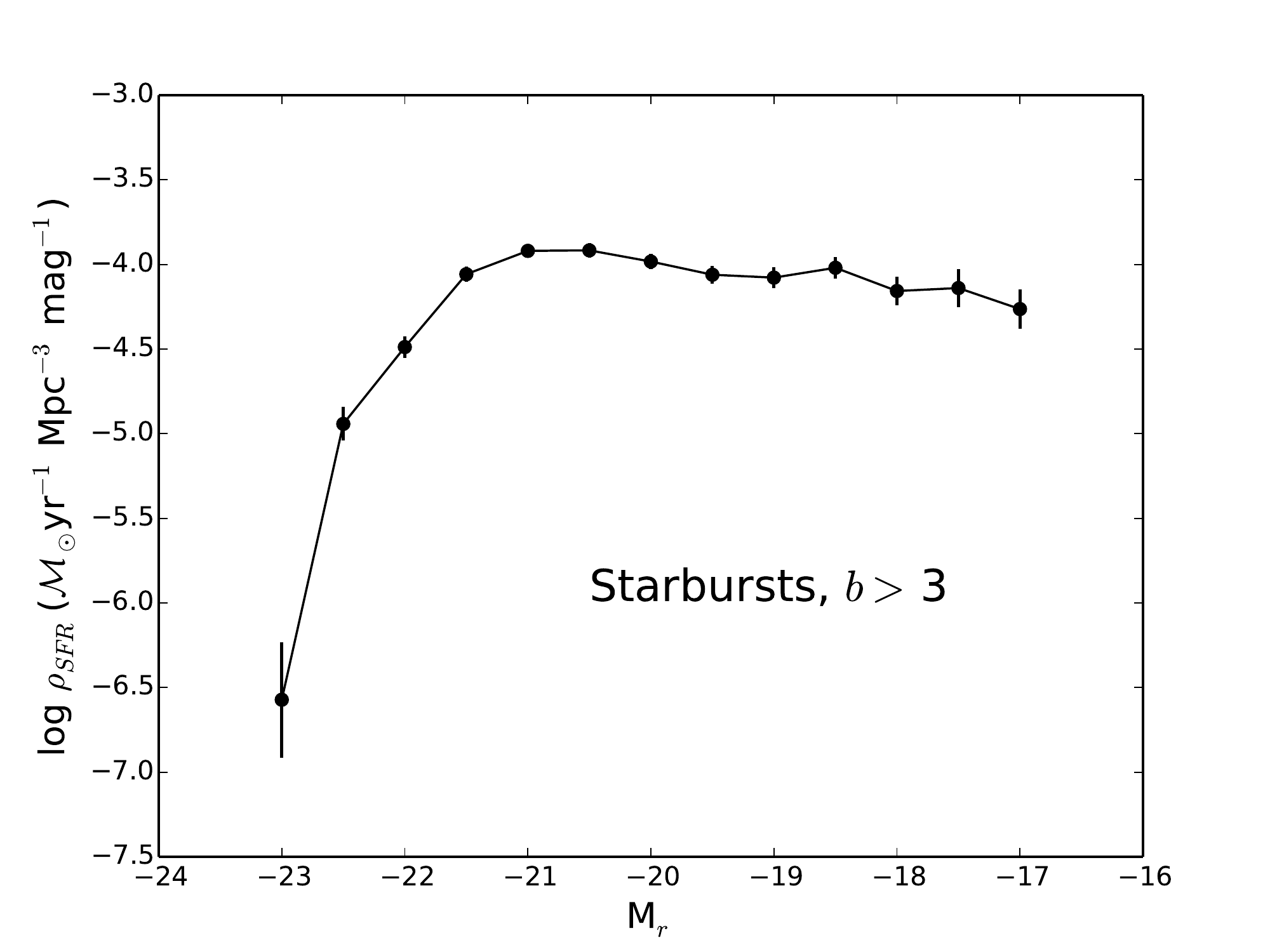}}
\caption{SFR density of the starburst $b>$3 sample vs. absolute $r$-magnitude. The errors bars are standard deviations.}
\label{sfr_mabs}
\end{figure}

An important question is how much starbursts contribute to the star formation in the local universe.  Fig. \ref{sfr_mabs} shows how the SFR density varies with luminosity. By integrating the SFR from $M_r$=-17 up to -23 we find that starbursts with $b$$>$3 have a SFR density of 4.43$\pm$0.14$\times$10$^{-4}${\msunyr} Mpc$^{-3}$. The errors given are random errors, assuming a Gaussian distribution. The systematic errors dominate and mainly stem from the uncertainties in the model, the dust correction and the aperture correction. We estimate these give rise to an additional $\sim$40\% uncertainty. The given SFR density only reaches down to $M_r$=-17. We see from the diagram that the SFR density distribution is only slowly declining towards fainter magnitudes. We know very little about the LF of starbursts at faint absolute magnitudes. If we extrapolate down to $M_r \sim$-13, assuming the starburst LF to rise slowly towards fainter magnitudes as do normal SF galaxies \citep{2015ApJ...810..108M}, we can obtain a rough estimate of the contribution to the SFR density from starburst galaxies at the faint end on the LF. From the diagram we estimate that this additional contribution between $M_r$ = -17 and -13 would amount to $\sim$ 3$\times$10$^{-4}${\msunyr} Mpc$^{-3}$. To this we should add the contribution hidden in AGNs. As we discussed above, the starburst LF is running in parallel to the postburst LF at intermediate to low luminosities. Postbursts are twice as common as starbursts. If we assume this to be true even at higher luminosities when starbursts are hidden by dust or outshone by bright AGNs we estimate that we miss $\sim$ 1.3$\times$10$^{-4}${\msunyr} Mpc$^{-3}$. The total contribution from starbursts in the range -23$<$$M_r$$<$-13 then becomes $\rho_{SFR}$=8.7$\times$10$^{-4}${\msunyr} Mpc$^{-3}$ of which $\sim$ 15\% are lost at high masses. Brinchmann et al., display in their Fig. 16 the redshift dependence of the SFR density on redshift based on different investigations. The median redshift our starburst sample is z=0.07. At this redshift the SFR density is $\sim$ 1.7$\times$10$^{-2}$\msunyr Mpc$^{-3}$. Their lower redshift limit restricts the sample to galaxies brighter than $M_r$=-14. For our starburst sample this corresponds to $\rho_{SFR}$=7.7$\times$10$^{-4}${\msunyr} Mpc$^{-3}$. Thus starbursts contribute $\sim$ 4.4$_{-1.2}^{+1.8}$\% to the total stellar production in the local (z$\sim$ 0.07) universe.

It is also interesting to compare our results to results from the $Herschel$ mission \citep{2011ApJ...739L..40R,2012ApJ...747L..31S,2015A&A...575A..74S}, although these investigations are based on near-IR data and do not include low mass galaxies. In these studies the starburst phenomenon is discussed in relation to the ``galaxy main sequence" (hereafter GMS), i.e. the SFR versus the galaxy mass. One particular aspect concerns the distribution of the specific SFR (sSFR). Provided one assumes that the GMS has a Gaussian distribution in log sSFR, starbursts here appear as a separate population. \citet{2011ApJ...739L..40R} focus on the redshift interval 1.5$<$$z$$<$2.5, believed to be the epoch of peak star formation in the universe. They require a starburst galaxy to have a SFR more than 4 times the SFR on the GMS. This definition is close to the $b$$>$3 definition in the local universe considering that the SFR has decayed since z=2. Rodighiero et al. find that starburst galaxies represent only 2\% of mass-selected SF galaxies and account for only 10\% of the SFR density at z$\sim$2. These results are not too different from our results. Compared to the results from Rodighiero et al., \citet{2015A&A...575A..74S} find values about 50\% higher and essentially independent on redshift in the range 0.5$<$z$<$4.

\subsection{Masses}
\label{masses}~

There is an ongoing debate about the cause of the broadening of the Balmer emission lines, whether dominated by virial motions and/or regular rotation and thus potentially useful for mass determinations or by non-ordered motions caused by gas infall or supernova generated outflows. In a few papers \citep{1981MNRAS.195..839T,1987MNRAS.226..849M,1988MNRAS.235..297M} it was demonstrated that the velocity dispersion of \ion{H}{ii} regions derived from the width of the \hbx emission line, $\sigma_{H\beta}$, is correlated with their luminosities, thus indicating that there should also exist a similar relation between $\sigma$ and mass. \citet{1996ApJ...460L...5G}, looking at a small number of galaxies, took the idea further and demonstrated that emission line widths can be used to derive reasonably correct masses. In two papers by \citet{o99,o01}, Fabry-Perot spectroscopy was carried out on 6 blue compact galaxies and two companions. They found a good agreement between dynamical masses and photometric masses. This is further confirmed by our recent study of a larger sample of local starburst dwarf galaxies (Marquart et al., in preparation) and other investigations \citep{2013ApJ...764L...8B}. 

Here we continue the comparison between dynamical and photometric masses. To derive the dynamical estimates of the masses, we use the \hax emission line widths. The velocity dispersion was derived from $\sigma_{corr}^2 = {\sigma_{H\alpha}}^2-{\sigma_{IP}}^2$, where $\sigma_{H\alpha}=FWHM_{H\alpha}/2.35$ and $IP$ signifies the instrumental profile. We have assumed $\sigma_{IP}$=70 kms$^{-1}$, as obtained from the SDSS online documentation\footnote{http://classic.sdss.org/dr7/products/spectra/}.

The dynamical masses were derived under the assumption that we are working with dynamically relaxed systems. In some starburst galaxies, in particular in mergers this is probably not a valid assumption since we expect gas infall as well as outflows caused by supernovae activity. Thus, we have poor knowledge about the kinematics of the ionised clouds in a specific galaxy, but can expect to obtain important information from a statistical sample. For simplicity we assume that the clouds move either in circular orbits in a disc like structure or in viral motions in a spherically symmetric system. A simple approximation of the relation between the dynamical mass and the width of the \hax line is give by the equation \citep{1983A&A...125..394L,o01}:

\begin{equation}
{\cal M}_{dyn} = f\times 10^6\times R\times \sigma_{corr}^2
\end{equation}

 $f$ is a constant which has the value $f\approx$ 0.79 for a disc and $f\approx$ 1.1 for a spherical distribution. In the first case it is assumed that the inclinations of the disks are randomly distributed. $\cal M$$_{dyn}$ is the dynamical mass in solar units, $R$ is radius in kpc and $\sigma_{corr}$ is given in km s$^{-1}$. For disks we assume that $R$ is equal to the scale length. In the spherical case we assume that $R=R_{eff} = R_{50}$. 
 
\begin{figure}[t!]
\centering
 \resizebox{\hsize}{!}{\includegraphics{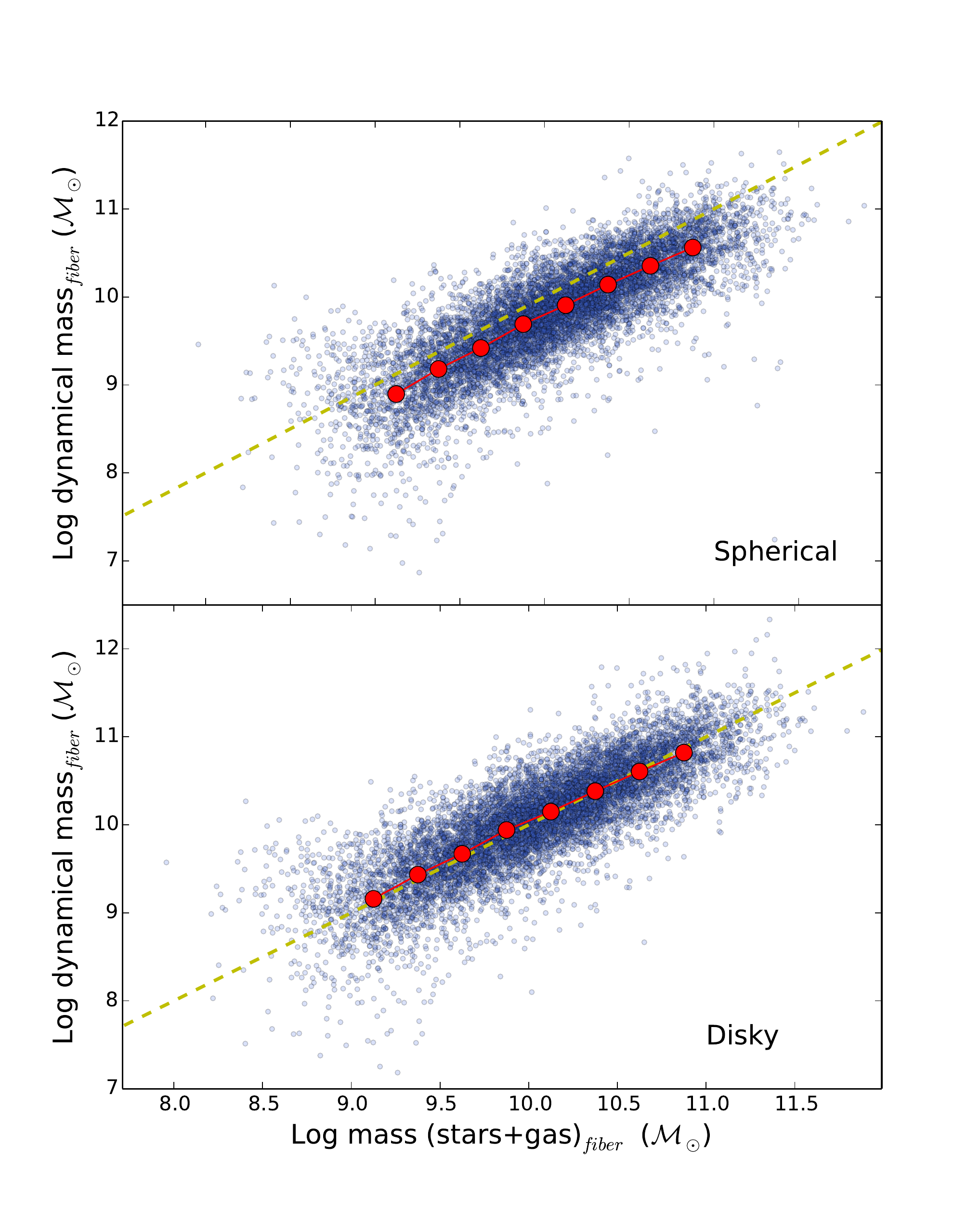}}

\caption{Dynamical mass, derived from the width of the \hax emission line versus the baryonic mass (stars+gas), derived from the spectral fit.  The two diagrams display the results emerging from two different assumptions about the kinematics of the \hax emitting gas. In the spherical case it is assumed that virial motions take place in a spherically symmetric structure and in the disc case it is assumed that the gas clouds move in circular orbits in a disk. Overplotted circles are the median values after binning along the x-axis. The hatched line marks the 1:1 relation.}
\label{mdyn_mphot}
\end{figure}

 In Fig.~\ref{mdyn_mphot} we show how our baryonic masses correlate with the dynamical masses for the two cases. To calculate the total masses we have added a crude estimate of the gas mass based on observations of star-forming galaxies of various masses. The total gas mass was estimated to be \m$_{gas}$=\m$_{H I}$+\m$_{He}$+\m$_{H_2}$. The {\hi} relation was obtained from a linear fit to various data in the literature giving  
 
\begin{equation}
log\m_{H I}=-0.35\times M_B+2.64 
\label{himass}
\end{equation}

(Bergvall et al., in prep.), where $M_B$ is the absolute magnitude in the Cousins $B$ band. To this was added a helium component of 25\% in mass (\m$_{He}$). The molecular masses were obtained from \citet{1992A&A...265...19S} and \citet{2000MNRAS.318..124G}.  We assumed a linear relation giving 

\begin{equation}
log\m_{H_2}=-0.45\times M_B+0.35. 
\label{molmass}
\end{equation}

Masses are in solar units. We applied these relations assuming $M_g$=$M_B$ where $M_g$ is the absolute magnitude in the SDSS $g$ band. The total baryonic mass is then \m$_{tot}$ = \m$_{gas}$+\m$_{stars}$, where \m$_{stars}$ includes stellar remnants.

\begin{figure}[t!]
\centering
 \resizebox{\hsize}{!}{\includegraphics{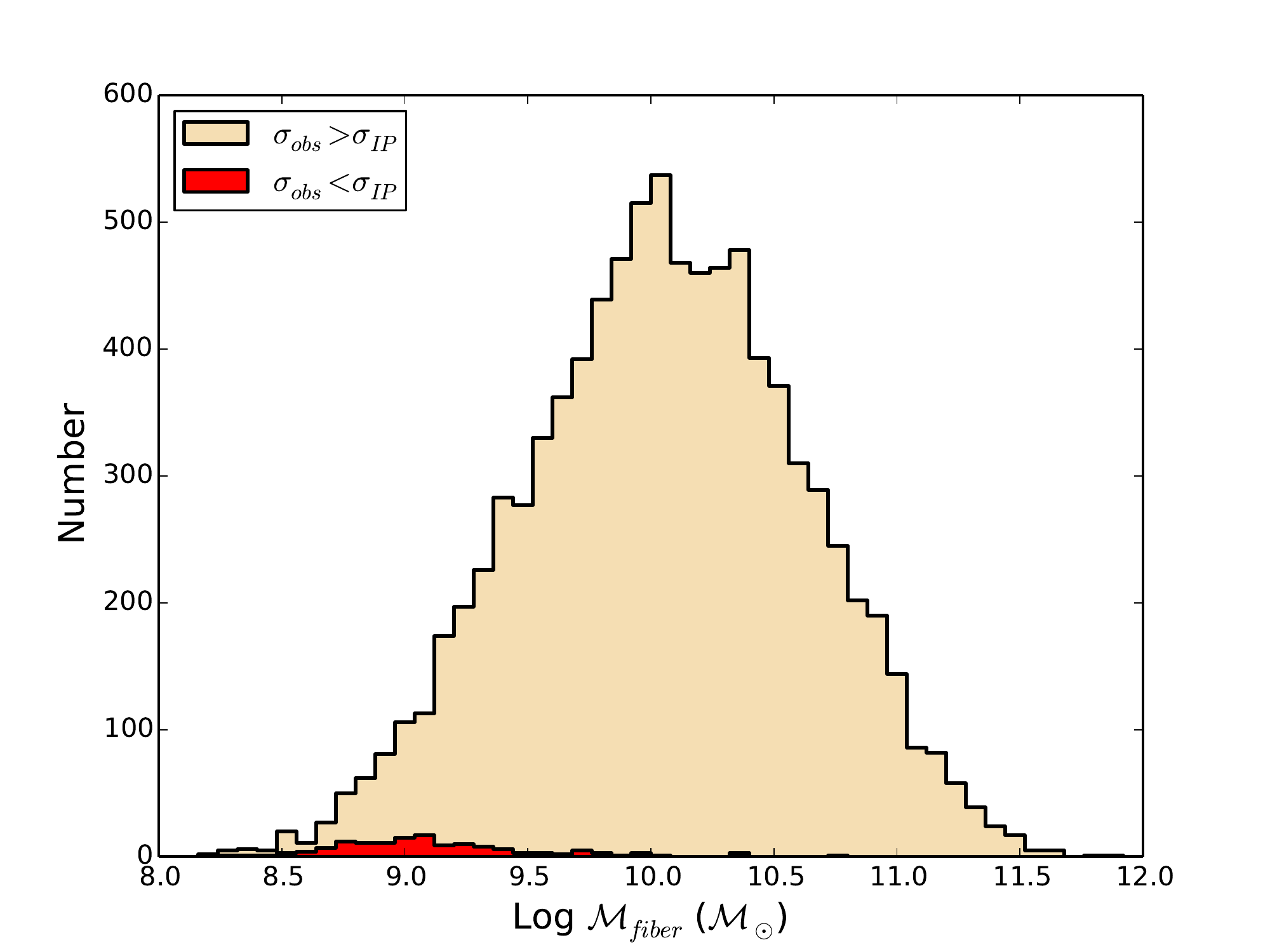}}
\caption{Distribution of the baryonic (stars+gas) masses inside the fiber aperture. The small region at low masses, marked in red, corresponds to cases where $\sigma_{obs}<\sigma_{IP}$.}
\label{hist_mbary}
\end{figure}

Which of the two cases, spheroid or disk, is most probable? It depends on the mass but as has been shown from luminosity profile fitting \citep{2013A&A...556A..10M} and the central concentration index $R_{90}/R_{50}$ \citep{2004MNRAS.351.1151B}, most of the star forming galaxies should be flattened systems. For a few others, e.g. mergers, the spherical case may be more appropriate. As is seen in Fig.~\ref{mdyn_mphot}, showing the fiber masses based on the velocity dispersion vs. the photometry, the disc option fits nicely to the data. The spherical case allows for a significant portion of dark matter ($\sim$ 60\%) to be added to the baryons. Fig. \ref{hist_mbary} shows the distribution of baryonic masses indicating in red the cases where $\sigma_{obs}<\sigma_{IP}$. About 2\% of all galaxies have this problem, most of them below masses of $\sim$10$^9$ \msun.

\begin{figure}[t!]
\centering
 \resizebox{\hsize}{!}{\includegraphics{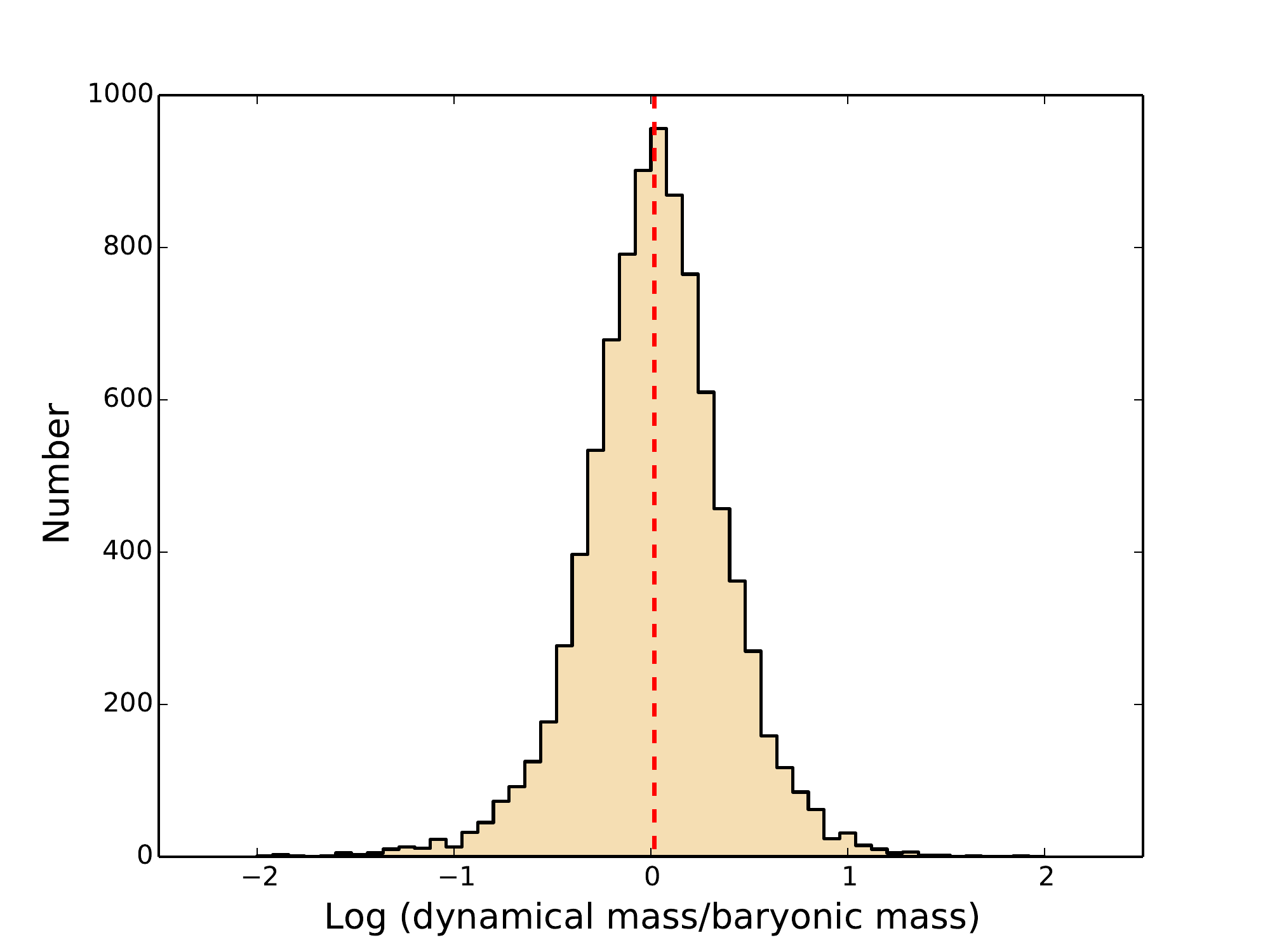}}
\caption{Histogram of the ratio between the dynamical and baryonic (stars+gas) masses inside the fiber aperture. The dynamical masses are based on the disc approximation. The hatched vertical line is the median. }
\label{massdiff}
\end{figure}

From Fig. \ref{massdiff}, we see that the distribution of the ratios between the dynamical and baryonic masses is nearly Gaussian around a value close to 1. Here the dynamical masses are based on the disc approximation. This could indicate that outflows are not prominent enough to influence the ionised gas motions in which case we would see a skewed distribution towards the positive side. The similarity between the two types of mass determinations seems to indicate that the contribution from dark matter (hereafter DM) to the the dynamical masses is insignificant {\sl within the relatively compact starburst region}. There are no broad investigations of the mass distributions in starburst galaxies. If we have a look at normal galaxies it is well known that gas rich dwarf galaxies with baryonic masses below $\sim$10$^9$\msun ~ are DM dominated over most of the stellar main body \citep[e.g][]{2011AJ....141..193O}. In more massive disc galaxies, $\sim$10-50\% of the mass is in baryonic form inside the peak of the rotation curve \citep{2013A&A...557A.131M}, i.e. at distances from the centre $<$2.2 scale lengths, which is larger than the typical spectroscopic aperture of our sample. Moreover, in starburst galaxies we expect the gaseous component to be more centrally concentrated due to angular momentum transfer from the gaseous component to the old stellar component \citep{1996A&A...314...59P,2001AJ....122..121V,2012A&A...544A.145L,2014A&A...563A..27L}. Therefore the baryonic component traced by \hax emission, tightly following the optical profile \citep{2013AJ....146..104H}, should be more prominent than in quiescent galaxies and may account for the low influence of DM.

From the fit to the disc approximation in Fig. \ref{mdyn_mphot} we derive a relation between the dynamical ($\cal M$$_{dyn}$) and baryonic ($\cal M$$_{bary}$) masses based on the fiber data:

\begin{equation}
log({\cal M}_{dyn})=0.93 \times log({\cal M}_{bary})+0.83. 
\end{equation}

The agreement at intermediate-high masses is remarkably good. A common problem with observations of starburst galaxies in the optical region is that the light from the young population completely dominates the emission and therefore makes it difficult to determine the contribution from the old stars. It seems that we have managed to take the old population into account in a way that allows us to use the masses at least for statistical purposes. The tight correlation also seems to give support to a practical use of the emission line widths at high redshifts to determine the baryonic masses inside the optical disk.

\begin{figure}[t!]
\centering
 \resizebox{\hsize}{!}{\includegraphics{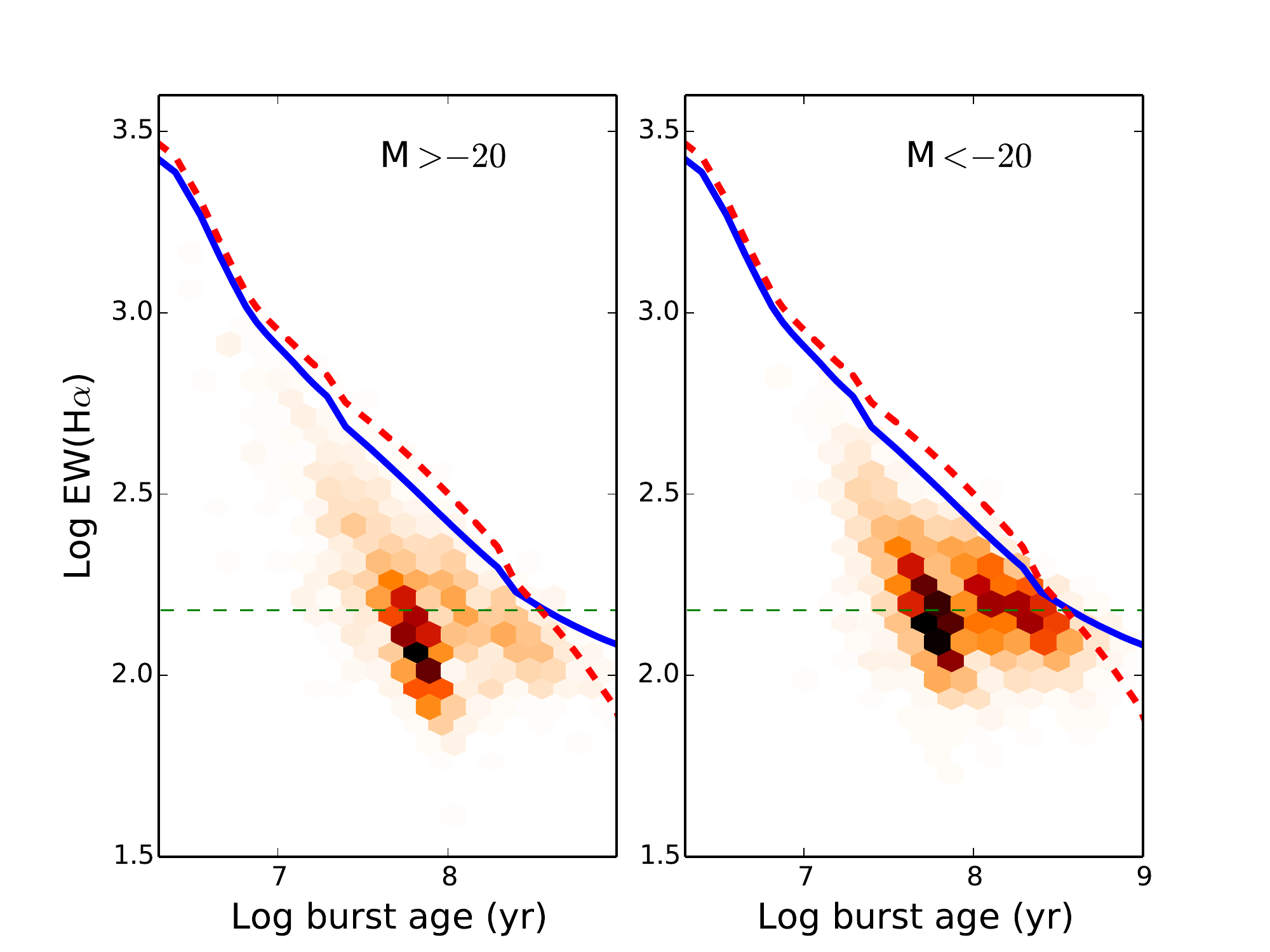}}
\caption{Dust attenuation corrected \ewha versus the age of the young population in the starburst candidate sample. The sample is divided according to absolute magnitude into `dwarfs' and `giants' at $M_r$=--20. The solid and dashed lines are identical to those in the left part of Fig.~\ref{ewha_dur} with a 3\% relative mass fraction of the young population. The dotted line at \ewha =150\AA ~ corresponds to the minimum value (but not sufficient)  of \ewha ~to qualify as a starburst with the b$>$3 criterion and an age $<$ 1 Gyr {\it if} the SFR is assumed to be constant during the burst. The hexagon symbols represent the surface density of the data points, where darker colours means more data.}
\label{ewha_age2}
\end{figure}

\begin{figure}[t!]
\centering
 \resizebox{\hsize}{!}{\includegraphics{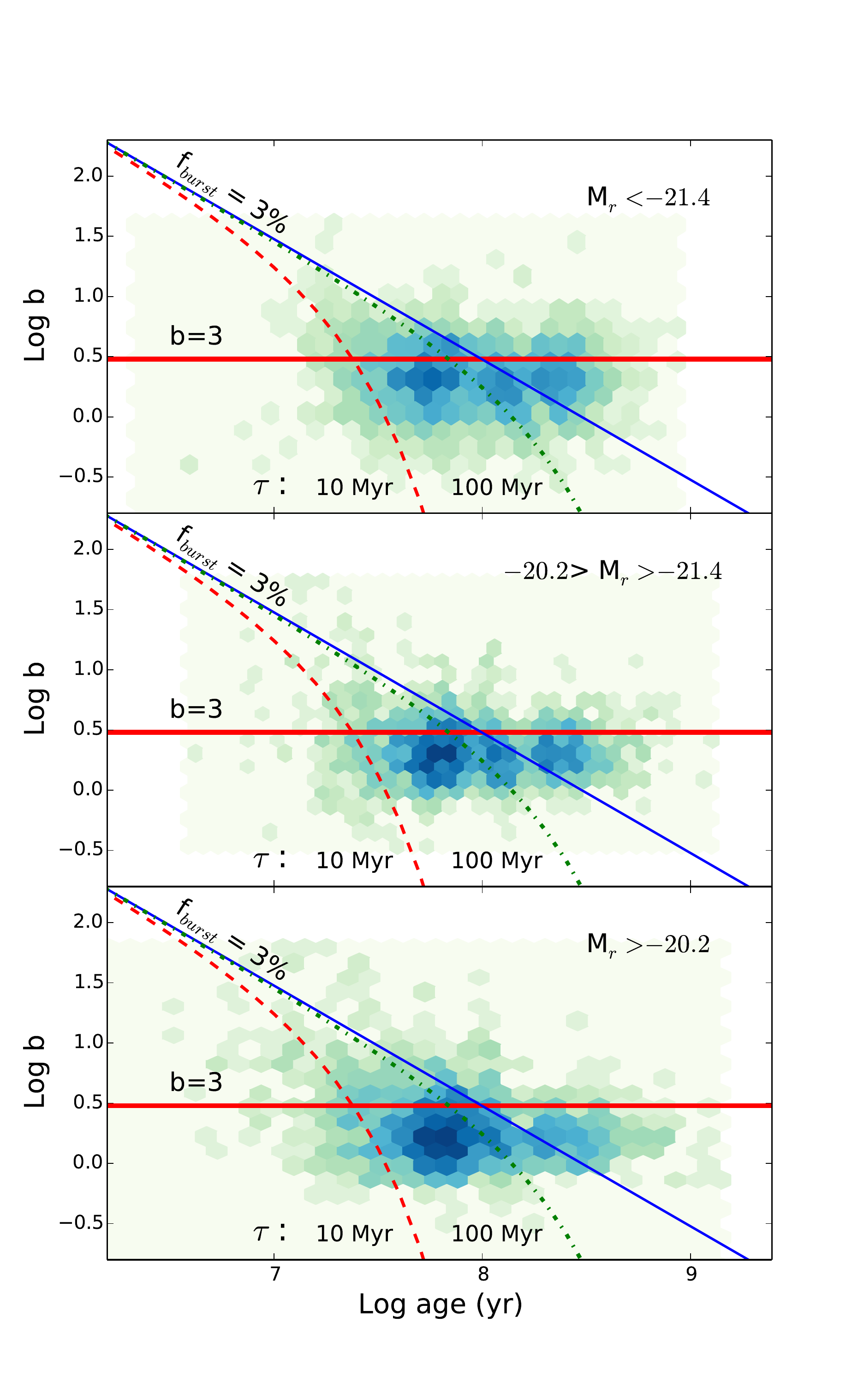}}
\caption{$b$--parameter versus age of the young population in the starburst candidate sample divided into three groups in luminosity indicated at the upper right.  Each group contains about 2000 galaxies. The starburst criterion $b$=3 is shown with a red horizontal line. Also displayed in three tracks are the minimum $b$ parameter values needed to reach $f_{burst}$=3\% at the age in question. In this case three alternatives of the SFH are displayed:  A constant SFR (solid blue line) and exponentially decaying SFR:s on timescales of 10 and 100 Myr (red dashed line and green dash-dotted line respectively). The hexagons show the surface density of data points. Darker - more data.}
\label{bpar_age}
\end{figure}

\subsection{Basic properties of the candidate samples}

\begin{figure}[t!]
\centering
 \resizebox{\hsize}{!}{\includegraphics{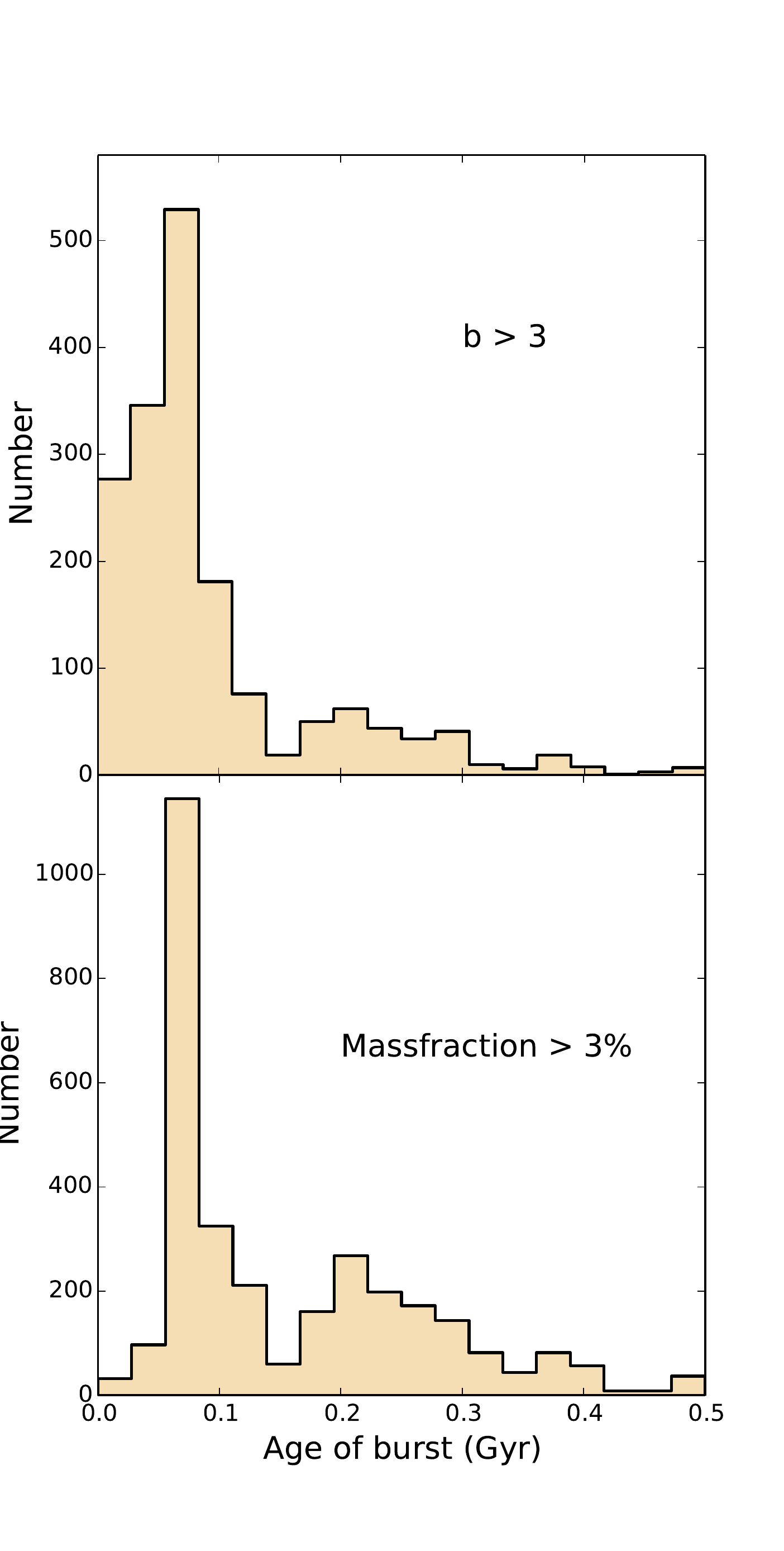}}
\caption{Ages of the starburst galaxies using the b$>$3 and the $f_{burst}>$3\%  criteria.}
\label{hist_sbages}
\end{figure}

\begin{figure}[t!]
\centering
 \resizebox{\hsize}{!}{\includegraphics{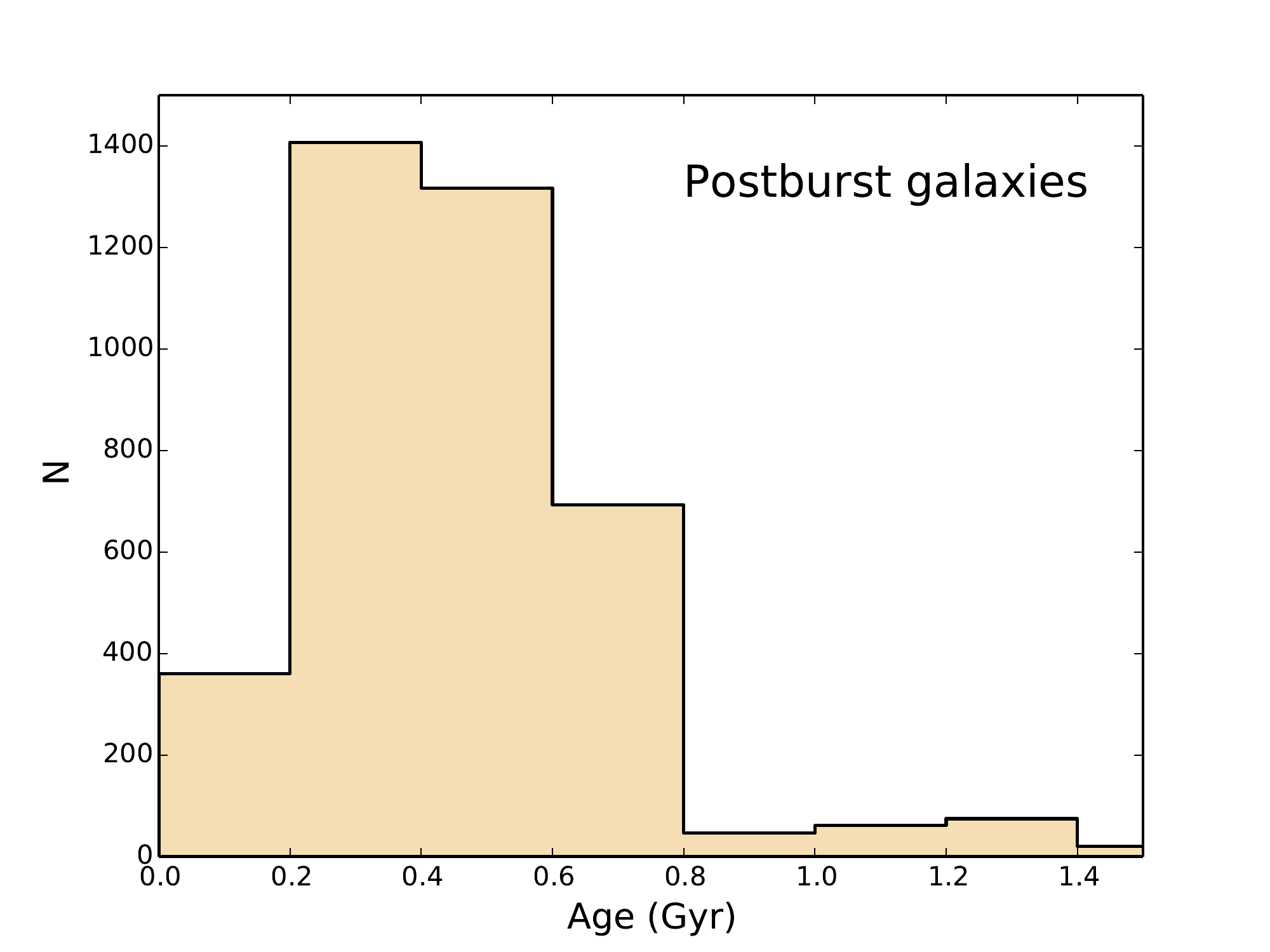}}
\caption{Ages of the postburst galaxies.}
\label{hist_postages}
\end{figure}

Here we have a look at the basic characteristics of the sample constrained by \ewha$\geq$60\AA ~for starburst candidates and \ewhd $\leq$--6\AA ~ for the postburst sample. Then we investigate the more constrained samples defined by the $b$--parameter and the $f_{burst}$ parameter. It is enlightening to begin by looking at where our selected galaxies end up if we would place them in the theoretical diagram seen in Fig.~\ref{ewha_dur}, remembering however that this diagram is based on the assumption that the SFR is constant, while we assume exponentially decaying SFR in the models. How \ewha~correlates with age is seen in Fig.~\ref{ewha_age2}. The horizontal line corresponds to the minimum (necessary but not sufficient) value of \ewha~ if a star formation epoch of an age lower than 1 Gyr should be regarded as a starburst if we apply the $b$$>$3 criterion.

The situation becomes more clear if we have a look at Fig.~\ref{bpar_age} showing the $b$-parameter versus age, which is fairly similar to the lower part of Fig.~\ref{ewha_dur}. Three different subsamples of different luminosity classes, are presented. They have approximately the same number of galaxies so one can have an impression of how the distribution changes with luminosity. Not unexpectedly, the most luminous galaxies have a lack of low ages. Except for displaying the $b$=3 limit (red solid horizontal line) between starbursts and non-starbursts we also show the line above which the mass of the burst is more than 3\% of the total mass. We do this for three different star formation histories: constant SFR (solid blue line), an exponentially decaying SFR with times scales 10 and 100 Myr (red dashed and green dot-dashed lines). For an exponentially decaying SFR the minimum value of the $b$-parameter necessary to fulfil the $f_{burst}$ criterion at a certain burst age $t_*$ (in years), assuming (as we do here) that the age of the old population is 10 Gyr, can be expressed as,

\begin{equation}
\label{eq:b}
b = \frac{SFR}{<SFR>}=\frac{SFR\times10^{10}}{{\cal M}_{tot}}=\frac{e^{-t_*/{\tau}}\times10^{10}\times f_{burst}}{\tau(1-e^{-t_*/{\tau}})}
\end{equation}

If we look at the distribution of the $b$-parameter values in Fig. \ref{bpar_age}, in particular in the lower diagram, we notice that the maximum $b$-parameter decreases with age from $b\sim$60  down to 3 at an age slightly below 1 Gyr.  The upper envelope of the distribution in the lower diagram corresponds to $f_{burst}$=0.10--0.15. It seems that the strongest starbursts are short-lived. The reason may be that the gas is consumed in a shorter time or that the minimum mechanical energy (generated by young stars and SNe) needed to eject the fuel for the starburst is reached in a shorter time in strong starbursts. The upper limit of the ages is in reasonable agreement with the maximum gas accumulation time scale in a massive merger \citep{1996ApJ...464..641M}. We also note that on the righthand side of the diagram, below the $b$=3 line, we find galaxies that very likely fulfil the $f_{burst}$$>$3\% criterion but few with the $b>$3 criterion. We show below that this is confirmed by the modelling. In a simple scenario,  the maximum SFR allowed in a starburst is equal to the rate at which the total gas content is converted to stars over the dynamical timescale \citep{1998ARA&A..36..189K}. For a Milky-Way type of galaxy it is convenient to express it:

\begin{equation}
SFR_{max}= 100 {\cal M}_\odot~yr\mone \frac{{\cal M}_{gas}}{10^{10}{\cal M}_\odot}\frac{10^8years}{\tau_{dyn}}
\end{equation}

where $\tau_{dyn}$ is the dynamical time scale. Consequently, the maximum $b$-parameter value is

\begin{equation}
b_{max} = \frac{SFR_{max}}{<SFR>}=\frac{{\cal M}_{gas}}{{\cal M}_{stars}}\frac{age}{\tau_{dyn}}
\label{bmax}
\end{equation}

where $age$ is the age of the galaxy, assumed to be 10$^{10}$ years. The total gas mass of the Milky Way is ${\cal M}_{gas}$=7$\times$10$^9$ \msun. Approximately half of this is in molecular form \citep{1991ARA&A..29..195C}. Since stars are formed in molecular clouds this number is what we enter into the equation. Then we assume ${\cal M}_{stars}$=6$\times$10$^{10}$ \msun ~ and $\tau$=10$^8$ years. We then obtain $b_{max}$$\sim$6, about half of the approximate upper limit in the diagram at a burst duration of 100 Myr. The content of H$_2$ in SF galaxies is typically 5-10\% of the total baryonic mass (equations \ref{himass} and \ref{molmass} and \citet{2005ApJ...625..763L}). As we noted above, the upper limit, running parallel to the $f_{burst}$=3\% line corresponds to $f_{burst}$=10--15\%. In the simple scenario we have discussed, this limit therefore corresponds to an almost complete consumption of the available molecular gas available for star formation. Depending on age, is also sets the limit of the maximum value of the $b$-parameter. At the lowest ages, $\sim$ 10 Myr, this corresponds to $b$$\sim$ 60. These estimates might be too crude however. Simulations of mergers \citep{1996ApJ...464..641M}, taking into account orbital parameters, bulge/disk ratios and feedback processes, adds complexity to the interpretation and molecular hydrogen may be formed or destroyed \citep[see also][]{2014MNRAS.444.1615B,2004ApJ...612..825H}. We return to this issue in Sect. \ref{discussion:evolution}.

\begin{figure}[t!]
\centering
 \resizebox{\hsize}{!}{\includegraphics{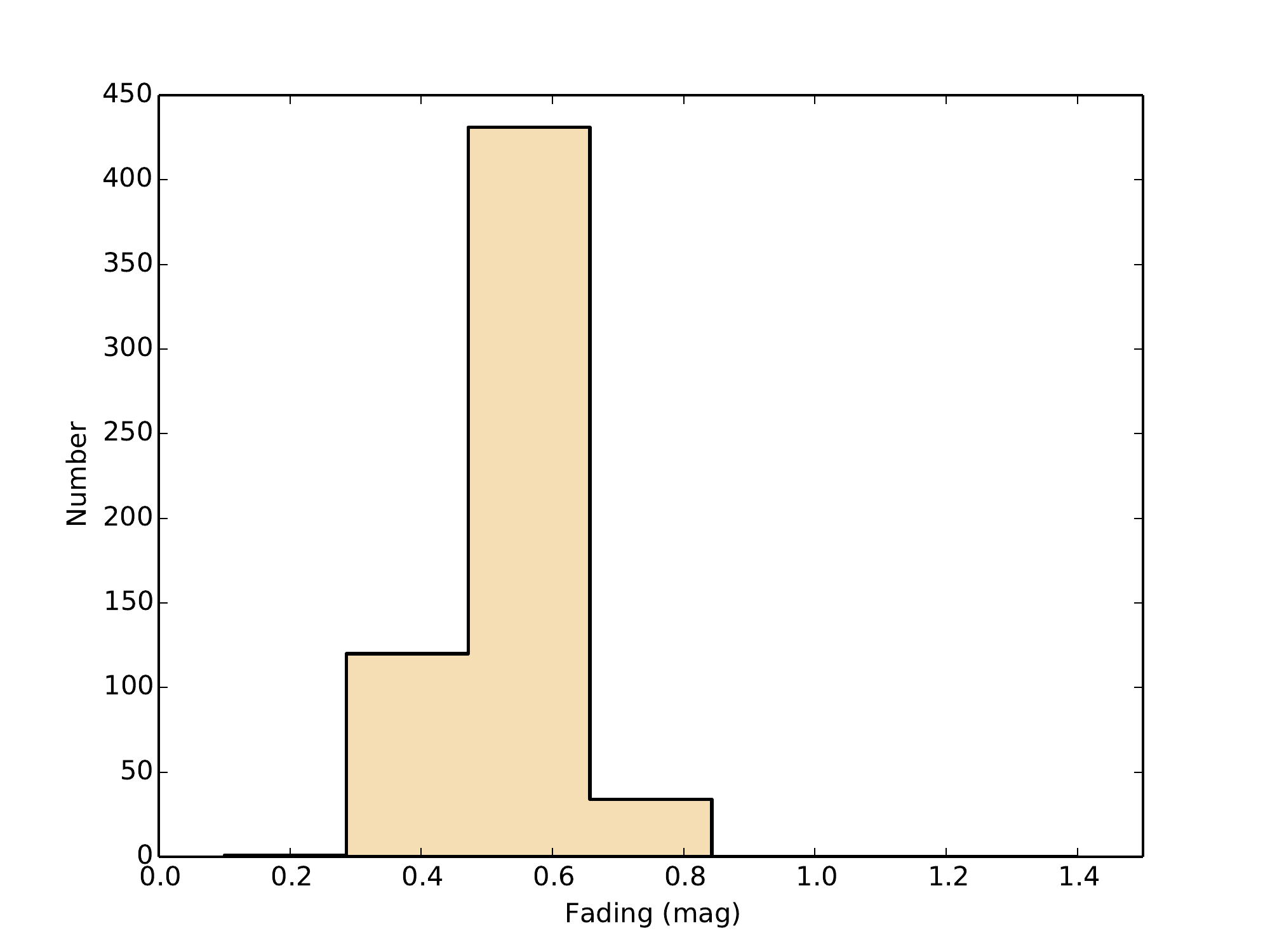}}
\caption{Difference between the luminosity at the end of the burst and the mean luminosity of the postburst phase assuming an exponentially decaying SFR on a time scale between 100 Myr and 1 Gyr.}
\label{hist_fading}
\end{figure}

In Fig.~\ref{hist_sbages} we show the age distribution of galaxies fulfilling the $b>$3 and the $f_{burst}$$>$3\% criteria. A prominent peak is seen at 50 - 90 Myr. The relative fraction of the decay rates chosen by the SEM as the best fit, and the median age of these subselections for the $b$$>$3 case are as follows: $\tau$, relative fraction, median age: 30 Myr, 53\%, 57 Myr; 100 Myr, 13\%, 82 Myr; 300 Myr, 19\%, 213 Myr; 1 Gyr, 15\%, 68 Myr. We can understand the sharpness of the peak if we consider how the visibility of the galaxy increases with age if the SFR is constant. In their simulation of gas rich mergers \citet{1996ApJ...464..641M} found a prominent peak in SFR at an age of 50 Myr in fair agreement with our empirical results. Another conclusion (as was also indicated by Fig.~\ref{bpar_age}) is that galaxies that fulfil the mass fraction condition tend to fall below the $b$--parameter condition at high ages. A significant fraction, essentially the tail below $b$=3 at log(age)$>$8 in Fig.~\ref{hist_sbages}, do not have genuine starburst progenitors but are produced through a long epoch of SF at a $b$--parameter between 1 and 3.

We may compare the age distribution of the starburst sample to the age distribution of the postburst galaxies shown in Fig.~\ref{hist_postages}. A peak, corresponding to the $\sim$ 100 Myr peak in the starburst sample, is seen about 300 Myrs later, more or less as we expect according to our models. Also in the Bruzual \& Charlot model \citep{2003MNRAS.344.1000B}, the peak in \hdx occurs about 400 Myr after an instantaneous burst.

We can proceed a bit further with this discussion. Fig.~\ref{hist_fading} shows how much fainter the postburst galaxies are expected to be if the star formation history continues with the exponential decay time scale chosen by the programme. It should be the minimum change but could be a factor of 2 higher for short bursts or rapidly quenched bursts. The change in luminosity is roughly a factor 2--3. At the same time the duration of the visible postburst phase is a few times longer than the starburst lifetime, depending on the star formation history and the starburst mass fraction. The consequence is that if we would follow the evolution of the $f_{burst}$$>$3\% LF from the active to the passive postburst phase after the SF has ceased we would see a shift of about 1 magnitude towards fainter luminosities and then a shift upwards to compensate for the longer lifetimes of the postbursts. This agrees with what we observe if the lifetime of the postbursts is 3-4 times longer than the burst phase (see Fig. \ref{lumfunction}) and the dust attenuation is about the same.

\subsection{Mass trends in the starburst and postburst samples}
\label{trends}

We now focus on mass trends in the two samples. We look at starbursts obeying the $b$--parameter criterion and the postburst sample obeying the \hdx criterion. First we look at the starburst sample. 

\begin{figure}[t!]
\centering
 \resizebox{\hsize}{!}{\includegraphics{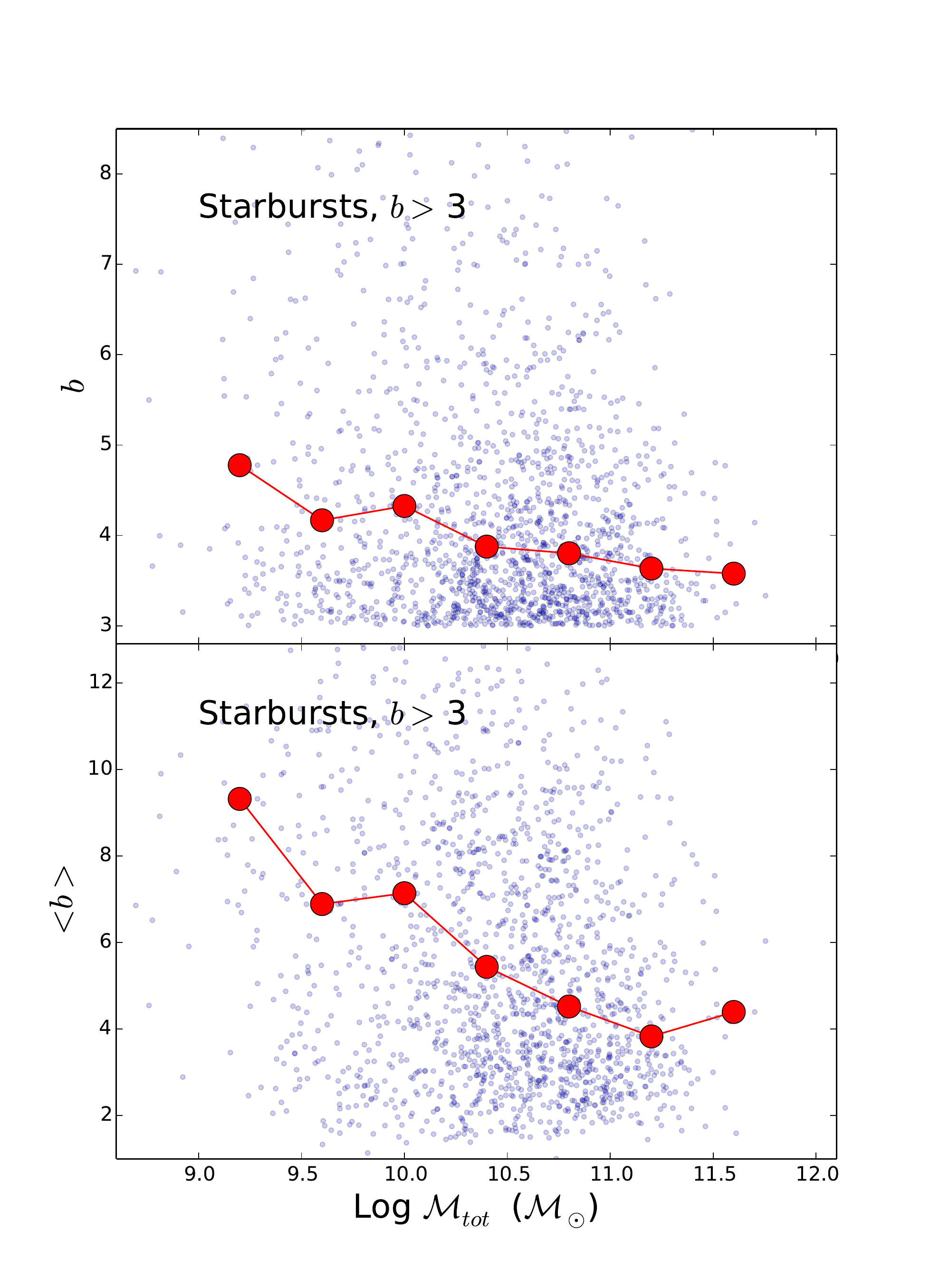}}
\caption{Upper part: The median $b$--parameter vs. the total baryonic mass (stars+gas) of the starburst ($b>$3) sample. Lower part: The median of the mean value of the $b$-parameter during the starburst epoch, as function of the total mass.}
\label{bpara_mtot}
\end{figure}

Fig.~\ref{bpara_mtot} shows how the $b$--parameter relates to the mass. In the upper part of the diagram we display the present values of the $b$ parameter. In the lower part we show how the mean value of the $b$ parameter, i.e. $<$$b$$>$=$f_{burst}\times$(age of galaxy)/(age of burst), varies as function of total mass. We can see that the median value of the $b$-parameter decreases with increasing mass, as a result either of an increasing collapse time or/and a decreasing amount of gas available to feed the burst (cf. Eq. \ref{bmax}). $<$$b$$>$ shows the same pattern but the mean $<$$b$$>$ is about 50\% higher than $b$. This can be understood if the SFR is decreasing with time. In the modelling of the spectra we assumed exponentially decaying SFRs. The preferred decay time scale is 100 Myr. The median age of a starburst is 68 Myr. Under the assumption of an exponentially decaying SFR and no problems with dust attenuation we would expect the ratio between $<$$b$$>$ and $b$ to be $\tau(1-e^{-t_*/\tau})/t_*/e^{-t_*/\tau}$. Adapting $t_*/\tau$=68/100 we obtain $<$$b$$>$ /$b$ = 1.4, which is close to what we observe.

\begin{figure}[t!]
\centering
 \resizebox{\hsize}{!}{\includegraphics{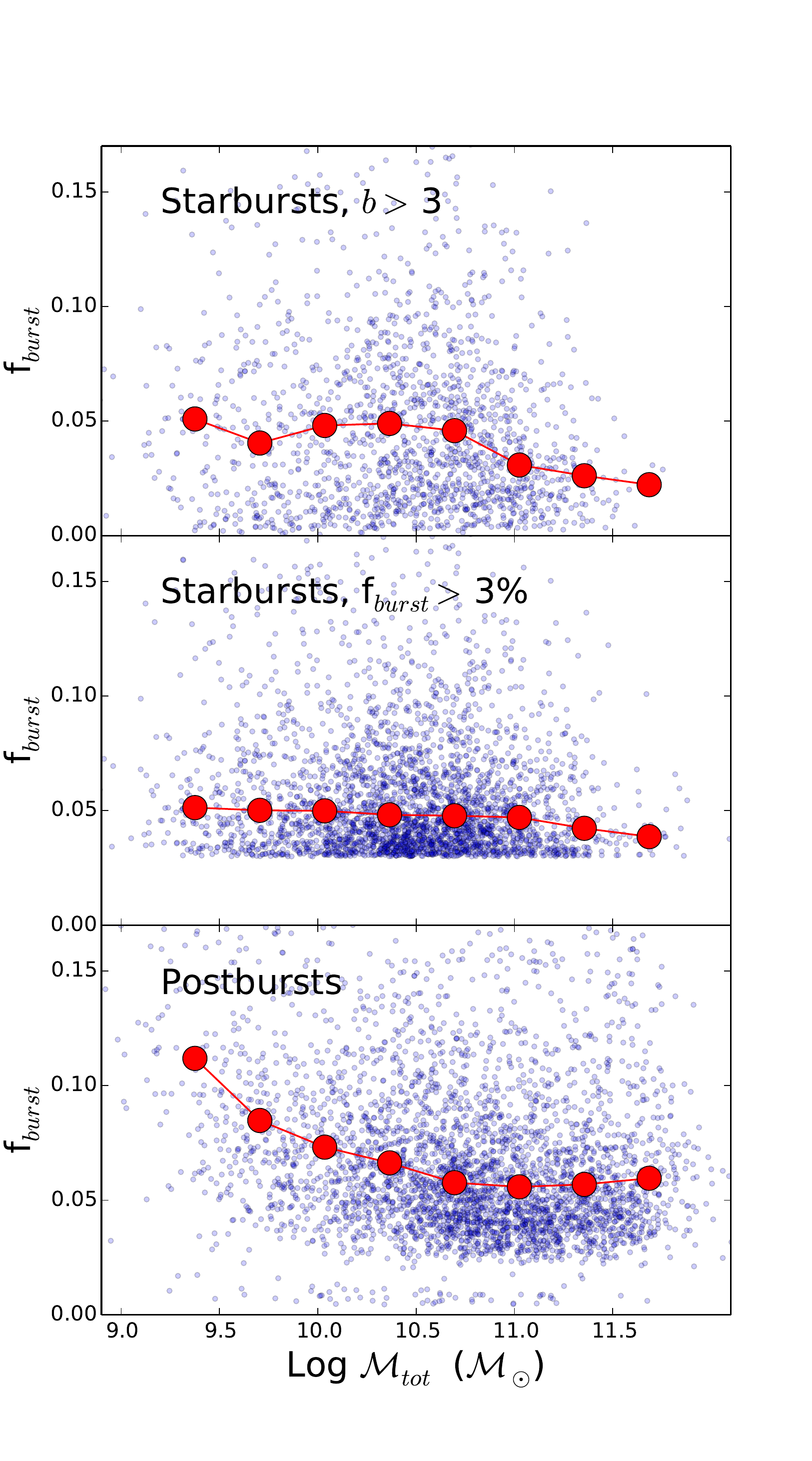}}
\caption{Median burst mass fraction vs. the total baryonic mass (stars+gas) of the starburst sample with $b>$3,  the starburst candidate sample with $f_{burst}>$3\% and the postburst sample (\ewhdx $<$ --6\AA).}
\label{mf_mass_pbsb}
\end{figure}

Fig.~\ref{mf_mass_pbsb} displays how the mass fraction relates to the total mass. As a consequence of the decrease of $<$$b$$>$ with mass we also find a decrease of the mass fraction with mass. In the mid part of the diagram is shown the medians of the burst mass fraction. If the burst is shut down abruptly at its peak, this diagram should agree fairly well with the postburst diagram at the bottom.  The agreement is fair at the mid-high mass end but deviates strongly at the low mass end. Using the same reasoning as in the preceding subsection we can estimate the maximum ratio between the total mass produced by the burst to the mass produced up to the observation of the burst (typically at an age of 68 Myr). We then obtain ${\cal M}_{tot}/{\cal M}_{*}$=(1-e$^{-t_*/\tau}$)$^{-1}$=2.0, in strong agreement with the low mass part of the diagrams. Another reason for the difference between the mass fractions is the metallicity dependence of $f_{burst}$ versus the \hdx selection criterion. From Fig.~\ref{bc_ewhd_age} we see that the \hdx lines are weaker at low metallicities and that we would need more mass in the burst population to produce a 6\AA ~absorption line. The fact that the lower envelope of the postburst distribution in Fig. ~\ref{mf_mass_pbsb} is bending upwards at smaller masses indicates that the metallicity effect dominates in explaining the difference in the mass fractions. Generally speaking we can conclude that the mass fractions we derive for the postburst sample is significantly smaller than found by some other groups \citep{1996ApJ...458L..63L,2001ApJ...557..150N,2013MNRAS.431.2034M} and, although still smaller, more in agreement with \citet{2012MNRAS.420..672S} and \citet{2007MNRAS.375..381N}.

\begin{figure}[h!]
\centering
 \resizebox{\hsize}{!}{\includegraphics{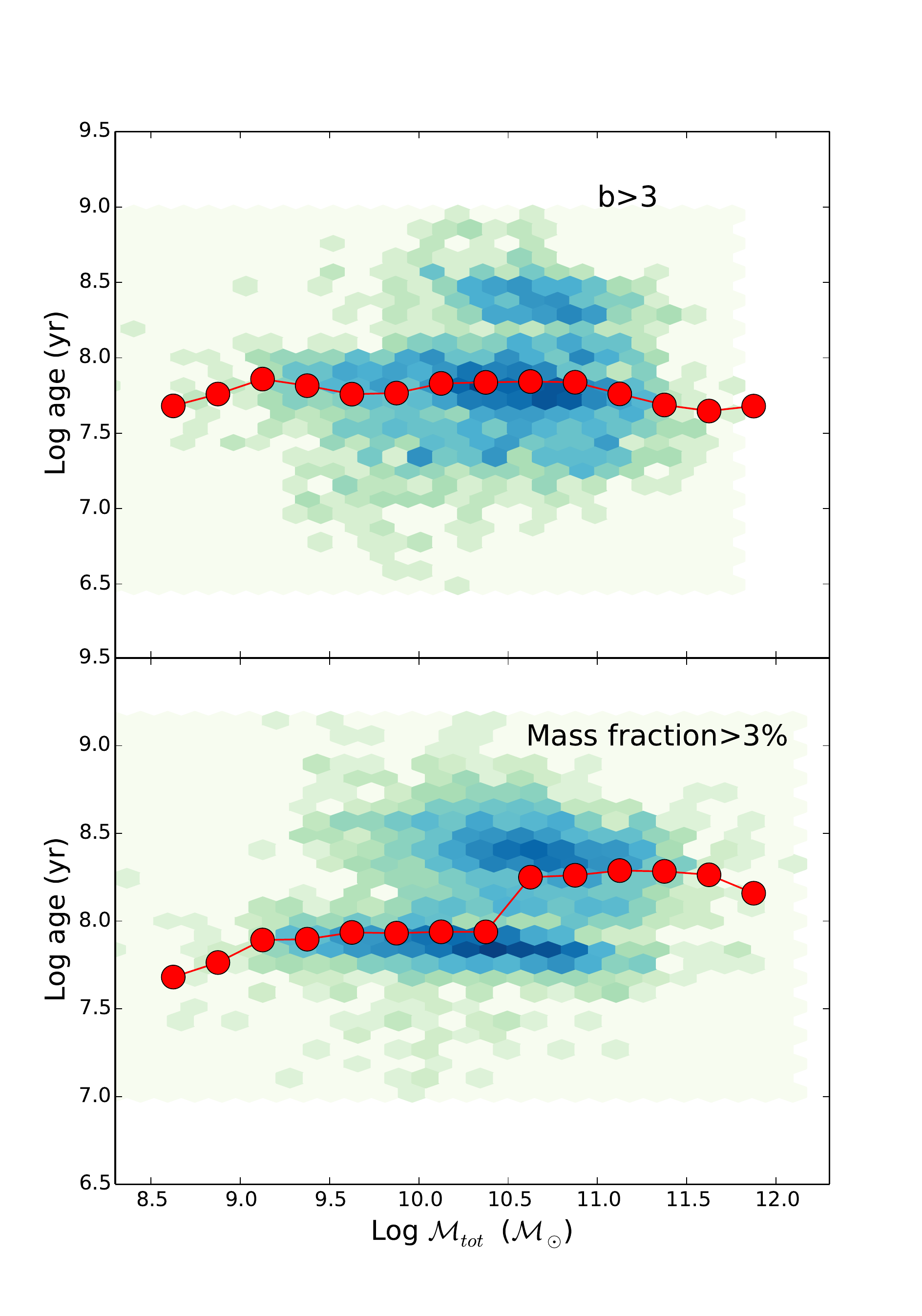}}
\caption{Total mass of stars, stellar remnants and gas versus the starburst age. Two different restrictions are applied. In the top diagram the condition is that the $b$ parameter should be $b>$3. In the lower diagram the restriction is that the burst mass fractions should be higher than 3\%. The green markers are medians in log age.}
\label{mtot_age}
\end{figure}

Fig. \ref{mtot_age} shows how the starburst age relates to the total baryonic mass of the galaxies under two different conditions, $b>$3 and $f_{burst}$$>$3\%. In the first case the ages are almost independent of mass. This is in agreement with the analysis of ellipticals by \citet{2010MNRAS.402..985H}. They investigate the properties of `burst relic' populations and try to recover the properties of the burst when it was active. They conclude that the starburst time-scale is $\sim$ 100 Myr and is nearly mass-independent, in agreement with our results.

In the bottom figure we show the mass as function of age under the condition that the burst mass fraction should be larger than 3\%. We can clearly define two branches in lifetimes. Part of this bimodality may be caused by the discreteness in SFR decay rates we use in the models. However, there is a prominent change in lifetime starting at a baryonic mass of log\m$\sim$10.5. The shift occurs at the same mass as the break observed by \citet{2003MNRAS.341...33K} in the colours of SDSS galaxies. The break signals a transition from star forming dwarfish galaxies to galaxies dominated by old populations, high mass densities and a high fraction of bulges. There is also an increasing presence of bars in massive disc galaxies, starting to dominate the star-forming galaxy population at a stellar mass log\m$\sim$10.2 \citep{2010ApJ...714L.260N,2012MNRAS.423.1485S}. Bars drive inflows of gas towards the central areas \citep[e.g][]{2007A&A...468...61D}. Could perhaps bars in the massive disks explain the shift at log\m$\sim$10.5? In such a scenario we can assume that the gas inflow rate  $\dot{\cal M}\sim$SFR. From Fig. \ref{mtot_age} we derive the minimum $<$SFR$>$  for a galaxy of mass 10$^{11}$\msun~  to $\sim$ 15 \msun yr$^{-1}$. Assuming we are dealing with inflows towards the central kpc region, our inflow rate is a magnitude higher than what models normally predict \citep{2004MNRAS.354..892M}. However, inflow rate and burst duration are both in fair agreement with models of star formation occurring in nuclear rings formed in exponential disks with extremely high surface density  \citep{2012ApJ...758...14K}.  But according to the authors, inflows driven by the bar potential is not enough to explain the fact the star formation in nuclear rings occurs over a few Gyrs \citep[e.g.][]{2006MNRAS.371.1087A,2007MNRAS.380..949S}. Another mechanism, as for example the dynamical influence of spiral arms, seems to be needed to sustain the SF activity over a longer period of time. It thus appears that the increased lifetime of the SF regions occurring in our sample at high masses could at least partly be explained by SF in nuclear rings. However, these galaxies do not often fulfil the $b>$3 criterion. An additional mechanism that may be important emerges from the simulations by \citet{2013MNRAS.430.1901H}. They study starbursts in mergers of equal mass galaxies. In low mass mergers that burst is rapidly quenched by stellar feedback and gas is expelled to large distances. In more massive galaxies gas is expelled from feedback-driven winds in the centre but much of the material is not unbound but falls back towards the disc after a short time. This can prolong the starburst phase over a few 100 Myr.

\subsection{Cold outflows}
\label{cold}

Galactic outflows are known to be quite common in galaxies of high star formation activity \citep{1993ASSL..188..455H,1994A&A...291L..13P,1996ApJ...462..651L,2000ApJS..129..493H,2005ApJ...621..227M,2012ApJ...754L..22A} and are found at all cosmic epochs \citep[e.g][]{1985A&A...149..475B,1988A&A...192...81J,2001ASPC..240..345H,2001ApJ...554..981P,2002ApJ...568..558F,2003ApJ...588...65S,2013MNRAS.429.2550E,2012A&A...540A..63N}. As was mentioned in the introduction, under certain conditions superwinds generated by SN activity can drive cold ($\la$ 10$^4$K) gas out to large distances from the starburst before the gas experiences instabilities and/or is heated by conduction \citep[e.g.][]{1985Natur.317...44C}. These events may have a strong influence on the global properties of starburst galaxies as well as the ambient intergalactic medium. A few observations of galaxies in the local universe show that the condition for `blowout', i.e. a significant part of the cold gas being accelerated to velocities higher than the escape velocities may be fulfilled \citep{2001AJ....122.3070O}. Another aspect is that mass outflows may open channels for Ly$\alpha$ and Ly continuum radiation to leak out. The importance of starburst dwarfs as sources of the cosmic reionisation and Ly$\alpha$ emitting starburst galaxies as beacons in the distant universe have often been discussed \citep{1998AJ....115.1319C,2000ApJ...545L..85R}.

Different methods are employed to study cold flows, e.g. broad emission lines and \hax morphologies indicating outflows in the minor axis direction \citep{1996ApJ...462..651L,2015A&A...576L..13B} and blueshift of absorption lines in the cold gas \citep{2000ApJS..129..493H,2009ApJS..181..272G,2010AJ....140..445C}.  Next we briefly discuss the problem and see what the SDSS data can tell us.

Observations show that cold outflows occur in galaxies with a high SFR per surface area, $\Sigma_{SFR}$. Typically, outflows are observed if $\Sigma_{SFR}>$0.1\msun yr$^{-1}$kpc$^{-2}$ \citep{2002ASPC..254..292H,2010AJ....140..445C}. This limit is supported by theoretical modelling \citep{2011ApJ...735...66M,2013ApJ...763L..31S}. We calculated $\Sigma_{SFR}$ using the area $\pi R_{50}^2$, in line with \citet{2010ApJ...719.1503R}. \citet{2012ApJ...758..135K} argue for a slightly different way of deriving the area $A = \pi R_P^2/3.7$, where $R_P$ is the Petrosian radius, which will result in slightly lower values of $\Sigma_{SFR}$.

\begin{figure}[t!]
\centering
 \resizebox{\hsize}{!}{\includegraphics{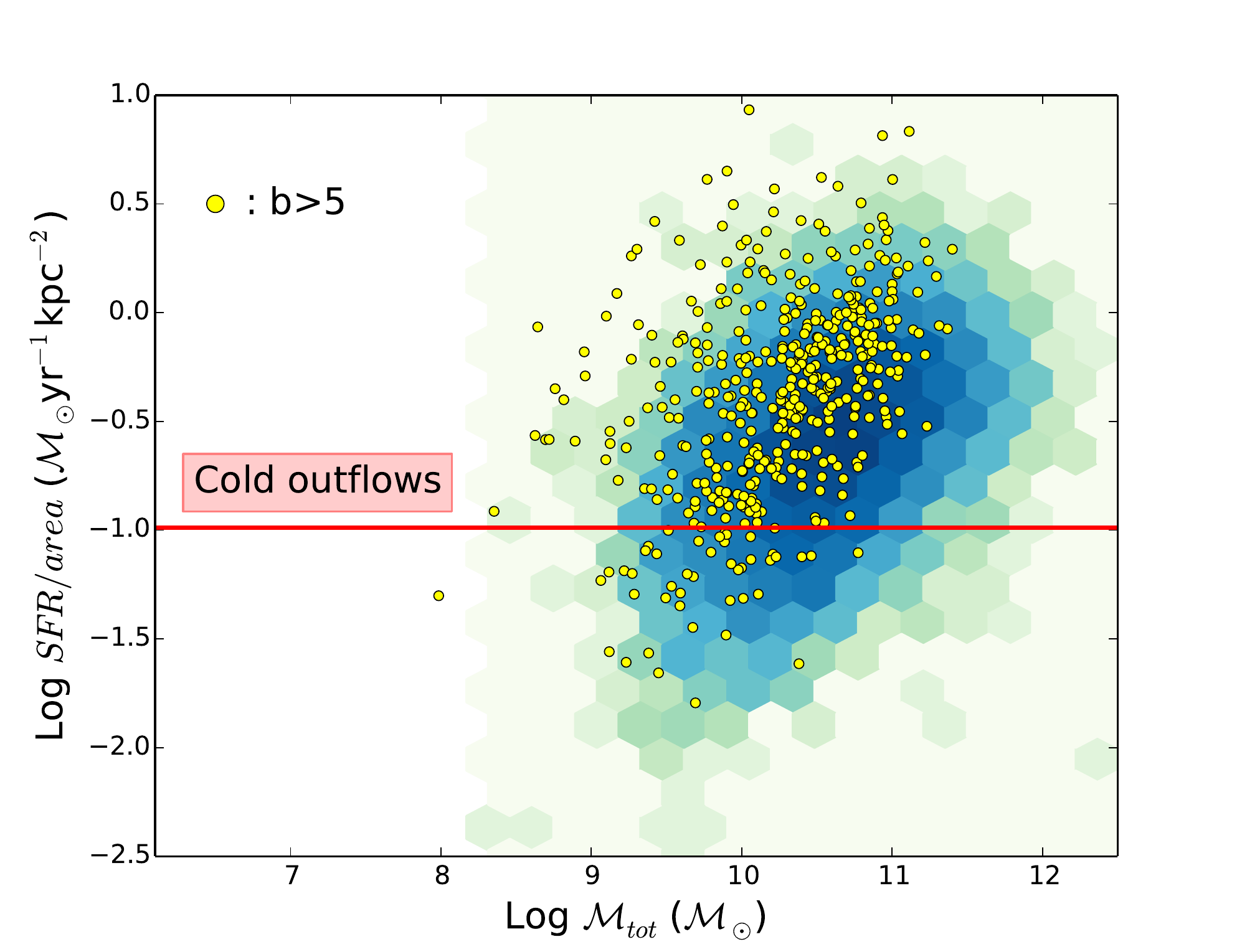}}
\caption{SFR per area as function of baryonic mass of the galaxies in the starburst candidate sample. Galaxies with $b>5$ (yellow circles) are superposed on the distribution of the remaining galaxies (blue triangles). The horizontal line marks the transition between when the energy produced by the starburst is sufficient to generate outflows, both according to models and to observations.}
\label{sfrarea_mtot_log}
\end{figure}

Fig. \ref{sfrarea_mtot_log} shows the $\Sigma_{SFR}$ as function of mass. The upper limit of the envelope agrees with the conclusion drawn by \citet{1996ApJ...462..651L}, namely that ``no starburst" seems to have a surface density of star-formation above $\sim$ 20 \msun yr$^{-1}$kpc$^{-2}$. This was interpreted as an indication of self-regulation in the star-formation process. In the diagram we also see the sample divided into galaxies with high $b$-parameters ($b$$>$5) superposed on galaxies with low b-parameters ($b$$<$5). We see that there is a correlation between mass and $\Sigma_{SFR}$. Another interesting result is that outflows in low mass galaxies often seem to be linked to strong starbursts while outflows in massive galaxies happen under more quiescent conditions.

How important are AGNs as drivers of the outflows? If they are important, one would expect to see them in action at high masses. Presumably this would lead to an anti-correlation between starburst age and  galaxy mass. We find no such relation (cf. Fig. \ref{mtot_age}).  \citet{2014MNRAS.441.3417S} investigate the outflows found in 12 massive galaxies, half of which contain AGNs, and conclude that the major drivers are actually starbursts. But a complete blow-out seems rare. \citet{2015ApJ...801....1F} recently discovered large amounts of molecular gas in postburst galaxies that ``.. rule out complete gas consumption, expulsion, or starvation as the primary mechanism that ends the starburst in these galaxies". We believe more information is needed in order to reach a firm conclusion.

\section{Discussion}
\label{discussion}

\subsection{Comparison with other similar investigations}

From the analysis above we can conclude that starburst galaxies, defined as galaxies with $b$$>$3 {\it or} $f_{burst}$$>$3\%, contribute marginally to the star formation history in the local universe, comprising only $\sim$1\% of the star-forming galaxies. Although we do not know of any investigation that argues for such a low frequency of starbursts as we do here, the paucity of starbursts has been confirmed by previous studies both in the local universe and at high redshifts \citep[e.g.][]{1987AJ.....93.1011K,2009ASPC..419..298N,2009ApJ...692.1305L,2011ApJ...739L..40R,2012ApJ...747L..31S,2013AJ....146...46K,2015A&A...575A..74S}. Rodighiero et al. were investigating the redshift interval 1.5$<$z$<$2.5 and found that starburst galaxies represent only 2\% of star-forming galaxies and account for only 10\% of the cosmic SFR density. As demonstrated by e.g. \citet{2015A&A...575A..74S} the fraction of starbursts does not change much with redshift in the range 0.5$<$z$<$4. To compare to our results we have to first look at the starburst selection criterion they used. The criterion they used was SFR$_{starburst}>$4$\times$SFR$_{GMS}$. How does this compare to the $b$ parameter? By comparing the GMS SFR, based on the data from \citet{2007A&A...468...33E}, to our median SFR for starbursts for masses $>$10$^{10}$\msun, we find that $b$$>$3 corresponds to 3$<$SFR/SFR$_{GMS}$$<$8, i.e. significantly higher than their criterion. Therefore it seems, if we take the more stringent definition we use into account, that our and the results from Rodighiero et al. agree quite well. 

One may argue that using the SFR/SFR$_{GMS}$ to define a starburst may be a better approach than using the $b$ parameter. We think that both methods have their pros and cons. As we mentioned above, one has to consider the fact that the relative gas content decreases with mass. We know that galaxy merging is the most probable starburst triggering mechanism in the local universe. We also know that the SFR follows the Kennicutt-Schmidt law for the \ion{H}{I} gas \citep{1998ARA&A..36..189K} and similar for the H$_2$ gas \citep{2013AJ....146...19L}. Both laws concern the relation between SFR and surface density. On the whole, the SFR is simply related to the amount of fuel available \citep[e.g.][]{2013AJ....146...46K,2014A&A...563A..27L}. Imagine a merger between 3 normal equal-mass galaxies, after the merger resulting in an $L_*$ galaxy. The merged galaxy will contain about twice as much gas (both \ion{H}{I} and H$_2$) as an isolated galaxy of the same mass. Most probably the SFR will increase by a factor of at least $\sim$2  {\sl without} the need of referring to an {\sl increased star formation efficiency} placing it above the GMS. A merger between low mass galaxies will not result in such a large increase in relative gas mass fraction. Taking this into account shows that using a constant value of the $b$-parameter is in fact well motivated. Although this thinking is valid in the local universe it may however fail at higher redshifts \citep{2014A&A...562A...1P}.

In \ref{subsec:connecting} we examined the frequency of the past burst strength in postburst galaxies and found that 95\% had $f_{burst}$$>$3\%. We concluded that this is lower than the $f_{burst}$$>$5\% \citet{2003MNRAS.341...33K}  found in their sample. Both were selected on the criterion \ewhd$<$-6\AA. In our study this conflict is of some concern since we use the postbursts to link the starburst population. In particular we use the LF of the postburst population at high luminosities to estimate the loss of starbursts due to heavy obscuration and contamination from AGNs. If the minimum burst mass fraction in postburst galaxies is in fact higher than what we found in our investigation, it means we should compare the postbursts to starbursts that are limited by $f_{burst}$$>$5\%. Since this leads to fewer bursting galaxies, one consequence is, in order to match the relative number of the $f_{burst}$$>$5\% sample to the postburst sample, that the duration of the postburst phase must become  longer.

We now look at the possible causes of the discrepancies between our result and that of Kauffmann et al.
As we show in the Appendix, the burst
strength we derive is quite uncertain for spectra with low S/N ($\leq$ 10). The
trend at low S/N is however to overestimate the mass fraction so the difference
between our result and Kauffmann et al. persists. Kauffmann et al. derive
maximum-likelihood estimates of \mlx and mass based on two indices - the
4000\AA ~break and \hdx in absorption. These indices were calibrated from a
Monte Carlo simulation of  32000 spectra with different star formation histories (SFHs). In these
simulations it was assumed that the starburst had a constant SFR over a
distribution of timescales in the interval 3$\times$10$^7$-3$\times$10$^8$
years. The starburst was then mixed in various proportions with an old stellar
population with decaying SFR of various timescales. The dust attenuation was
determined from the amount by which the observed $gri$ colors deviated from
those predicted. 

There are several details that differ between our procedure and that of Kauffmann et al.. Some may help to explain the difference between our results. While Kauffmann et al. use two parameters to determine \ml-ratios and masses, we compare more or less the full optical spectra with our model predictions, all in all about 3000 model spectra corresponding to the postburst phase. The corrections for dust attenuation is another detail that differs between our procedures. The way we do it for postbursts is however not too different from theirs so this should be a minor cause of the discrepancy. But what about the SED models we use? Our model contains an integrated gaseous emission component while they subtract the nebular component before comparing to the Bruzual-Charlot `stars-only' model data. This difference should however play a minor role for postburst galaxies. What may be more important is the SFH adopted for the starburst population. While we assume the old component comes from a 100 Myr burst, Kauffmann et al. allow a variety of SFHs for the old component. However, this should have a minor impact. If we compare the \mlx today between a 100 Myr burst and an exponentially decaying SFR on a timescale of 1 Gyr, the difference is $\sim$ 4\%. What may be more influential is the SFH of the young component. While Kauffmann et al.~adopt a constant SFR of variable duration we allow for a range of both exponentially declining and constant SFHs. In the final selection of the best fits to our model we only find 11\% SFHs with constant SFR. As we argue later on, and what is also stated by Kauffmann et al. as concerns the older stellar population, star formation in bursts most likely occur in a decaying mode. Yet they assume a constant SFR for their burst model. We have, among 57 models, the option to compare models based on constant SFR with those of exponentially decaying bursts to look for significant differences. Could that possibly give a  hint as to why our and Kauffmann's et al. results do not quite agree? When we do this comparison it appears the different SFHs cannot explain the disagreement. After a more detailed investigation of the error propagation generated in the processing steps of the different methods used we may possibly find an explanation but such a test is outside the scope of this paper.

 \citet{2004MNRAS.351.1151B} work with SDSS data and use a similar method as we do here to derive the $b$-parameter. Assuming a starburst to have $b$$>$2--3 they find that starbursts contribute about 20\% of the star formation in the local universe. If we apply $b$$>$2 as a criterion, the number of starbursts increases by a factor of 2 and we arrive at a contribution of  $\sim$ 9\%.  This is significantly lower than the result by Brinchmann et al..   
 
 We conclude that, although many studies support our result regarding the paucity of starbursts, others claim a significantly higher frequency of starbursts and a higher mass fraction. What is the cause of the discrepancy? One apparent problem is how to define a starburst galaxy. In many investigations the starburst criterion is based on the \ewha. We have demonstrated that all these studies must fall short because of the lack of correction for time dependent dust attenuation. Likewise, many teams use \ewha$>$100\AA ~as a starburst criterion. We show in Sect. \ref{subsec:connecting} that this criterion corresponds to $b\sim$1, i.e. hardly a starburst. 
 
But now let us look more closely at the two other studies, based on SDSS data, that seem quite related to ours. We have already commented on the work by \citet{2003MNRAS.341...33K} concerning postburst galaxies. We continue the comparison with this paper below. The second paper is the work by \citet{2004MNRAS.351.1151B}. They find higher values of the birthrate parameter than we do. One may suspect (cmp. eq. \ref{eq:b}) that significant differences exist between the determination of the SFR and/or the masses. Alternatively the data may be plagued by selection biases at low redshifts, where low mass galaxies tend to dominate. The lower redshift limit we use is z=0.02 while Brinchmann et al.~use z=0.005. As we demonstrated above, significant aperture problems occur below our redshift limit. Moreover, the risk of confusion between galaxies and \ion{H}{ii} regions in nearby galaxies also increases dramatically. However, relatively few galaxies, most of them at masses lower than our low limit, should be affected. Besides, the aperture corrections applied by Brinchmann et al. seem to work fine in most cases. 

For galaxies with strong emission lines, the SFR calculated by Brinchmann et al. is based on the \hax flux, corrected for dust attenuation. The method they apply for the dust correction \citep{ 2000ApJ...539..718C} is different from ours. One difference that may be important here is that their model is based on the assumption that the SFR of the burst is constant. We have demonstrated in Sect. \ref{trends} that most galaxies have short decay rates (30 Myr in $\sim$ 50\% of the cases). In such cases the effect of the age dependent dust attenuation will differ from when the SFR is assumed to be constant. In order to have an idea of the effect of different SFHs on the derived present SFR we compared the best results from spectral fits based on a decay rate of 3$\times$10$^7$ yr with one with a decay rate of 1 Gyr.  A 1 Gyr decay rate may be regarded as a nearly constant SFR for a starburst with a duration $\sim$100 Myr. We used our approach to correct for dust attenuation and found that in the latter case, the deduced SFR based on the \hax flux was $\sim$10\% lower than in the former. Thus, the difference between the corrections to the \hax flux because of the different assumptions as regards the SF history should not be important.

What about the derived masses? Brinchmann et al. use masses from the modelling by \citet{ 2003MNRAS.341...33K}. These are available from the MPA Garching homepages \footnote{http://www.mpa-garching.mpg.de/SDSS/DR7} and a comparison between our masses and Kauffmann's et al. masses indicates that the mismatch increases towards lower masses. Kauffmann et al. use two parameters - the observed 4000\AA ~break and \hdabs to derive \mlx ratios.  The models include starbursts with masses 0.03-4 times the mass of the old stellar component. This definition of mass fraction is not too different from our $f_{burst}$. But we show above that starbursts can have $f_{burst}$$<$3\%. Such low-mass bursts are not modelled by Kauffmann et al. and this can lead to incorrect \mlx ratios as is also shown in their  map of  the uncertainties in derived \mlx ratios (see their Fig.~5).  Moreover, for large $f_{burst}$ (and, as we have shown, $f_{burst}$ increases towards lower masses) it becomes difficult to derive the contribution to the light in the optical region limited to the two parameters they use.  Our strategy, to base the modelling on more or less the entire optical spectrum will, in our opinion, lead to more robust results. On the other hand, our model is simpler than Kauffmann's et al. We use only two populations, while they work with multiple component populations which is very useful when working with star-forming galaxies in general. We claim however that, using our model, we manage to derive reasonably accurate masses from spectra of starburst galaxies, using relatively few parameters.

As was described in Sec.~\ref{trends}, we found that the majority of low-mass starbursts, in contrast to the high mass starbursts, have short decay rates. The Kauffman's et al. models are based on a constant SFR. What are the consequences of the different approaches? Most likely the age and $\cal M$/L of the burst population in the model with constant SFR would be highest. The burst mass would be overestimated.

Finally, we wish to mention an additional problem occurring when a fit is made to the stellar continuum at ages below a few 10$^7$ yr. If the starburst is strong, the free-free and free-bound emission from the ionised gas will make a significant contribution \citep[e.g.][]{2002A&A...390..891B, 2008ApJ...676L...9Z} and cannot be neglected in the modelling. Since we include the nebular component in our model this will not be a problem for us but will have an effect in the work by Kauffmann et al. since they make a fit to the \citep{ 2003MNRAS.344.1000B} SEM that contains no nebular component. 

We find confidence in the fact that our mass determinations are strongly supported by the tight correlation between our photometric and dynamical mass estimates as well as the correlation between our data and model results from STARLIGHT (see Appendix).
 
\subsection{Starbursts and galaxy evolution}
\label{discussion:evolution}

In our study we have chosen to define a starburst as an event that will drastically influence the evolution of the galaxy over a short time. If we prefer to use the concept {\it starburst} as a measure of a dramatic change in the SFR compared to the present state of `normal' star formation activity, the $b$--parameter may not be the optimum parameter to use since one would have to lower the criterion with increasing mass and decreasing gas content. This motivated \citet{2009ApJ...695..561M,2010ApJ...721..297M} to base their $b$--parameter on the stars formed during the latest 6 Gyr. Similarly \citet{2015A&A...575A..74S} chose to define a starburst as a 4-folded increase in SFR compared to the typical galaxy of the same mass on the GMS. This may be a valid approach but it is not possible to apply it to SDSS galaxies in general. The problem is that we have too little information about what causes the burst. As we argued above, if the starburst is caused by a merger then a large amount of gas may be transferred to the starburst host galaxy and the previous gas content becomes irrelevant.

There are many claims that starbursts are related to tidal interactions but there are also critical views of this approach. Indeed, there is statistical evidence of increased SF activity in interacting galaxies but starbursts are relatively rare  \citep{1980A&A....89L...1H, 1981A&A....96..111H,1987AJ.....93.1011K, 1988ApJ...335...74B, 2004MNRAS.349..357B, 2004A&A...422..941C,2007AJ....133..791S,2008A&A...492...31D,2009ApJ...698.1437K}. We \citep{2003A&A...405...31B} argued, from comparing SF properties of isolated pairs to isolated single galaxies, that interactions rarely gave rise to starbursts. We also claimed that most Arp galaxies had a modest enhancement in SFR, a result that has been confirmed in other studies \citep[e.g.][]{2007AJ....133..791S}. Many of the studies mentioned above found a weak enhancement of star-formation in disks of interacting disc galaxies but an enhanced activity in the nuclear region. We also found an increase in SFR in the very central regions of a factor of $\sim$2. Similar results were published later, some showing a systematic increase in SFR with decreasing distance between components in pairs \citep[e.g.][]{2004MNRAS.355..874N,2008MNRAS.385.1903L,2012MNRAS.426..549S}. In most cases however, the increase in SFR is very modest and does not influence the global SFR significantly. On the other hand, many observations \citep[e.g.][]{2012MNRAS.426..549S}, including also postburst galaxies \citep{2015ApJ...799...59H}, indicate that the strongest starbursts were triggered by mergers and most likely the merger rate is correlated with the mean distance between galaxies. When one finds a tight pair of galaxies, there is an increased probability compared to isolated galaxies that a merger took place quite recently. One may think it is the close encounter that caused the starburst when in fact it is a recent merger and it only {\it appears} as if the neighboring galaxies caused the starburst. 

Observations indicate that the merger rate increases with redshift \citep[e.g.][]{2003AJ....126.1183C,2006ApJ...652..270B,2014AJ....148..137L}. Does this mean that the starburst rate is also increasing with redshift? It seems not. Although the main mode of star formation in the early universe was through mergers \citep{2008ApJ...677...37O} and the general SFR increases, fluctuations from the mean seem to stay the same  \citep{2007ApJ...660L..43N,2012ApJ...747L..31S,2015A&A...575A..74S}.  The feedback effects nicely regulate the mean SFR. 

In a study of elliptical galaxies,  \citet{2010MNRAS.402..985H} tried to separate the burst component from the old component using the Kennicutt--Schmidt law ``in reverse." They came to the conclusion that single bursts typically contribute to the stellar production with about 10\% of the total mass and that the star formation proceeds in cycles on a time scale of $\sim$100 Myr. The time scale agrees nicely with what we have derived and also with \citet{2009ApJ...692.1305L} and determinations from CM diagrams of local galaxies \citep{1996A&A...313..713V,1998AJ....116.1227D,2008ApJ...689..160W}. It also agrees well with theoretical models \citep{1994ApJ...431L...9M,2013MNRAS.430.1901H} but longer than \citet{1996ApJ...464..641M}. The burst mass fraction (cf. Fig.~\ref{mf_mass_pbsb}) in general is lower in our postbursts but rises towards low masses to a level close to 10\%. In a simulation of the Antennae galaxy, \citet{2008MNRAS.391L..98R} propose that the extraordinarily good conditions for star cluster formation in the system could be explained by the formation of compressive tides. Again we find the same time scale for the survival of the compressive mode, about 100 Myr. A longer duration was found by \citet{2009ApJ...695..561M,2010ApJ...724...49M} but their definition of starburst differs from ours.  Regardless, it seems obvious that starbursts occur in a `breathing, episodic' mode \citep{2007ApJ...667..170S} and if the bursting mode is caused by mergers, models predict a rather weak burst efficiency \citep{2008MNRAS.384..386C}, allowing a recurrence. 

An important question is what determines the duration of the starbursts. Is it the local feedback phenomena or the amount of  available fuel?  The gas depletion timescales of neutral hydrogen is of the order of 1 to a few Gyr \citep{2014A&A...563A..27L}. But star formation is tightly coupled to the amount of H$_2$ available. From \citet{2013AJ....146...19L} we obtain typical gas depletion time scales of molecular hydrogen for normal SF galaxies. These are remarkably independent of mass and typically $\sim$2 Gyr. The shortest depletion timescales are $\sim$100 Myr, close to the median lifetimes of our starbursts. A starburst with $b$$\sim$20 could apparently consume the molecular gas over 100 Myr. As we see from Fig. \ref{bpar_age}, such cases exist but are rare. Most likely molecular clouds are dispersed by stellar feedback after a relatively small fraction of the gas has been consumed. But, since the $b$-parameter increases with decreasing burst age it seems that the H$_2$ content marks the upper limit to the $b$-parameter at a give age.

\section{Conclusions}

\label{sec:conclus}

We have used data from the SDSS DR7 release in the redshift range 0.02$\leq$z$\leq$0.4 to investigate the starburst properties of galaxies in the local universe and to establish a link between starburst and post-starburst (called postburst in our study) galaxies. We also had a first look at the role of AGNs in this context at high luminosities. 

In order to select the starburst candidates we assigned a lower limit to the \hax emission line equivalent width of  \ewha =60\AA. This sample contained both starbursts and non--starburst galaxies that were later separated after proper treatment of the effects of dust attenuation. The selection criterion for the postburst galaxies was based on the \hdx line in absorption, demanding that the equivalent width should be $\leq$-6\AA.  In our analysis, we used a spectral evolutionary model based on two stellar components -- a young and an old. The young component is allowed to vary in age and mixes with the old component until we obtain a best fit that also gives an \ewhax that agrees with the observations. AGNs were selected into a separate group. The AGNs start to mix with the starburst/postburst sample at $M_r$$\approx$--21. Therefore studies of starburst properties at the bright end of the luminosity function demand special care. In this study we focus on the sub-L$^*$ population and do not treat the AGN problem in detail. 

The corrections for dust attenuation are important and we have applied two different methods. In the emission line sample we have used the \ha/\hbx ratio and assumed that the attenuation is age dependent and dramatically changes over a few Myr during the young phase. In the postburst case we have tentatively modified the amount of dust until we get the best fit. From the model we derive SFRs, ages and masses of the young and old component.

We also derive dynamical masses from the width of the \hax emission line. These masses were compared  with the photometric masses, after adding an estimate of the gas mass. We find a tight relation between the two over the entire mass range which gives us confidence that our masses are reliable to within $\sigma$=0.35 dex. 

We defined a starburst as a galaxy with a birthrate parameter $\frac{SFR}{<SFR>}$$\geq$3. In the lower end of the LF this population corresponds to only 0.5-1\% of all galaxy types in the local universe. 1\% of sub-$L^*$ star-forming galaxies is a starburst. We estimate that starbursts contribute only 4.4$^{+1.8}_{-1.2}$\% of the total production of stars in the local universe. The median value of the birthrate parameter is $b$$\sim$ 4 and decreases slowly with increasing mass. The {\sl mean} birthrate parameter over the starburst epoch is higher than the present value.  We interpret this as if the starbursts had a more violent past and are declining in an approximately exponential mode. The typical mass fraction of the burst population is 5\% of the total mass, slowly decreasing towards higher masses.

In the analysis of the postburst sample we find that in 95\% of the sample, the decaying burst  component has a mass $>$3\% of the total mass. The median mass fraction of the postburst population is 5-10\%, which is significantly smaller than what has been reported by other groups. We can select a subsample of our postburst {\it candidate} sample and derive the properties of the future generation of postburst galaxies to compare to our observed postburst sample. We then find that the observed LF of the postburst sample should run slightly above and parallel to the $f_{burst}$$\geq$3\% sample at low-intermediate luminosities when difference in luminosities and lifetimes between starburst and postbursts are taken into account. This is what we see in the LF we derive. We therefore think we have established a link between the active starburst and most of its descendants. Short bursts may have high birthrates but low $f_{burst}$. They will not show up as postbursts. On the other hand we also find that a small part of the postburst galaxies stem from star--forming galaxies with $b$$<$3.

The median age of the starbursts is $\sim$70 Myr, which appears to be independent of mass. The lifetime of the burst appears to be regulated by the ratio of available gas mass and dynamical timescale. The age of the postburst population is a few 100 Myr higher as one would expect. The progenitors to the postburst galaxies, i.e. galaxies with $f_{burst}$$>$3\% display a bimodal age distribution with a transition from ages slightly below 100 Myr to twice this value at log$\cal M$(\msun)=10.5. This value is the same as found previously in the colour distribution of SDSS star-forming galaxies. We discussed possible explanations.

At high luminosities AGNs dominate the population while the starburst population diminishes. The LF of the postburst population closely follows that of the AGNs while the progenitor LF separates more and more, indicating an deficit of a factor dex at the highest luminosities. The link between the postburst LF and the AGNs tells us that starbursts are also closely linked to AGNs but are difficult to detect. There are probably two reasons for this - an increasing dust obscuration and AGN domination at high luminosities. While the starburst dust obscuration steadily increases with luminosity, the attenuation in postbursts is lower and reaches a maximum and then decreases at high luminosities. Although we have not tried to estimate the amount of obscuration in AGNs, that fact that the LF follows the postburst LF, indicates that they have similar dust properties. If so, the starburst phase would precede both the postburst $\sl and$ the AGN phase.

An interesting question is the ability for a starburst to create SN superwinds that can lead to the removal of a significant fraction of the gas in the system. It is a question that has an impact on our understanding of the cosmic reionisation. We used the SFR per area as a criterion for gas removal efficiency and found that massive galaxies with high $b$-values are the most probable drivers of mass ejection. This supports the downsizing scenario for star formation as function of redshift.

The processes that initiate starbursts, in the majority of the cases probably via mergers, seem to be balanced by feedback processes that hinder catastrophic events and ensures that the star formation efficiently has a maximum level. We find only small variations in the birthrate parameter, mass fraction and age over the entire mass range. There are reasons to believe that the situation does not dramatically change with redshift but that the change is in the mean SFR. This does not mean that starbursts are unimportant. Starburst galaxies represent the ultimate test bench on galaxy scales for the physical processes in the modern universe. By studying starbursts we learn to understand the distant universe when the occurrence of events we now call starbursts were active. The high power of starbursts makes it possible to inject lots of energy into the ISM and drastically change the conditions and properties of a galaxy. Superwinds, high neutron rate production, Lyman continuum and Lyman line leakage, AGN ignition and many other unsolved feedback processes are some of the very exotic consequences of true starbursts.
\begin{acknowledgements}

We are indebted to Dr. Polychronis Papaderos for stimulating discussions and assisting us with the use of the STARLIGHT code and CAUP computer facilities. The anonymous referee is thanked for useful comments on the draft. 
EZ acknowledges research funding from the Swedish Research Council (project 2011-5349), the Wenner-Gren Foundations and the Swedish National Space Board.
This research has made use of NASA's Astrophysics Data System Bibliographic Services.

Funding for the SDSS has been provided by
the Alfred P. Sloan Foundation, the Participating Institutions, the National
Aeronautics and Space Administration, the National Science Foundation,
the U.S. Department of Energy, the Japanese Monbukagakusho, and the Max
Planck Society. The SDSS Web site is http://www.sdss.org/.

The SDSS is managed by the Astrophysical Research Consortium for
the Participating Institutions. The Participating Institutions are The
University of Chicago, Fermilab, the Institute for Advanced Study, the
Japan Participation Group, The Johns Hopkins University, Los Alamos National
Laboratory, the Max--Planck--Institute for Astronomy, the
Max--Planck--Institute for Astrophysics, New Mexico State University,
University of Pittsburgh, Princeton University, the United States Naval
Observatory, and the University of Washington.

\end{acknowledgements}

\bibliographystyle{aa}
\bibliography{sloan}

\newpage
\appendix

\section{Model stability}

In this study our results rely heavily on the outcome of our spectral evolutionary modelling. The information contained in a low--dispersion spectrum of a young stellar population is quite limited and it is relevant to ask to what extent we can trust the results. We should also be concerned about our lack of information about what preceded a strong starburst, the metallicity of gas and stars and the stellar mass function. 

Our spectral evolutionary model has previously been used in different contexts. It has been tested on super star clusters, blue compact galaxies and  low surface brightness galaxies and seems to behave well in these cases \citep[e.g.][]{2001A&A...375..814Z,2002A&A...390..891B,2003A&A...408..887O,2005A&A...435...29Z}. In the present investigation we therefore feel that we may limit ourselves to a few tests focused on the determination of the age and mass fraction of the young component. 

Ideally we would like to test our model against a set of simulated galaxies with known properties of stars gas and dust, covering the range of parameters we expect in our sample. Here we do it simply (via $\chi ^2$ analysis), but this is sufficient to give us the information we need. Other studies may need more advanced statistical methods. We may compare our strategy to that of the much cited work of \citet{2004MNRAS.351.1151B}. In our case we do not need detailed information about the likelihood distributions of the derived parameters in the manner they prefer. Instead, we obtain the information we need from the tests discussed below. The reason is that  our sample of galaxies is more restricted than theirs. Brinchmann et al.  work with star forming galaxies in general while we are only interested in starburst galaxies where the light in the optical region is normally completely dominated by the young stellar population. The dust attenuation can be easily controlled since the \hax and \hbx emission lines, used for deriving the extinction coefficient, are strong. Therefore, for most of the starburst galaxies the evolution of the corrected spectrum with time is close to a single parameter function: spectral slope vs time. As the starburst gets older, the extinction corrected spectrum gets redder. As we show in the diagrams below, there are therefore very few options for ambiguity and a simple $\chi ^2$ test can be used to find the most adequate model of the mixed population. Similar approaches to analysing data sets by applying models with a limited number of component populations can be found elsewhere in the literature and have been demonstrated to give reasonably coherent results \citep[e.g][]{2010ApJ...709..644I,2011MNRAS.417..900D,2011ApJ...730...61K,2012MNRAS.425..540H,2013A&A...549A...4S,2014A&A...571A..75M,2015MNRAS.447.3442L}.

In order to test the reliability of our model we created a grid of simulated spectra with a mixture of an old and a young population. To this we added various amounts of random noise. We fed the resulting simulated spectrum into our SEM and compared the input spectrum with the solution proposed by the model, selecting the model with the minimum $\chi ^2$ value as the ``best fit". In the case of the simulated spectra representing the active starburst phase, we varied the metallicity, the star formation history and the relative mass fraction to test the stability of the solutions. For example, we wanted to know if the programme could reliably distinguish between a 100 Myr starburst mixed with an old stellar population of some mass fraction from a 300 Myr starburst mixed with an old component of smaller mass. We also wanted to know when we could start to rely on the mass fractions. Starbursts with large mass fractions completely dominate the light over that from the old generation so if the spectra are noisy, to what level can we retain the initial information?  We show in the following how the tests give us information about the uniqueness of the fits. Moreover, our tests not only show how unique the best fits are, but also how closely the models reproduce the input parameters. This part of the test is important, but is not often given proper emphasis in comparable investigations.

In the case of the postburst galaxies we tested the reliability of the dust attenuation correction, the derived mass fraction and the age determination.  As an additional test we compared the total masses, young+old, determined by our model and the STARLIGHT spectral synthesis model \citep{2005MNRAS.358..363C}.

\subsection{Starbursts}

\subsubsection{Ages}

\begin{figure}[h!]
\centering
 \resizebox{\hsize}{!}{\includegraphics{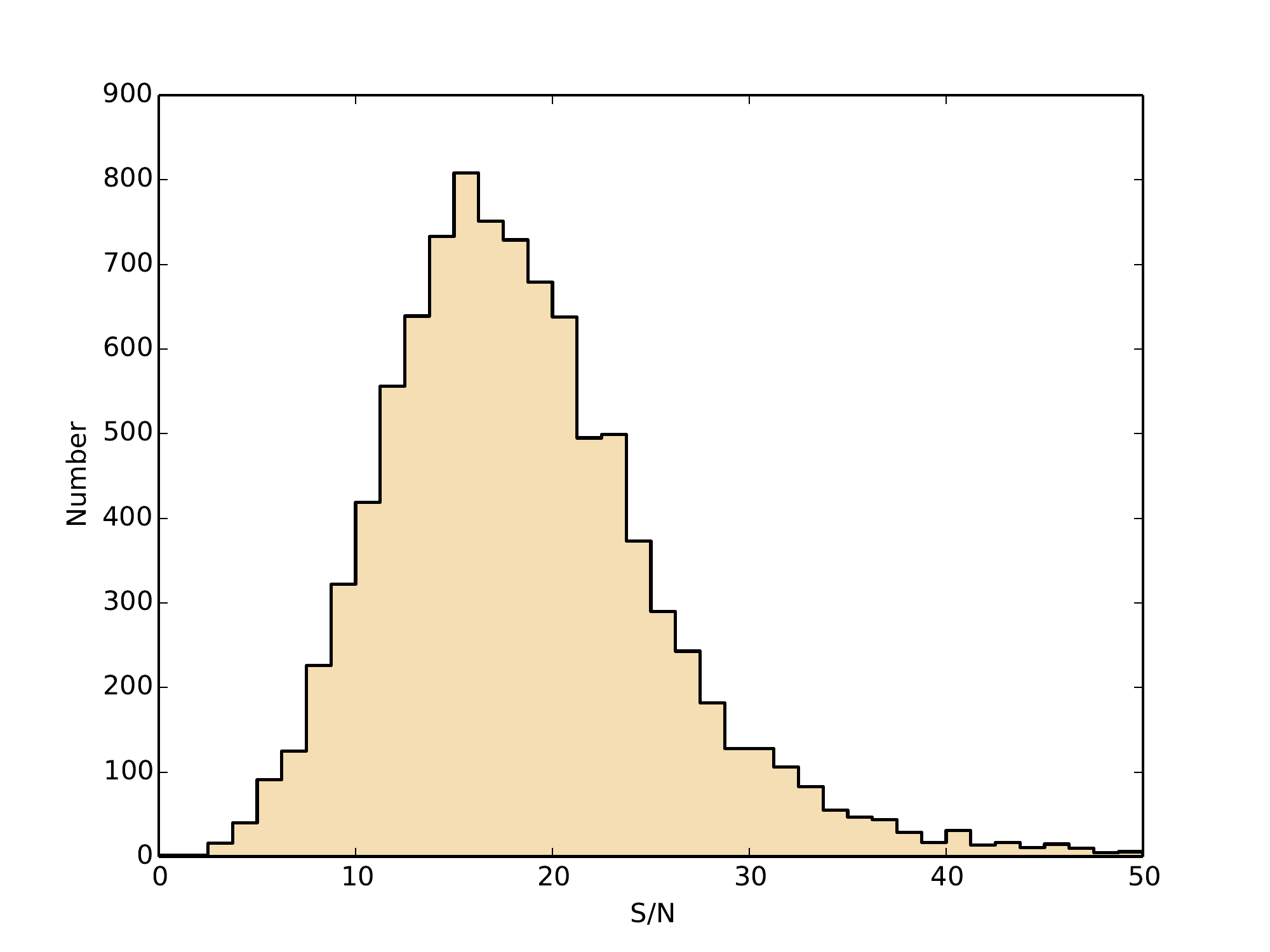}}
\caption{Distribution of the S/N in the region 5400-5800\AA ~of the SDSS starburst candidate galaxy spectra included in this study, as measured by us.}
\label{sdss_sn}
\end{figure}

Here we have a look at the correlation between the ages of the input synthetic galaxy spectra and the ages of the best fitting models to these spectra, as suggested by our SEM. We do the same with the mass fractions. The input synthetic spectrum is a mixture of a young starburst population and a 10 Gyr old component that was formed in a burst of duration 100 Myr. The synthetic spectra were produced at a few discrete ages as seen in the figures of this appendix. The metallicity was fixed to Z=0.008 ($\sim$40\% solar). In the SEM analysis of the synthetic spectra, the metallicities of the SEM were allowed to take on the values Z=0.004 or 0.008 and the star formation exponential timescale the values 100 Myr or 1 Gyr. The synthetic spectra were degraded with random Gaussian noise resulting in three different S/N per \AA: 10, 20 and 40. This corresponds well to the range in S/N of the SDSS spectra as shown in fig \ref{sdss_sn}. The S/N in these were derived in the region 5400-5800\AA ~and the median S/N was found to be 17.6.

In Fig. A.2 we show the age we recovered from our model as a function of the age of the starburst in the synthetic spectrum. Metallicities and decay rates are the same in the model producing the synthetic spectrum as in the model recovering the age and mass fraction. The correlation is strong even in the sample with the lowest S/N, with a maximum deviation of 0.1 dex. For each time step there are 10 models with different mass fractions: 2, 4, 6, 8, 10, 20, 40, 60, 80 and 100\%. So all in all we have 240 spectra in the diagram.

\begin{figure}[h!]
\centering
 \resizebox{\hsize}{!}{\includegraphics{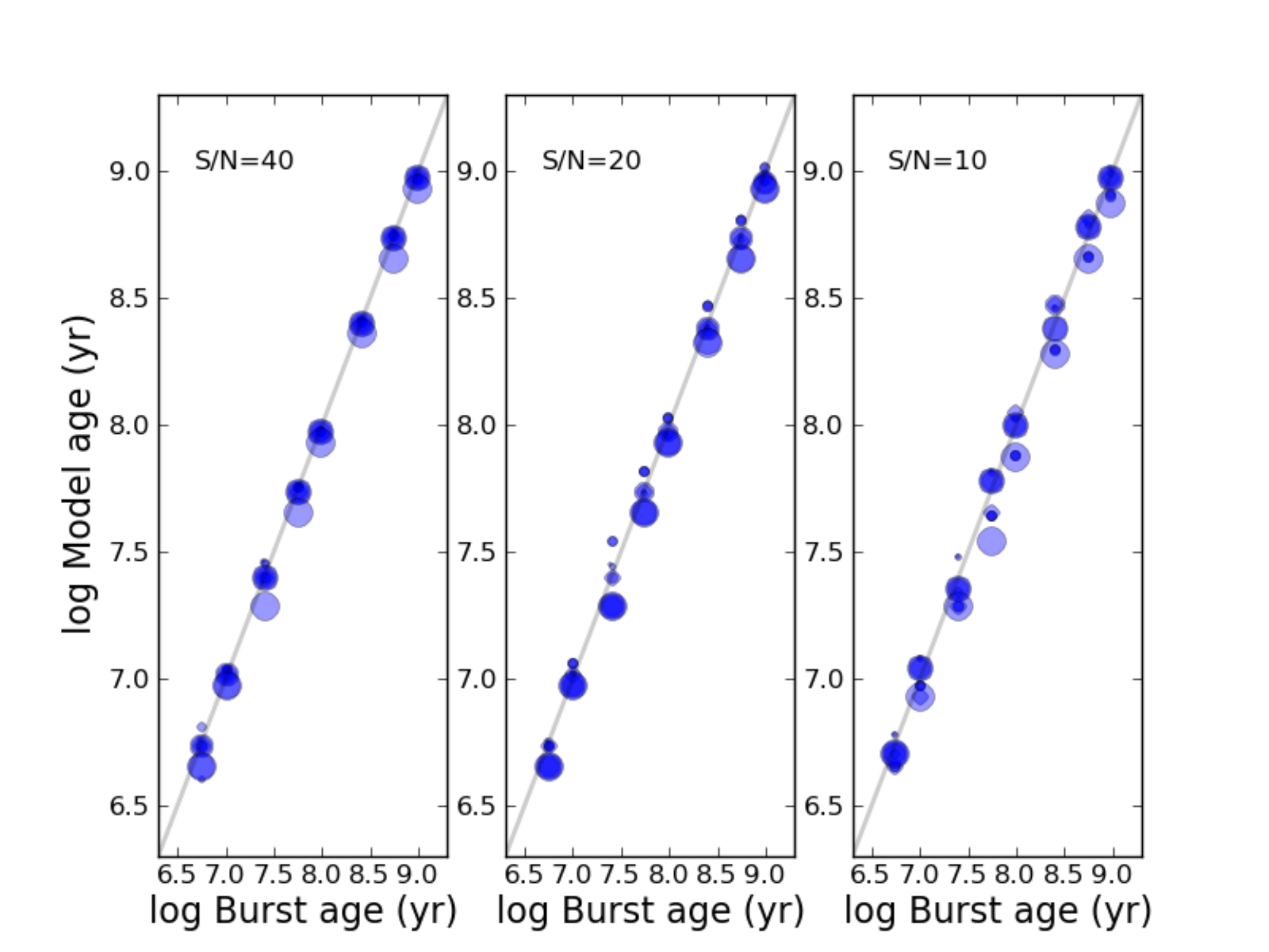}}
\caption{Test of the reliability of the SEM age determination. Synthetic spectra were fed into our SEM model and the derived age of the burst (ordinate) versus the input age (abscissa) are displayed. Here the star formation history is fixed to one single model with a decay rate of 1 Gyr and a metallicity of Z=0.008. At each time step there are 10 models with different mass fractions: 2, 4, 6, 8, 10, 20, 40, 60, 80 and 100\%. Larger symbol: larger mass fraction.}
\label{age_sn_mf}
\end{figure}

\begin{figure}[h!]
\centering
 \resizebox{\hsize}{!}{\includegraphics{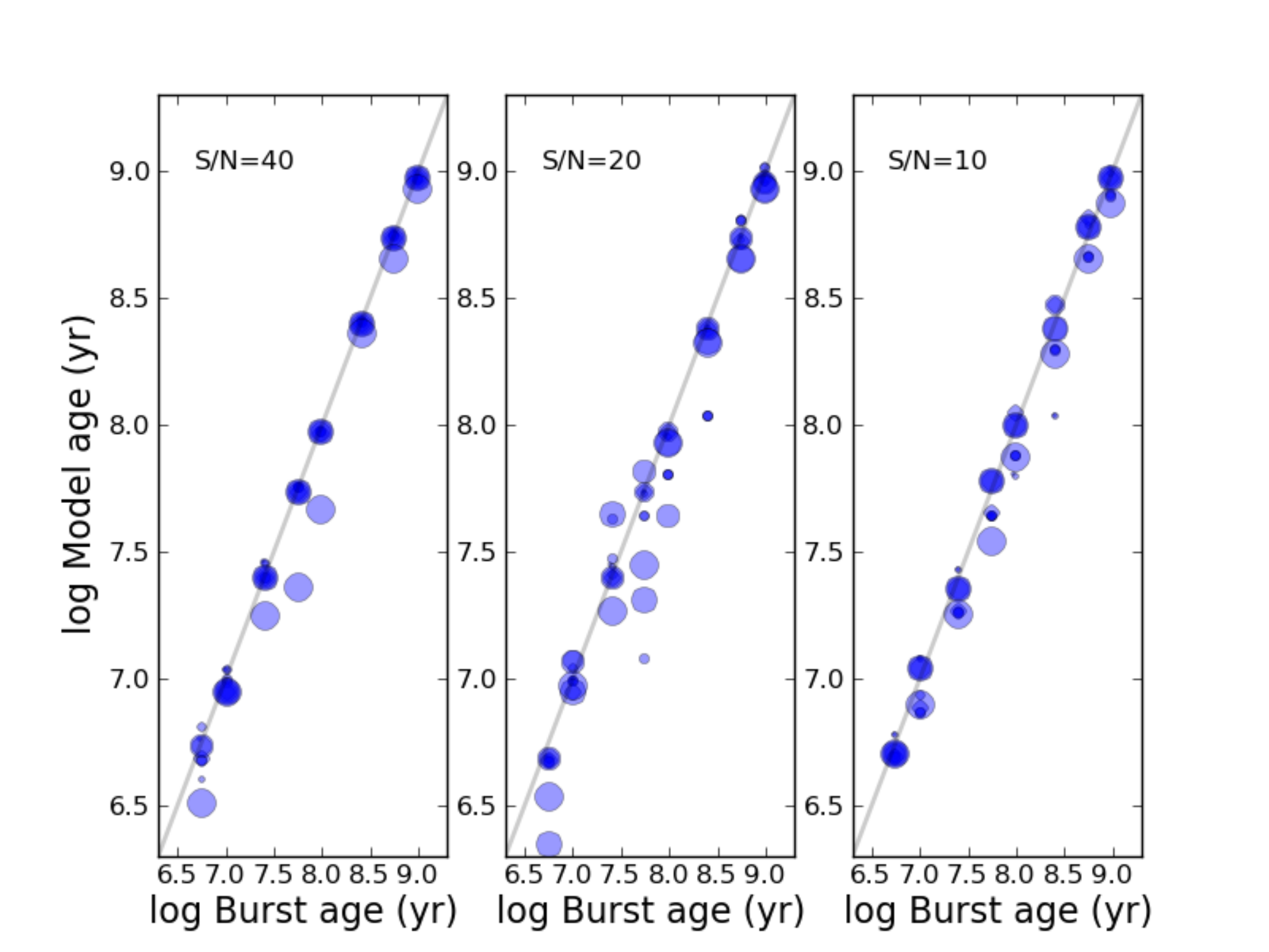}}
\caption{Test of the reliability of the SEM age determination. Synthetic spectra were fed into our SEM model and the derived age of the burst (ordinate) versus the input age (abscissa) are displayed. The input model has a decay rate of 1 Gyr but there are two different choices for the star formation history in the fit, either with a decay rate of 100 Myr or one with 1 Gyr.  The metallicity was fixed at Z=0.008. At each time step there are 10 models with different mass fractions: 2, 4, 6, 8, 10, 20, 40, 60, 80 and 100\%. Larger symbol: larger mass fraction.}
\label{age_sfh_mf}
\end{figure}

In Fig. \ref{age_sfh_mf} we show a similar relation but while the metallicity is fixed to Z=0.008, the decay rate is allowed to vary in the modelling ($\tau$=100 Myr or 1 Gyr). Here the fits are not as good as in the former case but large deviations only affect a few of the  spectra. In about 80\% of the cases the deviations are less than 0.1 dex.

\begin{figure}[h!]
\centering
 \resizebox{\hsize}{!}{\includegraphics{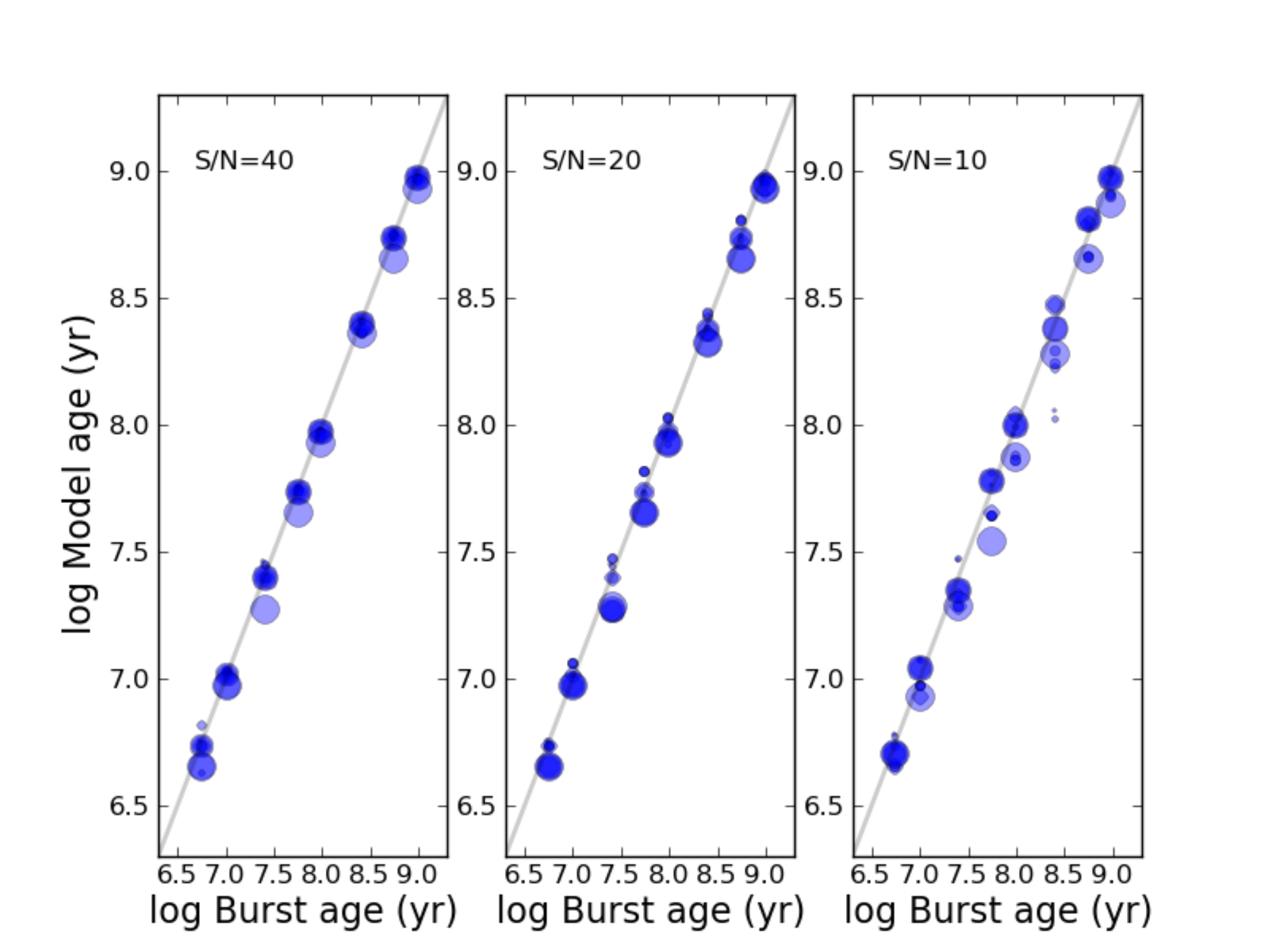}}
\caption{Test of the reliability of the SEM age determination. Synthetic spectra were fed into our SEM model and the derived age of the burst (ordinate) versus the input age (abscissa) are displayed. The input model has a decay rate of 1 Gyr but there are two different choices for the star formation history and metallicity in the fit, either with a decay rate of 100 Myr or one with 1 Gyr and a metallicity of Z=0.004 or Z=0.008. At each time step there are 10 models with different mass fractions: 2, 4, 6, 8, 10, 20, 40, 60, 80 and 100\%. Larger symbol: larger mass fraction.}
\label{age_met_mf}
\end{figure}

In Fig. \ref{age_met_mf} we fixed the metallicity and decay rate of the programme that produced the synthetic spectra (Z=0.008, $\tau$= 1 Gyr), but they are allowed to vary in the modelling (Z=0.004-0.020 and $\tau$=100 Myr or 1 Gyr). In this case the ages agree very well even if the spectra are noisy. Note that the timescale and the metallicity of the burst that the fitting code chooses to agree almost 100\% with the input data. 

\begin{figure}[h!]
\centering
 \resizebox{\hsize}{!}{\includegraphics{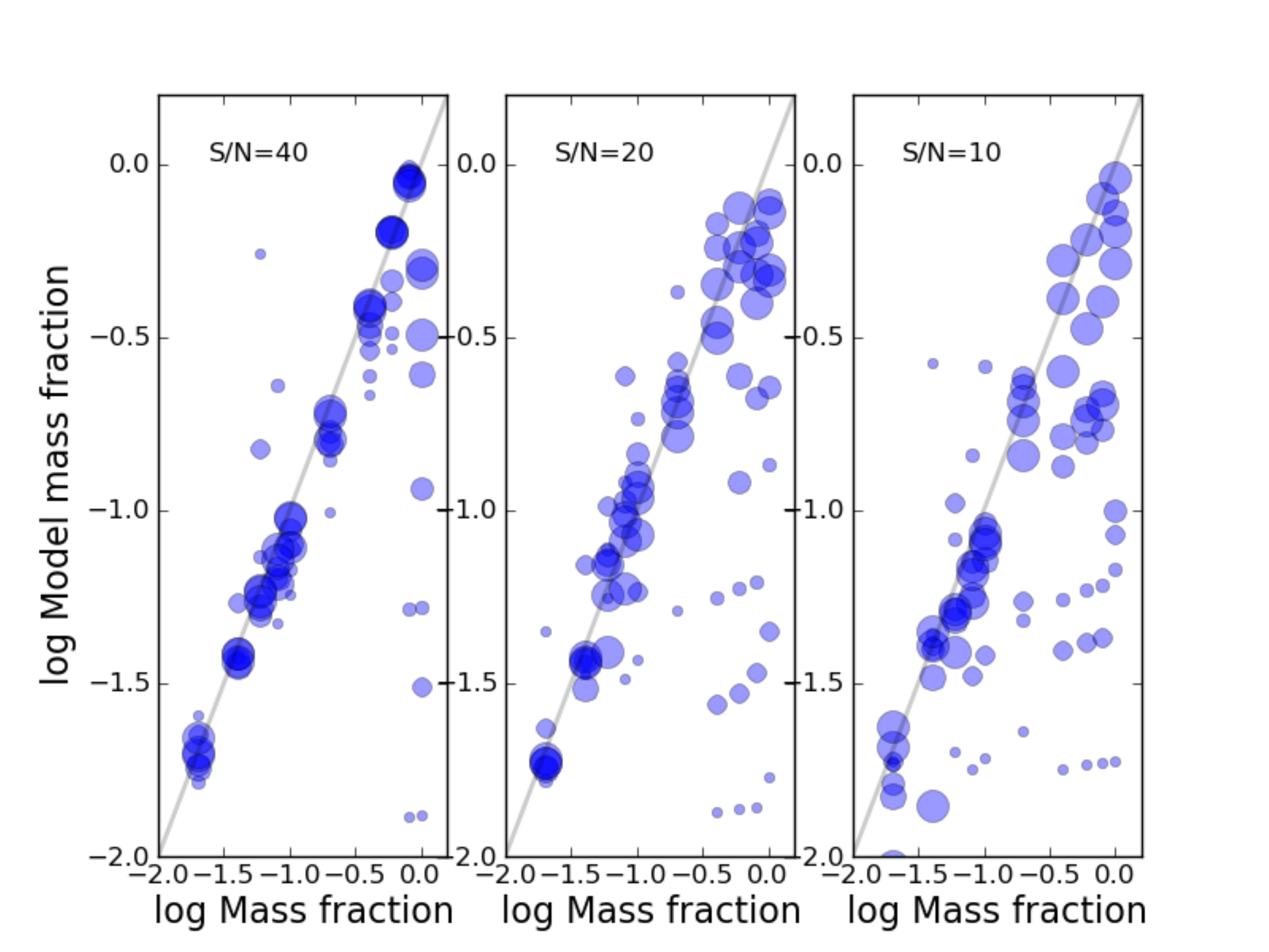}}
\caption{Test of the reliability of the SEM mass fraction determination. Synthetic spectra were fed into our SEM model and the derived mass fraction of the burst (ordinate) versus the input mass fraction (abscissa) are displayed. The same conditions for modelling holds as in Fig. \ref{age_met_mf}. At each time step there are 8 models with different ages of the starburst: 5.5, 10, 25, 55, 95, 250, 550 and 950 Myr.  Larger symbol: higher age.}
\label{mfrac_met_age}
\end{figure}

In Fig. \ref{mfrac_met_age} we show how the mass fraction can be recovered. We see that we are quite safe with spectra of high S/N but run into problems at S/N=20, in particular at high mass fractions and low ages. The mass fractions tend to be underestimated. We argue in our investigation that mass fractions of starbursts in general are low. Of course we cannot exclude that some of those galaxies, in case the S/N is low, in fact have a significantly higher mass fraction but continuity arguments tell us that this is very unlikely and should not affect the general results. Moreover, only 10\% of the SDSS galaxies have ages below 30 Myr where the problem occurs. From the diagrams we may say that mass fractions below 50\% typically agree to within 0.2 dex.

\begin{figure}[h!]
\centering
\includegraphics[width=9cm]{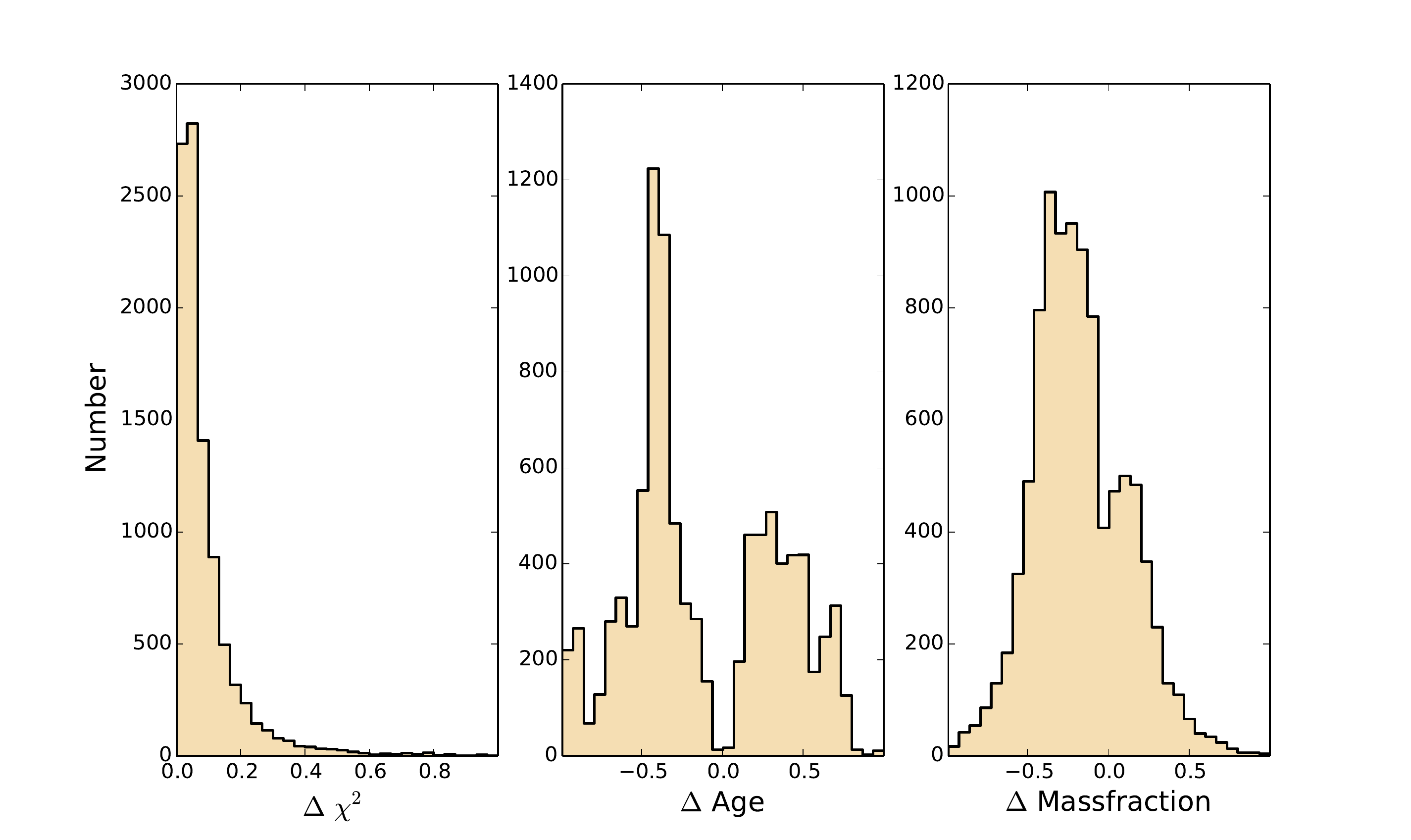}
\caption{Illustration of how well the agreement between the best and worst spectral fits to a single SDSS spectrum agree as model metallicities, dust and star formation histories are allowed to vary. The histograms show the deviations in log of the ratio between the predicted parameter values of the worst and the best fits.}
  \label{hist_deviations}
\end{figure}

As described above, applying the model parameters given in Table~\ref{modspectra} on each SDSS spectrum and varying the age and dust attenuation, will result in 36 (12 young $\times$ 3 old) best solutions. More information about the stability of the solutions presented here can be obtained by comparing the {\sl best} fits with the {\sl worst} fits among this set of 36 solutions. Figure~\ref{hist_deviations} displays histograms of the deviations between the best and worst fits. We plot the parameter $\Delta$ defined as the log of the ratio between the worst and  best fits among the 36 different model results, $\Delta$=log(worst/best) where we look at the parameters $\chi^2$, age and the mass fraction of the young component. We see from the diagram that ages (and to a lesser degree the mass fractions) can differ by a factor up to 5 but that the mass fractions are more stable and in general agree between the different solutions within a factor of 2. 

\subsection{Postbursts}

\subsubsection{Dust attenuation}

\begin{figure}[h!]

\includegraphics[width=9cm]{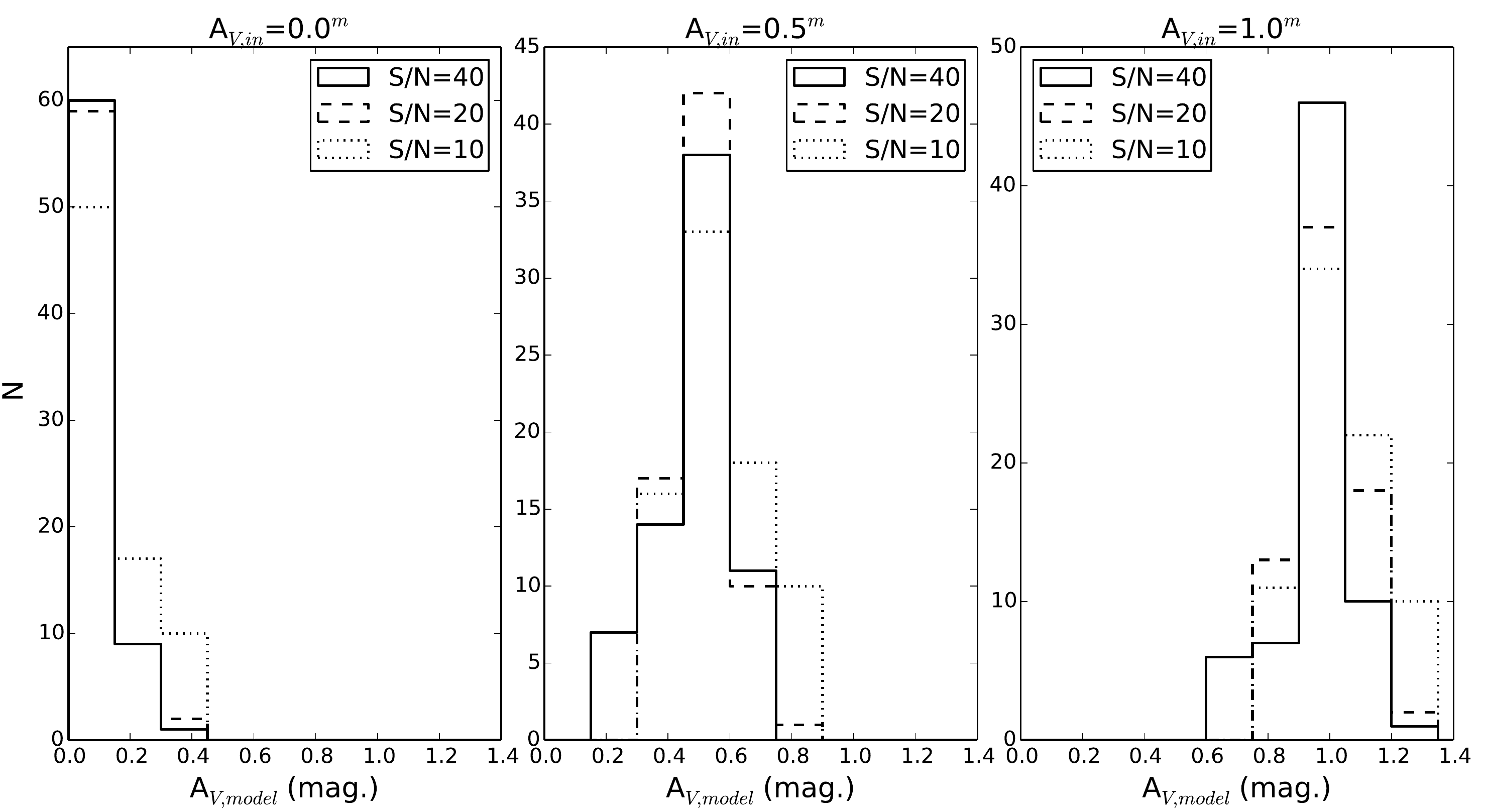}
\caption{Test of the reliability of our determination of the dust attenuation in the V band (A$_V$) based on a grid of 80 synthetic postburst spectra with ages 150, 250, 350, 450, 550, 750, 950 and 1500 Myr and mass fractions 2, 4, 6, 8, 10, 20, 40, 60, 80 and 100\%. The preceding starburst was assumed to have constant SFR over 100 Myr. The old component was assumed to have an age of 10 Gyr. Both have 40\% solar metallicity. To these spectra we added reddening corresponding to A$_V$ = 0, 0.5 and 1 magnitudes as titles in the diagrams. We then degraded the spectra with Gaussian noise corresponding to the S/N per\AA = 10, 20 and 40. Finally we used our SEM to derive A$_V$ from the noisy spectra, as shown on the abscissa. The result is displayed in the histograms with the number of spectra on the ordinate.}
  \label{hist_av}
\end{figure}

The amount of dust attenuation in the starburst sample can be derived from the \ha/\hbx ratio. In the postburst sample we are forced to use a more unreliable method. We calculated a grid of synthetic postburst galaxies, more or less in the same spirit as with the starburst galaxies. At each time step of the modelling of a target spectrum we varied the value of the dust attenuation as we asked the code to make a new fit. That gave us the value of the amount of attenuation. How reliable is this method, considering that the shape of the spectrum is also affected by the age of the stellar population and the mass fraction? In order to test the reliability of our determination the dust attenuation expressed in magnitudes in the Johnson V band (A$_V$) we produced a grid of 80 synthetic postburst spectra with ages 150, 250, 350, 450, 550, 750, 950 and 1500 Myr and mass fractions 2, 4, 6, 8, 10, 20, 40, 60, 80 and 100\%. The preceding starburst as well as the old population (age 10 Gyr) was assumed to have had a constant SFR over 100 Myr. Both have 40\% solar metallicity. To these spectra we added reddening corresponding to A$_V$ = 0, 0.5 and 1 magnitudes. We then degraded the spectra with Gaussian noise corresponding to S/N per\AA = 10, 20 and 40. Finally we used our SEM to derive A$_V$ from the noisy spectra. The result is shown in Fig. \ref{hist_av}. The figure shows what our SEM proposes to be the most probable value of the attenuation in A$_V$ magnitudes. The S/N is varied over the range 10-40 and as we see, our SEM manages to a obtain the correct value to high degree. A$_V$ is determined to an accuracy of $<$0.1 dex.

\subsubsection{Ages}

\begin{figure}[h!]

\includegraphics[width=9cm]{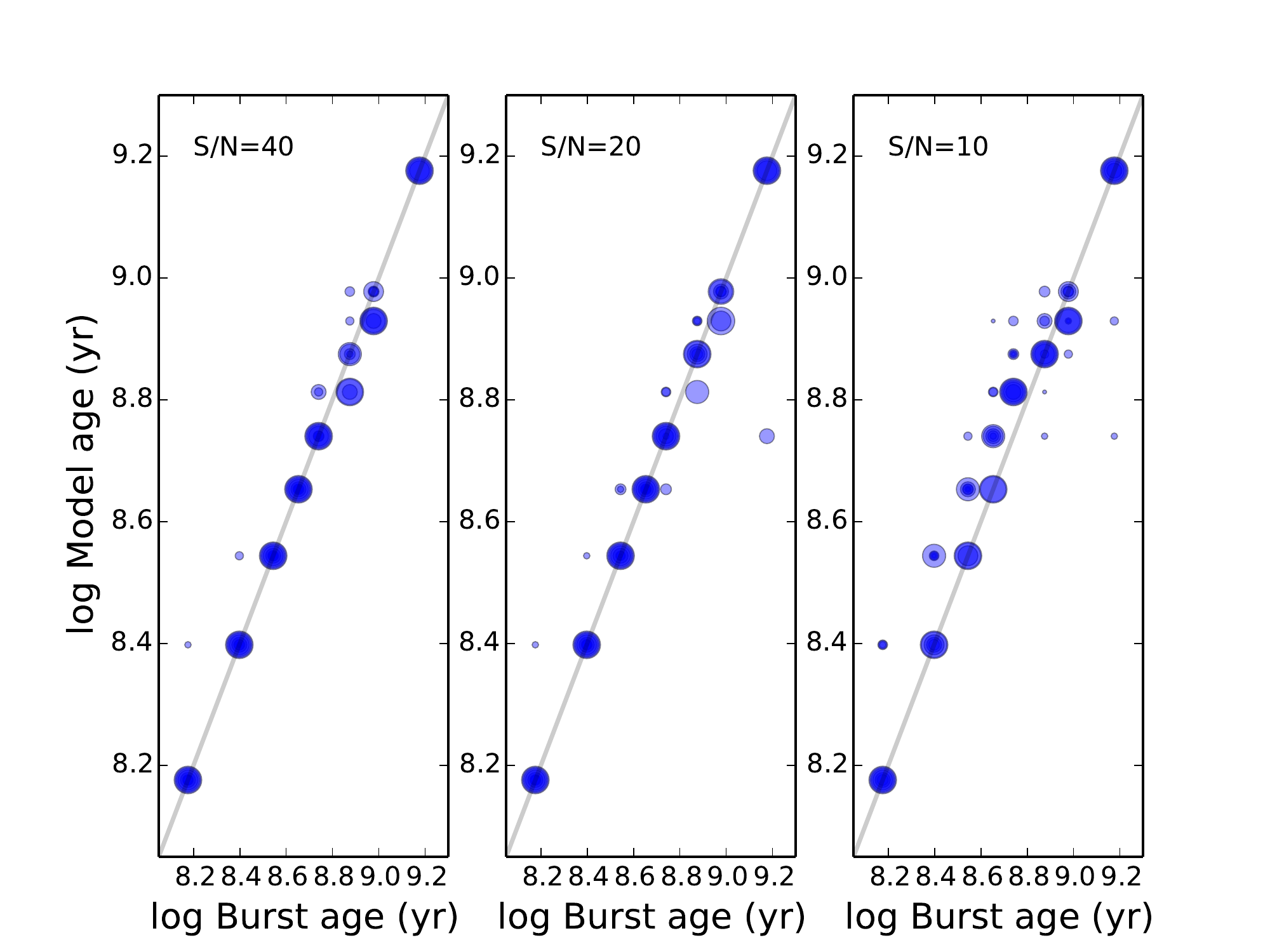}
\caption{Test of the reliability of our determination of the ages of the postburst galaxies. We loaded our SEM with the same set of spectra described in the previous section applying 3 different S/N reductions. Each time step contains 10 spectra with different mass fractions between 2 and 100\% as previously described. Larger symbols represent larger mass fractions.}
  \label{age_age_mf_exti_p5}
\end{figure}

Here we carried out a test of the age determination in a similar way as with the starburst sample. We used the same spectra as were discussed in the previous section assuming A$_V$=0.5 for all. Fig. \ref{age_age_mf_exti_p5} shows the result. The correlation is fairly high with 
only a few outliers.

\subsubsection{Mass fractions}

\begin{figure}[h!]

\includegraphics[width=9cm]{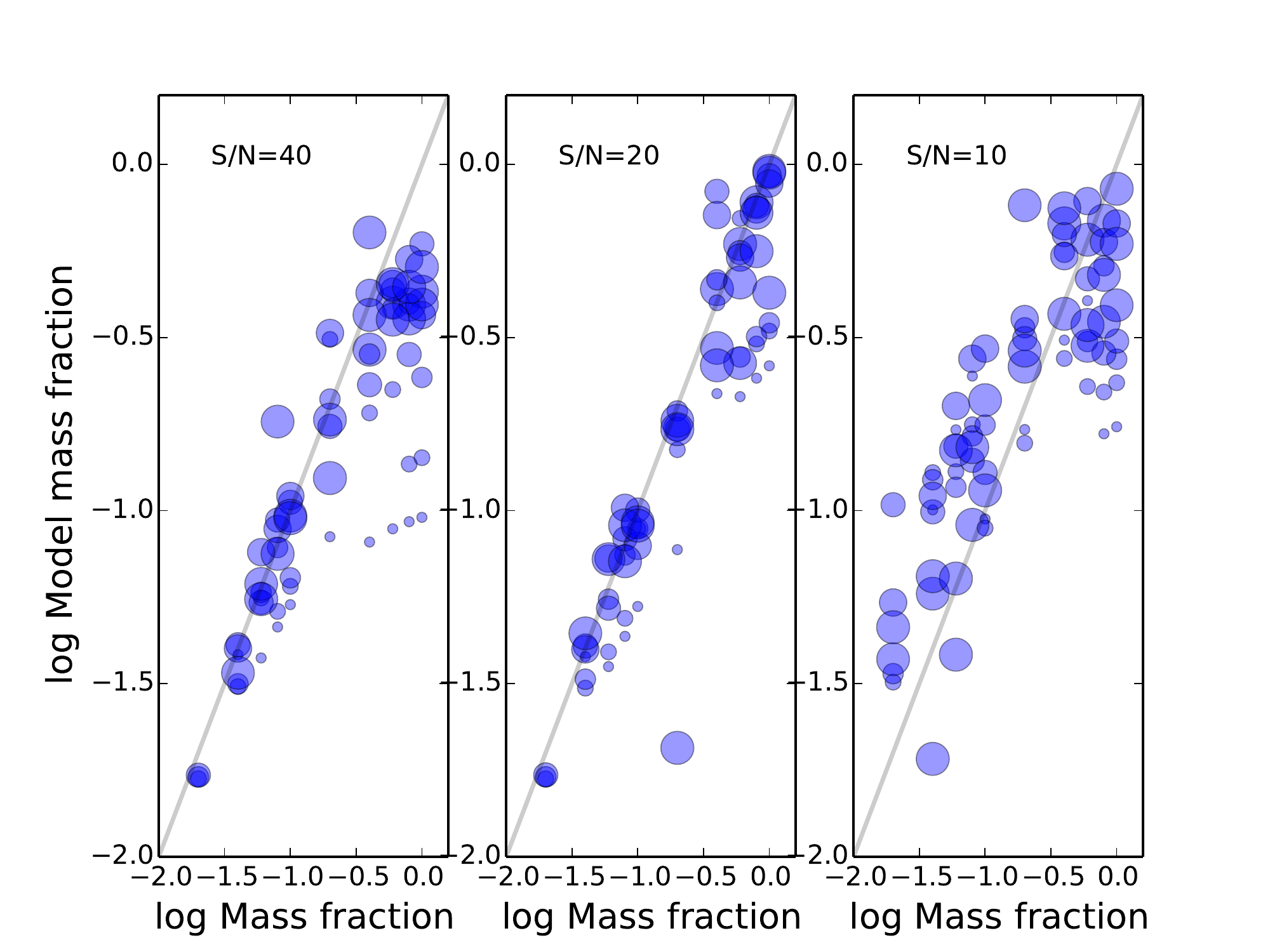}
\caption{Test of the reliability of our determination of the starburst mass fractions of the postburst galaxies. The procedure used was the same as described in Fig. \ref{age_age_mf_exti_p5}. The size of the symbols represent different ages. Larger symbols represent larger ages.}
  \label{mfrac_mfrac_age_exti_p5}
\end{figure}

In a similar fashion as described above, we tested the reliability of the calculation of the mass fractions of the starburst. Fig. \ref{mfrac_mfrac_age_exti_p5} shows the results. We used the same synthetic spectra as described above. Clearly the determination of the mass fractions have significant problems at low S/N. We tend to overestimate the mass fractions by a factor of almost 2 at low S/N. The situation is better at higher S/N. At mass fractions below 10\% the mass fractions between the input and the model results tend to agree very well.

\subsubsection{Masses}

\begin{figure}[t!]
\centering
 \includegraphics[width=\columnwidth]{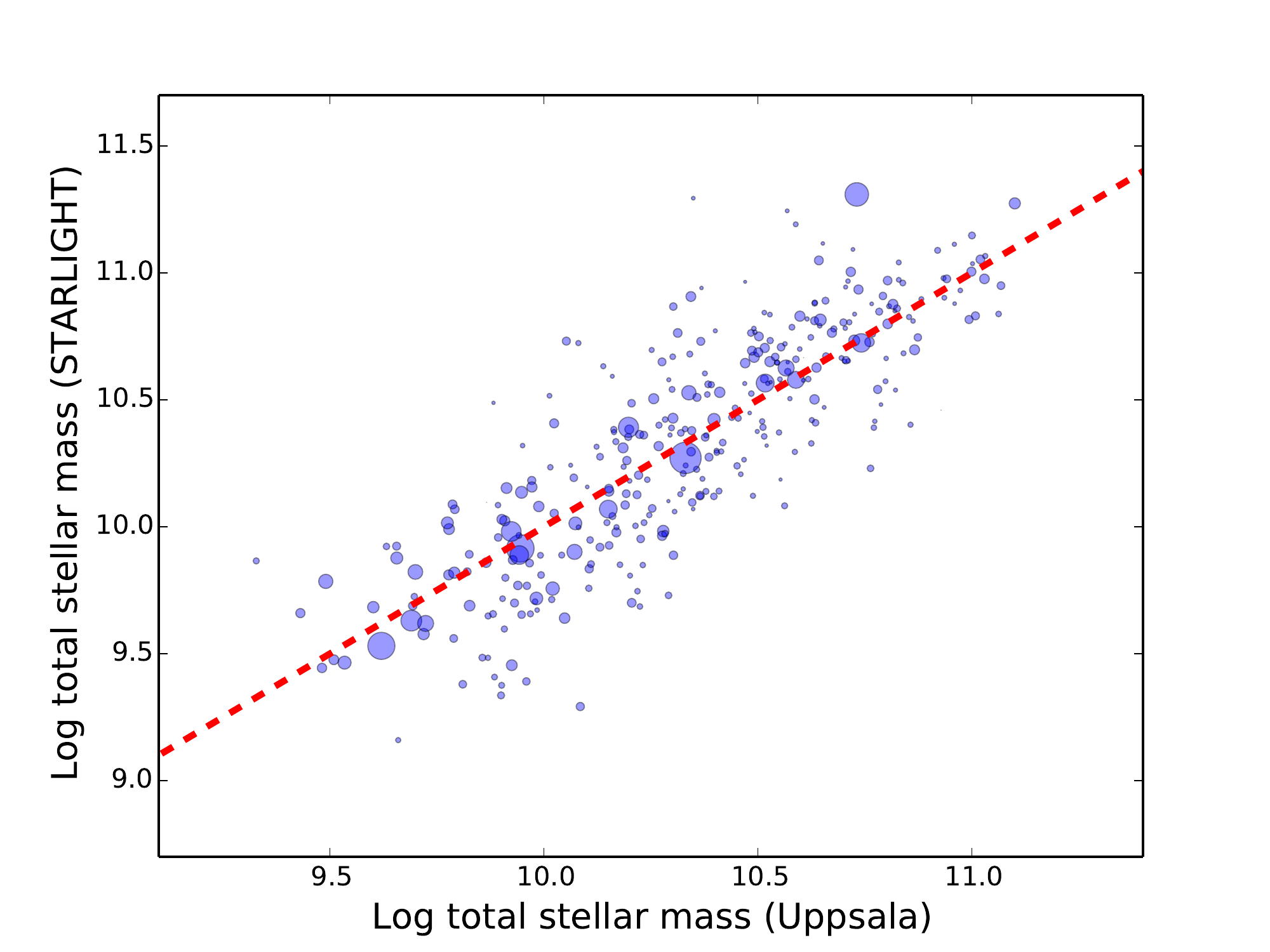}
\caption{Mass (based on fiber magnitudes) of the stellar (stars+stellar remnants) component of a set of intermediate/high-mass postburst galaxies as derived with our code (Uppsala) vs. the STARLIGHT code. The sizes of the symbols indicate the transparency of the starlight: small symbols - large dust attenuation. The hatched line marks the 1:1 relation.}
  \label{massu_masss}
\end{figure}

In Sect. \ref{masses} we demonstrated that there was a strong one-to-one correlation between dynamical mass and photometric + gas mass. We are also interested in testing our mass determinations against similar models. We have limited possibilities to do so as concerns the starburst galaxies since in almost all cases the nebular component is not included in other models. But we may compare postburst modelling. Here we have had a look at the STARLIGHT model \citep{2005MNRAS.358..363C}. Since our model is a bi-component model (young and old) and the STARLIGHT model is a multicomponent model, our concepts `postburst component' and `mass fraction' of the two components are non-existent in the STARLIGHT model.  But we can compare the masses derived from the spectral fits, including both young and old stars. We ran about 500 galaxy spectra with STARLIGHT and then compared the derived spectro-photometric masses.  Fig. \ref{massu_masss} shows the relation between our fiber stellar masses and those derived from STARLIGHT. As mentioned above, we used the diet Salpeter IMF to derive our masses while STARLIGHT uses the Chabrier IMF \citep{2003PASP..115..763C}. Considering that the models have such different approaches in the derivation of the age and mass distribution of the stellar content one must say that the fit is quite satisfactory (1$\sigma \sim$ 0.13 dex). The size of the symbols are roughly proportional to the inverse of the dust attenuation. Large symbols thus means the the correction for dust attenuation is small. It seems that dusty galaxies with low masses systematically deviate so that our masses tend to be higher that those derived with STARLIGHT. 

\subsection{Metallicities}

\begin{figure}[h]
\centering
 \includegraphics[width=0.85\columnwidth]{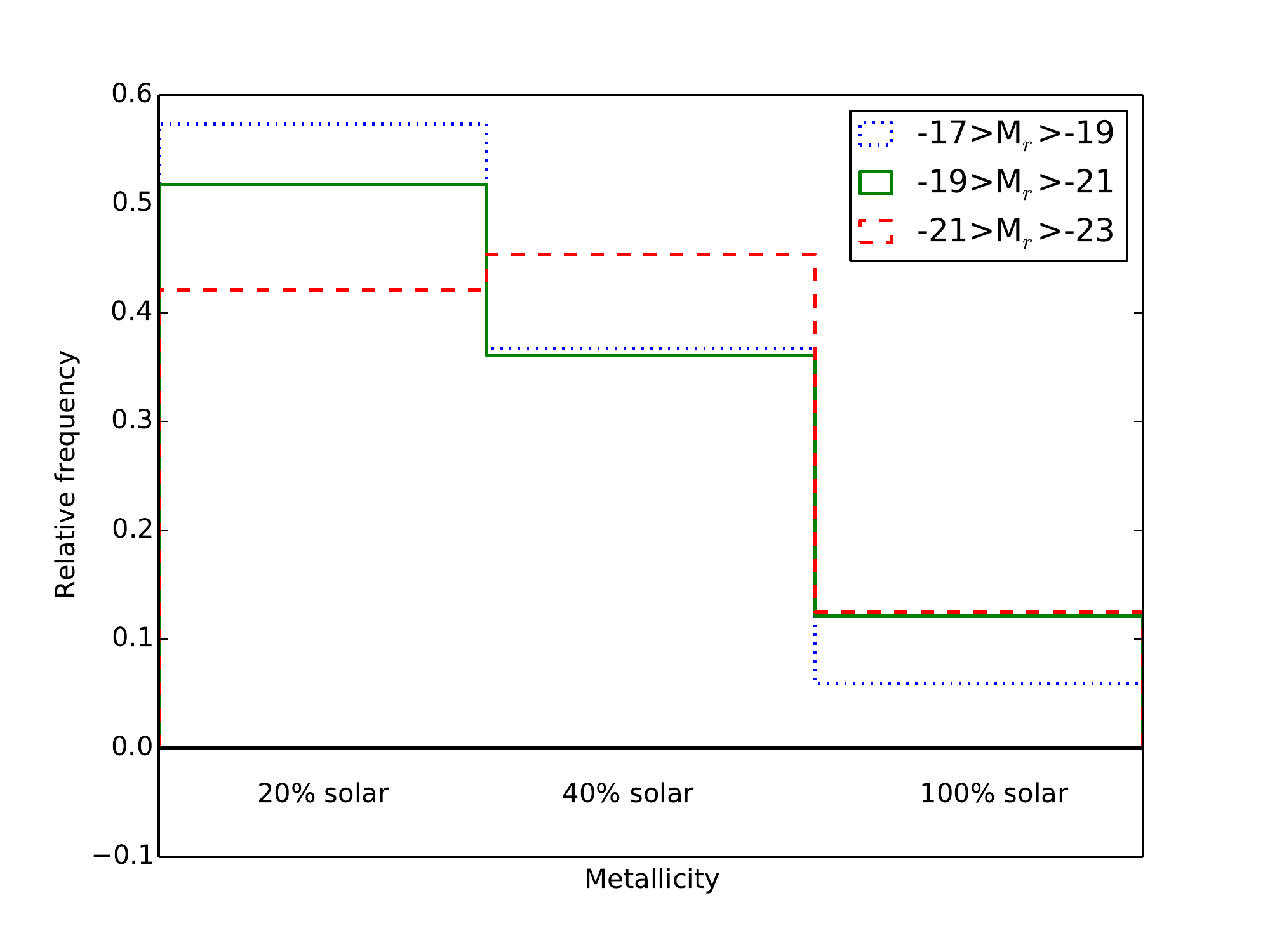}
\caption{Metallicities of the models that gave the best fits to the observed spectra. The metallicities available in the model are given at the bottom of the figure in solar metallicity units. The sample has been divided into three bins with respect to absolute magnitude in the $r$ band. The galaxies tend to become more metal rich with increasing luminosity.}
\label{metal}
\end{figure}

We have used three different metallicities in the model, 20\% solar, 40\% solar and 100\% solar. First we wish  to see if the given metallicities correspond reasonably well to the metallicities obtained from the analysis of emission lines. In Fig.~\ref{metal} we see histograms of the distribution in metallicity for three different luminosity bins. Even though we cannot really make rough determinations of metallicities from our results, it is nevertheless interesting to see it as a check of how reliable our models are. As one can see from the diagram, the model mean metallicities increase with luminosity, as expected. 

\end{document}